\documentclass[useAMS,usenatbib,usegraphicx,uselongtable]{mn2e}
\usepackage{lscape}

\title[The radial distribution of dust species in young brown dwarf disks]{The radial distribution of dust species in young brown dwarf disks}
\author[Riaz et al.]
{B. Riaz,$^{1}$ M. Honda,$^{2}$
H. Campins,$^{3}$ G. Micela,$^{4}$
M. G. Guarcello,$^{5}$ T. Gledhill,$^{1}$
 \newauthor J. Hough,$^{1}$ E. L. Mart\'{i}n$^{6}$ \\
$^{1}$Centre for Astrophysics Research, Science \& Technology Research Institute, University of Hertfordshire, Hatfield, AL10 9AB, UK \\
$^{2}$Department of Information Sciences, Kanagawa University, 2946 Tsuchiya, Hiratsuka, Kanagawa 259-1293, Japan \\
$^{3}$Physics Department, University of Central Florida, Orlando, FL, 32816, USA \\
$^{4}$INAF - Osservatorio Astronomico di Palermo, Piazza del Parlamento 1, 90134 Palermo, Italy \\
$^{5}$Smithsonian Astrophysical Observatory, MS-3, 60 Garden Street, Cambridge, MA 02138, USA \\
$^{6}$Centro de Astrobiolog\'{i}a (CSIC/INTA), 28850 Torrej\'{o}n de Ardoz, Madrid, Spain\\
 }

\begin{document}

\date{}

\pagerange{\pageref{firstpage}--\pageref{lastpage}} \pubyear{2002}

\maketitle

\label{firstpage}

\begin{abstract}

We present a study of the radial distribution of dust species in young brown dwarf disks. Our work is based on a compositional analysis of the 10 and 20$\micron$ silicate emission features for brown dwarfs in the Taurus-Auriga star-forming region. A fundamental finding of our work is that brown dwarfs exhibit stronger signs of dust processing in the cold component of the disk, compared to the higher mass T Tauri stars in Taurus. For nearly all of our targets, we find a flat disk structure, which is consistent with the stronger signs of dust processing observed in these disks. For the case of one brown dwarf, 2M04230607, we find the forsterite mass fraction to be a factor of $\sim$3 higher in the outer disk compared to the inner disk region. Simple large-scale radial mixing cannot account for this gradient in the dust chemical composition, and some local crystalline formation mechanism may be effective in this disk. The relatively high abundance of crystalline silicates in the outer cold regions of brown dwarf disks provides an interesting analogy to comets. In this context, we have discussed the applicability of the various mechanisms that have been proposed for comets on the formation and the outward transport of high-temperature material. We also present {\it Chandra} X-ray observations for two Taurus brown dwarfs, 2M04414825 and CFHT-BD-Tau 9. We find 2M04414825, which has a $\sim$12\% crystalline mass fraction, to be more than an order of magnitude brighter in X-ray than CFHT-BD-Tau 9, which has a $\sim$35\% crystalline mass fraction. Combining with previous X-ray data, we find the inner disk crystalline mass fractions to be anti-correlated with the X-ray strength.

\end{abstract}


\section{Introduction}

A study of the compositional alteration of dust grains in protoplanetary disks allows us to understand the conditions that initiate the formation of planets. The most dominant dust species in the circumstellar material are oxygen-rich silicates, that are either olivines (Mg$_{2x}$Fe$_{2(1-x)}$SiO$_{4}$), ranging from fayalite ($x = 0$) to forsterite ($x = 1$), or pyroxenes (Mg$_{x}$Fe$_{(1-x)}$SiO$_{3}$), ranging from ferrosillite ($x = 0$) to enstatite ($x = 1$) (e.g., Kessler-Silacci et al. 2005). The spectra of silicate dust have vibrational spectral features near 10$\micron$ (covering $\sim$8-14$\micron$ wavelengths), due to the Si-O stretching mode, and near 20$\micron$ (covering $\sim$15-35$\micron$ wavelengths), due to the O-Si-O bending mode. Several previous studies on silicate dust composition in circumstellar disks have been based on the 10$\micron$ Si-O feature, which probes the warm inner regions at a range of a few 100K to $\sim$1000 K and radii of within a few AU in a typical T Tauri disk. In comparison, the 20$\micron$ spectral feature probes cold dust in the outer zone of the disk, where temperatures are expected to be less than a few hundred kelvin. Both features probe the optically thin surface layers of the disk. This difference in the location of the silicate emission zone can be utilized to study any radial dependence of the dust chemical composition, as well as processes such as turbulent diffusion and large-scale circulation that would result in the mixing of disk material from the warm inner to the cold outer regions, thereby reducing any radial concentration gradients. 

Silicate dust in the interstellar medium is observed to be almost entirely amorphous in composition, with negligible crystalline fractions ($<$1-2\%; Kemper et al. 2004; Li et al. 2007). Amorphous dust can be thermally annealed into crystalline dust at temperatures above $\sim$800 K (e.g., Fabian et al. 2000). However, crystalline silicates have been found in high abundances in comets and meteorites, which are formed in regions much colder than 800 K. This suggests that some radial mixing took place between the inner and outer parts of the Solar nebula during the early formation phase of our Solar System (e.g., Nuth et al. 2000; Wooden 2008). The presence of crystalline grains in cold regions could also be due to in situ heating events, such as the disk shocks proposed by Harker \& Desch (2002). Recent compositional studies based on both the 10 and 20$\micron$ silicate features for large samples of solar mass T Tauri stars have indicated nearly equal, and in a few cases, even higher crystalline grain abundances in the cold component of the disk compared to the inner warm regions (e.g., Sargent et al. 2009; Kessler-Silacci et al. 2006; Watson et al. 2009). This has been explained by thermal annealing in the warm part followed by efficient radial transport to the cold outer disk. The analysis of Stardust samples from comet Wild 2 also support the view of processing of silicates near the Sun and then extensive mixing with much lower temperature materials such as cometary ices (e.g., Brownlee et al. 2006; Zolensky et al. 2006). Large-scale radial mixing, however, cannot explain radial gradients in dust chemical composition as observed for a few cases among solar-type stars and Herbig Ae/Be stars (e.g., van Boekel et al. 2004; Bouwman et al. 2008). A radial dependence in the relative abundances of the crystalline dust species instead suggests a localized crystalline formation mechanism. A local temporal heating process such as shock heating or grain-grain collision could raise the temperature sufficiently to cause crystallization, followed by slow cooling in an optically thick and turbulent disk environment (e.g., Molster et al. 1999). However, Harker \& Desch (2002) have addressed the cooling issue for heated grains in shocks in (ordinarily) cool disk regions. They conclude that making the region optically thick does not slow down the cooling enough. Desch \& Connolly (2002) conclude that it is the hot gas that keeps the grains warm, while the gas cools slowly, at rates comparable to the slow rates of cooling inferred for chondrules. Alternately, crystallization process could occur in low-temperature environments induced by exothermic chemical reactions (e.g., Tanaka et al. 2010). Such external processes could effect the timescales over which dust processing proceeds in the different regions of the disk. 

In the sub-stellar mass domain, Riaz (2009; hereafter R09) presented compositional fits to the 10$\micron$ silicate emission features for 23 brown dwarf disks in the Taurus-Auriga ($\sim$1 Myr; Kenyon \& Hartmann 1995) and Upper Scorpius ($\sim$5 Myr, Preibisch \& Zinnecker 1999) star-forming regions. Collectively, this sample provided an opportunity to probe any age dependence of the different dust processing mechanisms. We had found the detection rate for emission in the 10 $\micron$ silicate feature to decline with age, with only 20\% of the brown dwarfs at $\sim$5 Myr showing any detectable emission in the feature. The median crystalline mass fraction for the Taurus brown dwarfs was found to be a factor of $\sim$2 higher than the median reported for the higher mass stars in Taurus, indicating crystallinity to be more prominent in the warm inner regions of circum(sub)stellar disks. For most objects, we had obtained nearly equal mass fractions for the large-grain and crystalline grains, and only 5\% of the Taurus brown dwarfs were found to be dominated by pristine ISM-like dust. This indicated that both the grain growth and crystallization processes are active in these disks, and significant dust processing has already occurred for most systems by a young age of $\sim$1 Myr. The R09 work was focused on the 10$\micron$ feature only, and we could not study any radial dependence of the dust chemical composition. Among brown dwarf disks, there have been mainly two studies that have focused on the 20$\micron$ silicate emission feature. Mer\'{i}n et al. (2007) and Bouy et al. (2008) presented compositional studies of two young brown dwarf disks in the Lupus and Taurus star-forming regions, respectively, and found weak evidence of a decline in grain growth or the crystallinity levels when comparing the mass fractions in the inner and the outer parts of the disk. In the present study, we have analyzed 20$\micron$ silicate spectra for 13 brown dwarf disks in Taurus, which is the largest sample studied so far to probe the cold dust composition. This provides us with an opportunity to study the radial distribution of dust species for a large sample of brown dwarf disks, and to investigate if the crystalline formation and radial transport mechanisms are as efficient in the sub-stellar domain as observed among the T Tauri systems. 



\section{Observations and Data Reduction}
\subsection{{\it Spitzer} IRS Observations}

We searched the {\it Spitzer} IRS archives for 20$\micron$ silicate spectra for the 20 Taurus brown dwarf disks studied in R09. These sources are spectroscopically confirmed members of Taurus with spectral types (SpT) of $\geq$ M6, making these more likely to be substellar objects (e.g., Luhman 2004), and are known to show excess emission in the {\it Spitzer}/IRAC and MIPS bands, thus confirming the presence of disks around them (Luhman et al. 2010). We were able to find 20$\micron$ spectra for 18 out of the 20 sources. The {\it Spitzer} PIDs for these observations are 30540, 2, 248, and 20435. Five of these show `flat' features at both 10 and 20$\micron$, i.e., there is no detectable emission in the silicate features and only the underlying continuum is detected. We have not considered these flat spectra in our study. Our work is thus based on 20$\micron$ spectra for a sample of 13 brown dwarf disks. The spectra were obtained in the IRS low spectral resolution modules, Short-Low [SL] and Long-Low [LL] ($\lambda$/$\delta$ $\lambda$ $\sim$ 90). The SL module covers 5.2-14.5 $\micron$ in 2 orders, and the LL module covers 14-38 $\micron$, also in 2 orders. The observations were obtained at two nod positions along each slit. The Basic Calibrated Data (BCD) were produced at the Spitzer Science Center (SSC) by the S15.3 pipeline for the SL module and the S17.2 pipeline for the LL module. The sky background was removed from each spectrum by subtracting observations taken in the same module, but with the target in the other order. The spectra were extracted and calibrated using the Spitzer IRS Custom Extraction (SPICE) software provided by the SSC. The extraction was performed using a variable-width column extraction that scales with the width of the wavelength-dependent point spread function. We used the SPICE ``optimal'' extraction algorithm that yields a minimal-noise, unbiased flux estimate in each wavelength bin by weighting the extraction by the object profile and the signal-to-noise of each pixel. The optimal extraction provides greater improvement over regular extraction for low signal-to-noise ratio (S/N) data. Gains in S/N up to a factor of 2 have been achieved for sources with S/N $\sim$ 3, corresponding to an effective quadrupling of the exposure time. The rectified flux and uncertainty products written out by the optimal extraction algorithm include corrections for any offsets between different modules. Further details on the optimal extraction method are provided in the SPICE documentation. We have not applied any extinction correction since Taurus suffers from little reddening ($A_{V}<$1 mag; e.g., Luhman et al. 2010). 


\subsection{{\it Chandra} X-ray Observations}

The brown dwarfs 2M04414825 and CFHT-BD-Tau 9 have been observed with Chandra/ ACIS-S in the {\it very faint} mode, using the back-illuminated CCD S3 (OBSID 12335 and 12336; P.I. Riaz). Data reduction has been made with {\it CIAO 4.3} packages (Fruscione et al. 2006). The level 2 events files have been obtained from the level 1 events files retaining only events with grade=0, 2, 3, 4, 6, $status=0$, energy ranging from 0.3 keV to 8.0 keV, and removing the events position randomization introduced by the Standard Data Processing, using the tasks {\tt acis process events} and {\tt dmcopy}. We calculated the exposure maps with the {\tt merge all} task. 

The source detection has been performed with PWDetect (Damiani et al. 1997), using a threshold of $4.8 \sigma_{sky}$, corresponding to one spurious source for each exposure.  Both targets have an X-ray counterpart, with an offset equal to $0.22^{\prime\prime}$ (CFHT-BD-Tau 9) and $0.14^{\prime\prime}$ (2M04414825). CFHT-BD-Tau 9 is a weak source with a significance of 6, while 2M04414825 has a bright counterpart (significance $>$ 50 ). 

Photons extraction has been made using the IDL software ACIS Extract\footnote{http://www2.astro.psu.edu/xray/docs/TARA/ae\_users\_guide.html} ($AE$, Broos et al. 2002), which uses TARA\footnote{http://www.astro.psu.edu/xray/docs/TARA}, CIAO, FTOOLS\footnote{http://heasarc.gsfc.nasa.gov/docs/software/ftools} and MARX\footnote{http://space.mit.edu/CXC/MARX/}. AE calculates the Point Spread Function (PSF) for each source, and the source events are extracted in a region encompassing the 90\% of the PSF evaluated at 1.49
keV. Background events have been extracted for each source in circular annuli centered on the sources, wich have been masked first covering the 99\% of the local PSF and then redefining a more accurate mask region. 

The X-ray counterpart of 2M04414825 is bright enough to determine its parameters (the hydrogen column along the line of sight, $nH$ and the emitting plasma temperature, $kT$) fitting the observed spectra using XSPEC\footnote{http://heasarc.nasa.gov/xanadu/xspec/} v.12.6.0 (Arnaud 1996). We assumed a plasma in collisional ionization equilibrium thermal model, with one temperature. The best fit model ($\chi^{2}_{\nu}=0.93$) is a 1T thermal model with $kT=1.5\pm0.2\,keV$ and $N_{H}=0.36\pm0.09\times10^{22}\,cm^{-2}$, corresponding to a visual extinction of $A_V=2.01$, compatible with previous estimates of the extinction of Taurus members (e.g., Luhman et al. 2010).  Using the 1T model and adopting a distance of 2M04414825 equal to 140pc  (Kenyon et al. 1994), we obtained an X-ray luminosity equal to $log\left( L_X \right)=30.31\,erg/s$. 

The few counts detected from the X-ray counterpart of CFHT-BD-Tau 9 prevent us to perform a spectral fitting. In order to estimate the X-ray luminosity of this source, we set nH to the value derived from the typical extinction of Taurus members ($A_V=2\pm1$), corresponding to $N_{H}=0.36\pm0.09\times10^{22}\,cm^{-2}$. We adopt a coronal temperature of $kT=1\,keV$ and find a conversion factor (PIMMS) of $3.9 \cdot 10^{-12} erg cm^{-2}$, that translates in $log(L_{X})=28.30\,erg\,cm^{-2}$. Note that the conversion factor may vary not more than 20\% in the 2MK - 20 MK range, while the statistical error ($\sim 40\%$ ) dominate the final uncertainty in Lx. Table \ref{resume} lists the properties of the X-ray counterparts of the two brown dwarfs described in this section.



\section{Analysis}
\subsection{Modeling of Silicate Emission Spectra}
\label{silicate}

The 10 and 20$\micron$ features probe the surface layers at different radii in the disk. For a typical T Tauri disk, 95\% of the flux at 10$\micron$ comes from within $\sim$1 AU, while 95\% of the 20$\micron$ flux comes from within $\sim$10 AU (Kessler-Silacci et al. 2006). Thus with a factor of $\sim$2 increase in the peak wavelength of the silicate feature (10$\micron$ to 20$\micron$), there is a shift by a factor of $\sim$10 in the disk radii from which most of the emission arises. There is a difference in the location of the silicate emission zone for stars of different luminosities. For a brown dwarf disk, 95\% flux at 10$\micron$ arises from smaller radii of $\leq$0.1 AU, while 95\% flux at 20$\micron$ arises from radii within $\sim$1-2 AU (Kessler-Silacci et al. 2007; Bouy et al. 2008). For any system, both silicate features probe the surface layers of the disk. For a young brown dwarf disk, the location of the 20$\micron$ silicate emission zone at relatively larger disk radii implies cooler surface temperatures of $<$100 K, while smaller radii of $\leq$0.1 AU probed by the 10$\micron$ feature imply warmer surface temperatures of 300 K or higher (e.g., Mer\'{i}n et al. 2007). At the inner edge of the disk close to the central substellar source, dust sublimation temperatures of $\sim$1500 K can be reached (e.g., Whitney et al. 2003). Due to this, the silicate emission zone probed by the 10$\micron$ feature is usually referred to as the `warm' or `inner' component of the disk, while the region probed by the 20$\micron$ feature is referred to as the `cold' or `outer' component. The dust temperature, however, is not expected to be constant and could vary by more than a $\sim$100 K over a certain silicate emission zone. 

\begin{figure*}    
      \includegraphics[width=80mm]{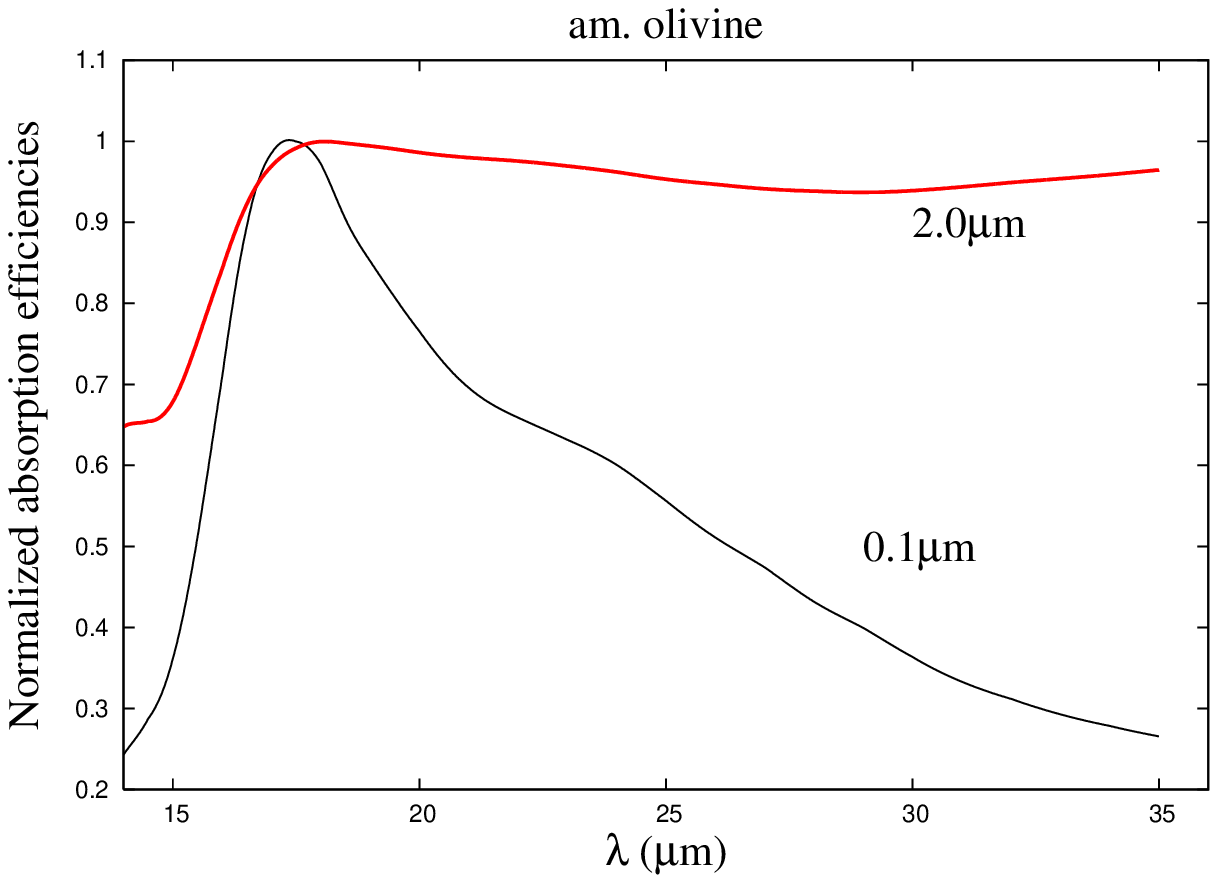} 
      \includegraphics[width=80mm]{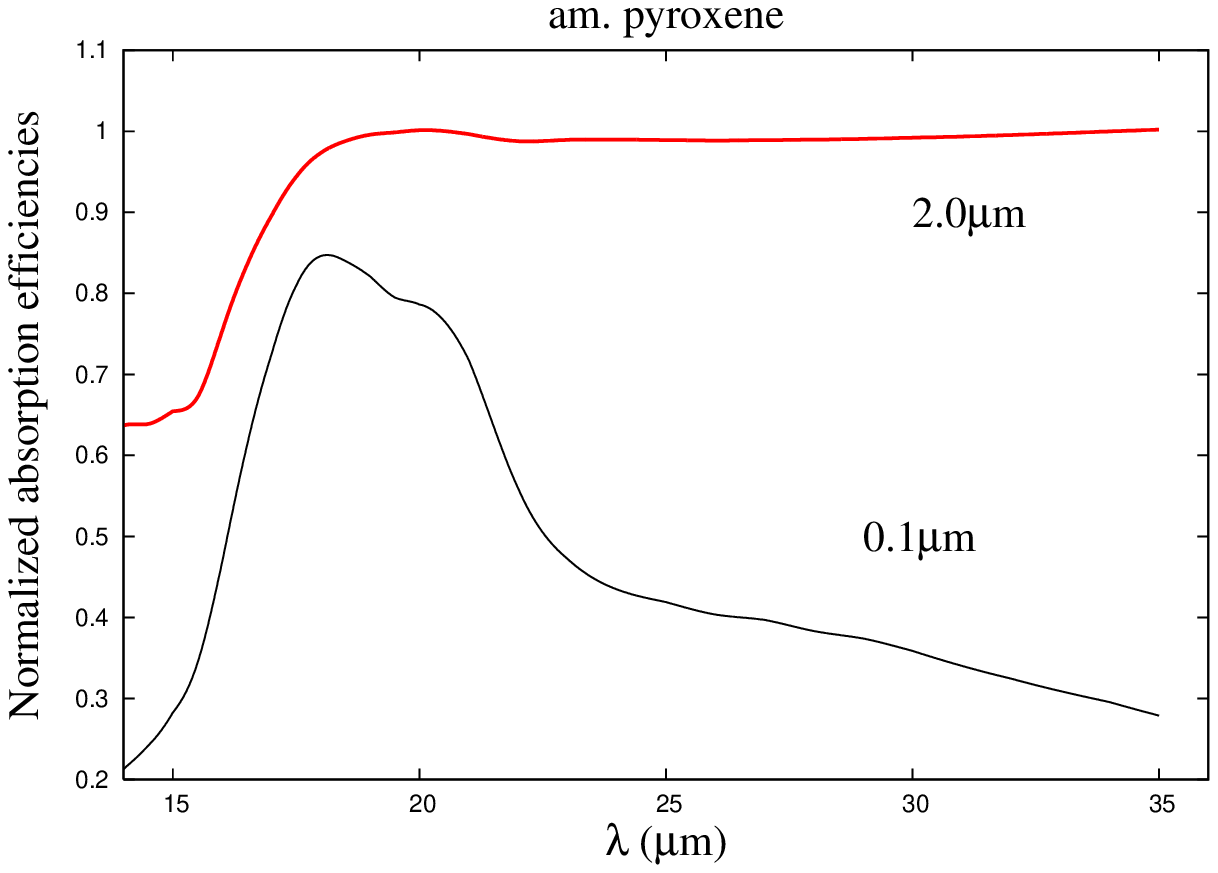} \\      
            \includegraphics[width=80mm]{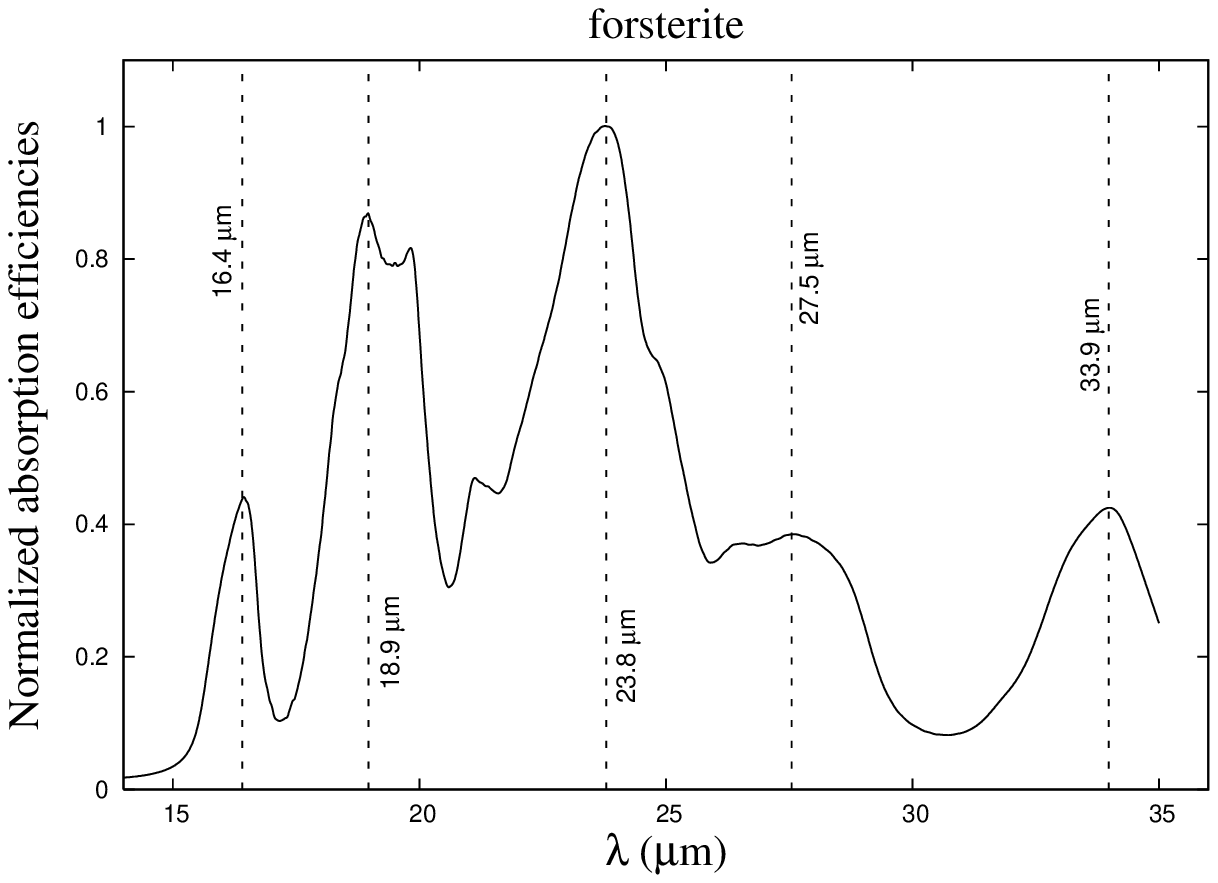} 
            \includegraphics[width=80mm]{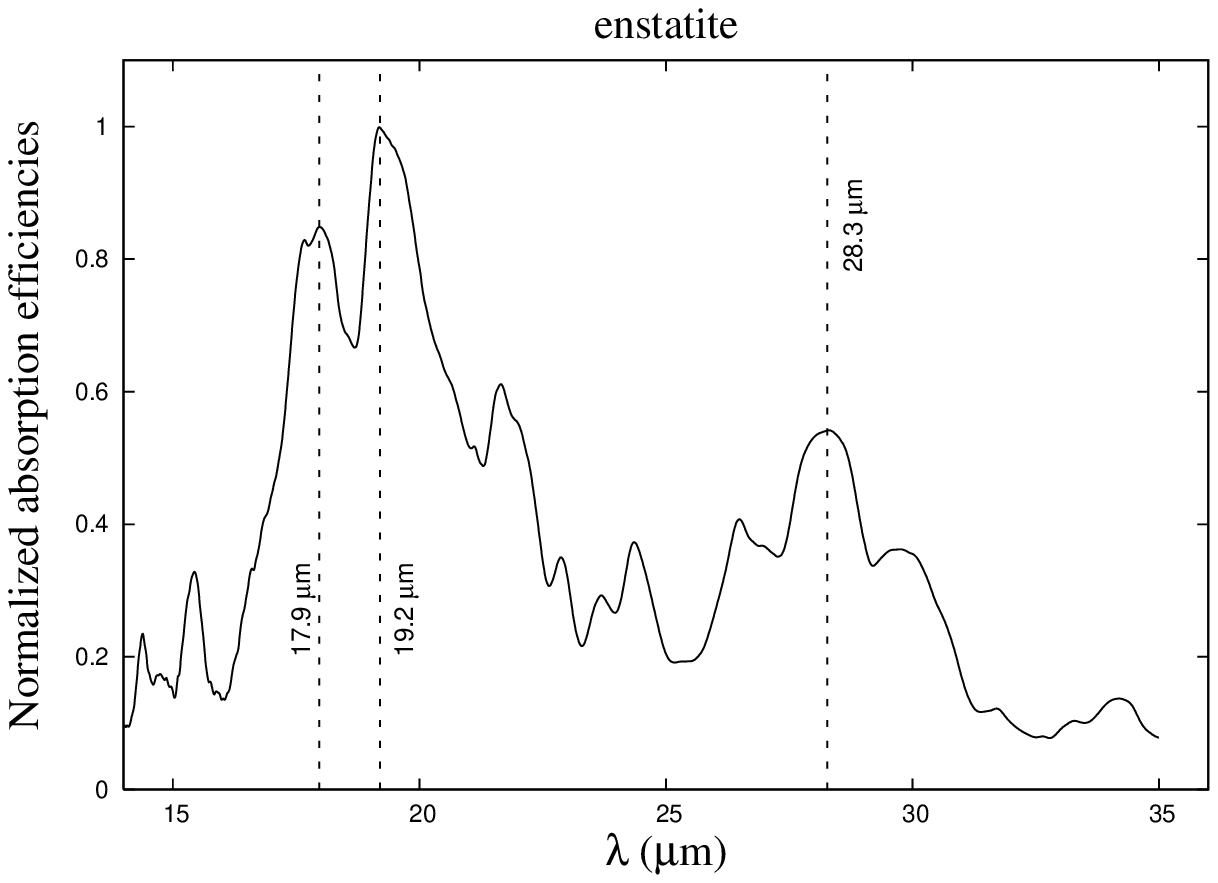} \\
            \includegraphics[width=80mm]{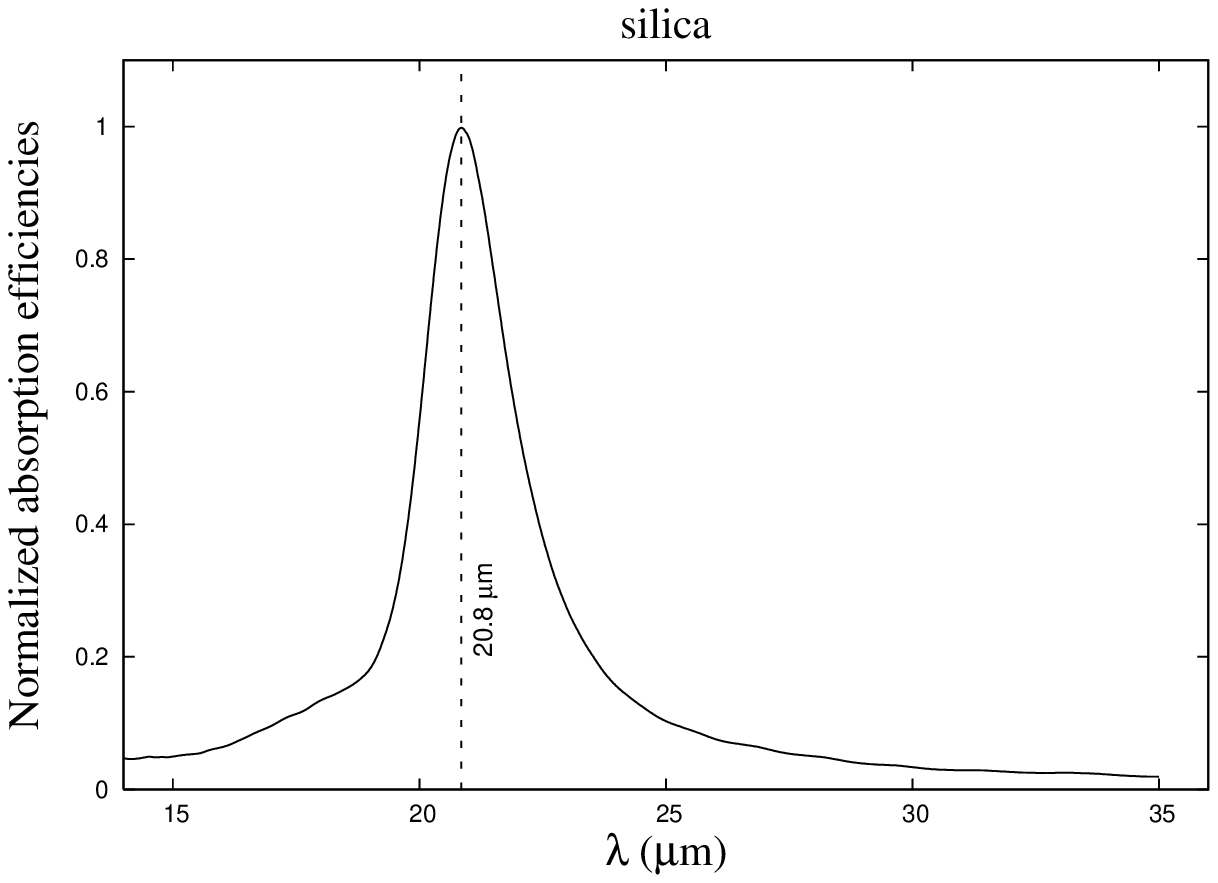} \\            
     \caption{Normalized spectral profiles for the five dust species considered for modeling. Upper panel shows the profiles for small and large amorphous olivine and pyroxene silicates for grain sizes of 0.1$\micron$ ({\it black}) and 2.0$\micron$ ({\it red}). Middle and bottom panels show the profiles for forsterite, enstatite and silica grains. Prominent emissivity peaks are marked by dashed lines.   } 
    \label{species}
 \end{figure*}

We used a similar method as described in R09 to model the 20$\micron$ silicate emission features for the targets. Five dust species were considered in order to obtain a compositional fit to the observed spectra: amorphous olivine and pyroxene, crystalline enstatite, crystalline forsterite, and silica. The characteristics of these are outlined in Table~\ref{density}, and the spectral profiles plotted in Fig.~\ref{species}. The bulk densities listed in Table~\ref{density} have been obtained from Bouwman et al. (2001). The amorphous grains show little dependence on shape (e.g., Bouwman et al. 2001), and thus can be assumed to be homogeneous and spherical so that the standard Mie theory can be applied to determine their spectroscopic properties. The properties of crystalline silicates however are sensitive to the grain shape which makes it important to consider shapes other than homogeneous spheres. Min et al. (2003) have studied the absorption and scattering properties of particles that are inhomogeneous in structure and composition, based on the assumption that the characteristics of such irregularly shaped particles can be simulated by the average properties for a distribution of shapes, such as ellipsoids, spheroids or hollow spheres. We considered both the distribution of hollow spheres (DHS) and the continuous distribution of ellipsoids (CDE) from Min et al. (2003), and found a better match both in terms of the location and width of the spectral profiles using the CDE routine. Therefore the absorption efficiencies of crystalline silicates (enstatite and forsterite) and silica have been calculated using ellipsoidal grains. Since the CDE method has been used, the grains are treated in the ``Rayleigh limit'', i.e. the grain sizes are assumed to be much smaller than the wavelength of radiation. Our model is therefore biased towards sub-micron (0.1 $\micron$) sized crystalline grains, and the effects of grain growth could not be studied for these species. For the case of amorphous silicates, two grain sizes of 0.1 and 2.0 $\micron$ have been used. As discussed in Bouwman et al. (2001), the optical properties of dust grains with a range in sizes between 0.01 and 5 $\micron$ can be characterized by two typical grain sizes: 0.1 $\micron$ for ``small'' grains with sizes $<$ 1 $\micron$, and 2.0 $\micron$ for ``large'' grains with sizes $>$1 $\micron$. 

To construct a model spectrum, we have followed the method outlined in R09 and Honda et al. (2003), wherein a power-law source function is assumed and the model flux $F_{\lambda}$ is the sum of the emission from a featureless power-law continuum and the emission from the different dust species. We have considered a power-law continuum since it represents a sum of blackbody emission at various temperatures. It is unlikely for the dust temperature to remain constant over a $\sim$20$\micron$ wide spectral range. A power-law continuum provides a better approximation than using a single blackbody function at a fixed temperature to fit the continuum. The model spectrum can be written as:

\begin{eqnarray*}
\lambda F_{\lambda} = a_{0} \left(\frac{\lambda}{20 \micron}\right)^{n} +~ (a_{1} Q_{0.1ao} ~+~ a_{2} Q_{2.0ao} \\ 
~+~a_{3} Q_{0.1ap} ~+~ a_{4} Q_{2.0ap} ~+~ a_{5} Q_{ens} ~+~ a_{6} Q_{fors} \\
~+~ a_{7} Q_{sil})~ \left(\frac{\lambda}{20 \micron}\right)^{m} ~~~~~~(1)
\end{eqnarray*}

\noindent where $Q_{0.1ao}$ and $Q_{2.0ao}$ are the absorption efficiencies for 0.1 and 2.0 $\micron$ amorphous olivine, $Q_{0.1ap}$ and $Q_{2.0ap}$ are the absorption efficiencies for 0.1 and 2.0 $\micron$ amorphous pyroxene, and  $Q_{ens}$, $Q_{fors}$, and $Q_{sil}$ are the absorption efficiencies for enstatite, forsterite and silica, respectively. The free parameters in the model are the multiplicative factors $a_{1}$-$a_{7}$ in $10^{-14}$ W $m^{-2}$, and two spectral indices of the source function, {\it m} and {\it n}. To correctly define the continuum level, we fixed the parameter $a_{0}$ to the observed flux density at 14-15.5$\micron$. We selected this wavelength range to be the continuum-fiting region as it is least affected by features in the spectra. To fit the model spectrum, we used the $\chi^{2}$-minimization method outlined in van Boekel et al. (2005). The best-fit values thus obtained for the multiplicative factors were used to derive the mass fractions. Table~\ref{results} lists the mass fractions for the amorphous and crystalline silicates. The 10$\micron$ mass fractions are from R09. The uncertainties for the fit parameters were obtained by considering the errors on the observed fluxes, as well as the S/N ratio. The errors on the observed fluxes are between $\sim$1-8\% of the observed flux density, but are higher (15-20\%) in the $\sim$30-35$\micron$ wavelength range. We used the Monte Carlo method for error estimation as outlined in van Boekel et al. (2005), and discussed further in Juhasz et al. (2009). In this method, random Gaussian noise is added to the spectrum, with 1000 synthetic spectra generated and fitted at each noise level. We considered four different noise levels with a $F_{\nu}^{error}/F_{\nu}^{observed}$ of 0.2, 0.1, 0.01 and 0.001, corresponding to a S/N ratio of 5, 10, 100 and 1000, respectively. The mean of the resulting distribution of all fit parameters then corresponds to the best-fit value, and the errors are derived from the standard deviation.

In order to validate our method, we modeled some known low-mass young objects and compared our results with those published in the literature. Our mass fractions are consistent with the fractions reported by Sargent et al. (2009a) for T Tauri disks in Taurus. These authors have also considered sub-micron sized crystalline grains in their model, and have used the CDE distribution for grain opacities, similar to the distribution used in our model. As discussed in R09 and Sargent et al. (2009a), large grains can be considered as heterogenous aggregates, made up of sub-micron sized amorphous silicates, forsterite or enstatite, with the highest abundance of amorphous grains. The opacities of the forsterite components of such large heterogenous grains have been shown by Min et al. (2008) to resemble the opacity of small homogenous forsterite grains, whereas the opacity of the amorphous component of the aggregate is found to resemble that of large amorphous homogenous grains, with a size comparable to that of the heterogenous aggregate. Therefore modeling the crystalline component of a large heterogenous aggregate with small crystalline grains should not result in largely erroneous mass fractions for the crystalline silicates. This is consistent, for e.g., with the findings of Bouwman et al. (2008), who have considered three different grain sizes of 0.1, 1.5 and 6.0 $\micron$ in their models, but find the typical sizes from the model fits to be sub-micron for the crystalline silicates, and 6 $\micron$ for the amorphous grains. We modeled some of the forsterite exemplars from the study of Sargent et al. (2009a), such as DK Tau and GN Tau. We find a crystalline mass fraction of 38$\pm$10\% and 55$\pm$12\% for DK Tau and GN Tau, respectively, with a small and large amorphous grain mass fraction of 12$\pm$8\% and 42$\pm$13\% for DK Tau and 20$\pm$10\% and 7$\pm$5\% for GN Tau. For sources such as DM Tau or DO Tau that show negligible signs of grain growth in the disk, we find a small grain mass fraction of $\sim$80-82\% and a crystalline fraction of 15$\pm$9\%, with a large-grain fraction of $<$5\%. These fractions are consistent within the uncertainties with the values listed in Sargent et al. (2009a) for the cold component of the disk. 


We also modeled the Taurus brown dwarf 2M04442713 discussed in Bouy et al. (2008), and obtained a large-grain and a crystalline mass fraction of 68$\pm$8\% and 30$\pm$11\%, respectively. These values are consistent with the mass fractions listed in Bouy et al. (2008) for the low-temperature polynomial continuum case. For the very low-mass star SST-Lup3-1 discussed in Mer\'{i}n et al. (2007), we find a crystalline and large-grain mass fraction of 38$\pm$13\% and 22$\pm$10\%, respectively, which are comparable to the fractions reported by these authors for the high-continuum case. We note that a direct comparison of our results with those reported in Bouy et al. and Mer\'{i}n et al. is not appropriate, since these works have considered a different set of grain opacities (the DHS model), and therefore the input optical constants are different than the ones considered in our model. 

\subsection{Disk Modeling}
\label{diskmodel}

We have used the 2-D radiative transfer code by Whitney et al. (2003) for disk modeling. The circumstellar geometry consists of a rotationally flattened infalling envelope, bipolar cavities, and a flared accretion disk in hydrostatic equilibrium. The disk density is proportional to $\varpi^{-\alpha}$, where $\varpi$ is the radial coordinate in the disk midplane, and $\alpha$ is the radial density exponent. The disk scale height increases with radius, $h=h_{0}(\varpi / R_{*})^{\beta}$, where $h_{0}$ is the scale height at $R_{*}$ and $\beta$ is the flaring power. For the stellar parameters, we used the $T_{eff}$-SpT relation from Luhman et al. (2003) to determine the stellar effective temperature. We have considered an age of $\sim$1 Myr for the Taurus-Auriga cluster (Kenyon \& Hartmann 1995). The stellar mass and radius were obtained by considering the 1 Myr isochrone from Baraffe et al. (2003). The NextGen (Hauschildt et al. 1999) atmosphere file for the appropriate $T_{eff}$ and log {\it g} = 3.5 was used to fit the atmosphere spectrum of the central sub-stellar source. A distance of 140 pc (Kenyon et al. 1994) was used to scale the output fluxes from the models to the luminosity and distance of the brown dwarfs. 

Figure~\ref{modelfits} (right panel) shows the disk model fits obtained for the targets. Also indicated are the separate contributions from the disk and the stellar photosphere. The {\it Spitzer} IRAC and MIPS data points shown in the fit are from Luhman et al. (2010), the near-infrared points are the 2MASS data, and the optical data has been obtained from SIMBAD. Detailed discussions on the fitting procedure and the variations in the model SEDs with the different disk parameters are provided in Riaz \& Gizis (2007). Here we provide a brief description of the best model-fits (based on the lowest reduced-$\chi^{2}$ value) obtained for these brown dwarfs. 

The disk models provide the capability to include different grains in different disk regions (see Whitney et al. 2003, Fig. 1). There are three grain models supplied by the code: `large' grains with a size distribution that decays exponentially for sizes larger than 50 $\micron$ up to 1 mm, the `medium' sized grains of sizes with $a_{max}\sim$ 1 $\micron$, and `small' ISM-like grains with $a_{max} \sim$ 0.25 $\micron$. All three models consider a minimum grain size of 0.0025$\micron$. The large-grain model used by Whitney et al. (2003) is the same as used by Wood et al. (2002) to fit the HH 30 disk SED. The medium grain size distribution is the same grain model that Cotera et al. (2001) have used to model the HH 30 near-IR scattered light images. The ISM grain model is the same as used by Kim et al. (1994) from their mass distribution for the fit to the canonical (R$_{v}$=3.1) average diffuse ISM. All three models are based on the method of a power-law with exponential decay (PED) above some cutoff grain size a$_{c}$, as discussed in Kim et al. (1994; Equation 3) and Wood et al. (2002; Equation 4). For the large grain model, the values for the exponents p and q of the PED are 3 and 0.6, respectively, and the cutoff and maximum grain sizes are a$_{c}$=50$\micron$ and a$_{max}$=1000$\micron$. For the medium grain size model, q=1, p=3.5, a$_{c}$=0.55$\micron$ and a$_{max}$=20$\micron$. For the diffuse ISM model, a$_{c}$=0.02$\micron$ and a$_{max}$=0.25$\micron$, with the PED exponent $\gamma$= -3.06 for silicate grains (Eqn. 3 of Kim et al. 1994). 

We have varied these grain models in the disk midplane and the upper atmosphere in order to obtain a good fit to the silicate emission features and the overall SED. The model used here assigns large grains to the high density regions. Due to this, when large grains are used in the disk midplane, similar grains are placed close to the inner wall since it is of high density. This affects the emission from the inner wall and hence the observed flux at 10$\micron$. Thus by using sub-micron sized grains in the disk midplane, we have indirectly reduced the inner wall grain size, and were able to obtain a good fit to the 10$\micron$ silicate feature. There could still be bigger grains in the really dense midplane region of the disk, the presence of which can be confirmed with far-IR/sub-mm observations. On the other hand, if this were a ``three-layered'' disk model instead of a ``two-layered'' one, large grains could be placed in the very dense regions without affecting the 10$\micron$ feature. For most sources, good fits are obtained if sub-micron sized grains are placed in both the midplane and the disk surface layers. This produces a model SED with a more peaked emission at 10$\micron$ and a steeper slope at far-infrared/sub-mm wavelengths (e.g., Riaz \& Gizis 2007). For MHO 5 that has a very flat disk structure, the large grain model with $a_{max}$ of 1 mm was considered in both the disk midplane and the atmosphere. This results in a flat silicate feature as well as a flat sub-mm slope in the model SED. Considering that we do not have any far-infrared/sub-mm observations for these objects, the silicate features provide the only way to constrain the different grain sizes considered by these disk models. 

There are degeneracies in the disk model fits presented here, considering that we have only $\sim$5-38$\micron$ data to model these disks. The uncertainties in the stellar parameters could also result in changes in the disk parameters. There are mainly six parameters related to the disk emission in these models, that can be varied to obtain a good fit. We had fixed the disk mass and the outer disk radius given the absence of longer wavelength data. The disk outer radius was fixed at 100 AU, and the disk mass for modeling was set to 1E-4$M_{\sun}$. This is the typical disk mass obtained by Scholz et al. (2006) from 1.3 mm observations for a sample of Taurus brown dwarfs. We had fixed the disk mass accretion rate at 10$^{-10} M_{\sun} yr^{-1}$, a typical value observed among accreting T Tauri stars (e.g., Muzerolle et al. 2003). It is, however, possible to constrain the amount of flaring in the disk, the inclination angle, and the inner disk radius, using the available observations. 

The disk scale height varies as $h=h_{0}(\varpi / R_{*})^{\beta}$, where $h_{0}$ is the scale height at $R_{*}$ and $\beta$ is the flaring power. We varied both $h_{0}$ and $\beta$ to determine the amount of flaring in these disks. The variations in the model SEDs as $\beta$ is lowered are more evident for $\lambda$ $>$ 10 $\micron$, and the 24 $\micron$ observation provides a good constraint to the flaring power. The model SED is highly flared for $\beta$=1.25, which is considered a typical value for models of T Tauri disks in hydrostatic equilibrium (e.g., Walker et al. 2004), while the structure is flat for $\beta$=1.0 or smaller values (e.g., Riaz \& Gizis 2007). For most sources, we were able to obtain good fits using $\beta$$\leq$1.1 and $h_{0}$$<$1, indicating flat structures for these brown dwarf disks. 

We have explored a range of inclination angles to the line of sight. Due to binning of photons in the models, there are a total of 10 viewing angles, with face-on covering 0-18$\degr$ inclinations. The mid- and far-IR fluxes increase with decreasing inclinations, as the emission at these wavelengths is from an optically thick disk, while the optically thin millimeter fluxes are independent of the inclination (e.g., Riaz \& Gizis 2007). We obtained good fits using intermediate inclinations of 50-60$\degr$ for most disks (with a $\pm$10$\degr$ uncertainty). A few disks including V410 X-ray 6 are at a larger inclinations of $>$70$\degr$ and can be considered as edge-on systems. Four disks including MHO 5 were fitted well using face-on inclinations of $<$40$\degr$, although MHO 5 has a very flat disk and the inclination would not really matter for such a flat system. 

The inner disk radius, $R_{in}$, was varied as multiples of $R_{sub}$, where $R_{sub}$ is the dust sublimation radius and varies with the stellar radius and temperature, $R_{sub} = R_{*} (T_{sub}/T_{*})^{-2.085}$ (Whitney et al. 2003). $T_{sub}$ is the dust sublimation temperature and was set to 1600K. Increasing the inner radius to 5 or 7$R_{sub}$ results in higher fluxes near the 10 $\micron$ silicate band and at longer wavelengths. For nearly all disks, we found a good fit using $R_{in}$ = 1 $R_{sub}$ ($\sim$0.0004 AU). For the transition disk V410 X-ray 6, a larger inner disk radius of $\sim$0.05 AU provided a good fit. The best-fit values for these parameters are listed in Table~\ref{results}.

\section{Results}


\subsection{The least processed disks: V410 X-ray 6 and 2M04141760}
\label{least}

The brown dwarfs 2M04141760 and V410 X-ray 6 show a very smooth 20$\micron$ feature, with a peak near $\sim$18$\micron$ that is indicative of small amorphous olivine grains (Fig.~\ref{modelfits}a). The 10$\micron$ spectra for both of these sources are also quite smooth with a peak at $\sim$9.8$\micron$ indicative of small amorphous olivine grains, and no clear crystalline peaks are observed. We find a high small-grain mass fraction of $>$65\% in these disks, from both the 10 and 20$\micron$ features, with a $<$5\% fraction of crystalline grains and $<$15\% fraction of large grains. The inner and outer regions of these disks thus show weak signs of dust processing.  


The amount of crystalline silicates in a disk could be dependent on the presence of dense disk material in the inner disk regions. If crystalline silicates form due to thermal annealing of amorphous grains in the warm inner disk regions, and then radially transported to the outer disk, then the presence of an inner hole in the disk could result in a reduction in the crystalline silicate production. This may be the case for V410 X-ray 6. This brown dwarf has a transition disk and displays photospheric emission up to wavelengths of $\sim$8$\micron$ (Fig.~\ref{modelfits}a). From disk modeling, we estimate an inner disk hole of $\sim$118 $R_{sub}$ ($\sim$0.05 AU). In comparison, 2M04141760 has a much smaller inner disk radius of $\sim$3 $R_{sub}$ ($\sim$0.001 AU), but exhibits a similar lack of crystallinity in the disk. No clear correlation is thus seen between the presence of a large inner hole and low crystallinity in the disk. 

The disk for V410 X-ray 6 flares up at wavelengths longward of $\sim$15$\micron$ (Fig.~\ref{modelfits}a). Watson et al. (2009) find a trend in the Taurus T Tauri sample of more flattened disks having greater crystallinity. Also, Sargent et al. (2009a) find a possible weak trend in a similar sample of flattened disks tending to have higher large grain mass fractions. The lack of crystallinity in V410 X-ray 6 is consistent with a flared geometry, indicating that the correlations noted by Watson et al. and Sargent et al. are also applicable to the sub-stellar cases. As mentioned, the sub-micron amorphous grain fraction for V410 X-ray 6 and 2M04141760 is quite high ($>$60\%), and the large-grain mass fraction is only 15\% or less. These disks are at an early stage of evolution and significant dust processing has not yet occurred. This is more apparent from the SED for 2M04141760 that shows a strong excess emission even at 3.6$\micron$. Among the 65 T Tauri systems in Taurus studied by Sargent et al. (2009a) and Watson et al. (2009), 5\% are found to have high mass fractions of amorphous sub-micron grains $>$70\%, with weak signs of crystalline or large silicates. We have a $\sim$15\% fraction of such systems in our Taurus brown dwarf sample that are still dominated by ISM-like grains. 


The one property that is found to be different for these two sources is X-ray variability. Both V410 X-ray 6
and 2M04141760 show strong variations in X-ray emission, and have been classified as X-ray flaring sources (Grosso et al. 2007). In R09, we had noted a weak anti-correlation between the X-ray emission strength and the extent of crystallinity in the disk, suggesting that X-ray activity could affect the crystalline structure of the dust grains. We discuss this further in Section \S\ref{X-ray}.

\subsection{The mixed composition disk: 2M04141188}

2M04141188 shows an interesting 20$\micron$ spectrum, with weak emission between $\sim$25 and 30$\micron$. The spectrum shows a broad peak near $\sim$20$\micron$, and then a rise in emission longward of $\sim$30$\micron$. There may be a smaller peak near $\sim$33$\micron$ due to crystalline forsterite, although it may well be consistent with the point-to-point noise around that wavelength range. The broad feature near 20$\micron$ is well fit using a model with a mixture of small amorphous and crystalline enstatite and forsterite grains, along with some contribution from amorphous silica. The 10$\micron$ spectrum for 2M04141188 shows a broad, trapezoidal shape with a crystalline forsterite peak near 11.3$\micron$. Such shapes are indicative of grain growth and crystallization in the disk. We find roughly equal mass fractions of $\sim$30-50\% for the small and large amorphous, and the crystalline silicate grains. The fractions are similar in the inner and outer disk regions. This disk thus exhibits a mixed composition in both the warm and cold disk regions. 


The `dip' observed in the 20$\micron$ spectrum between $\sim$25 and 30$\micron$ could be due to obstruction by the bright edge of the disk. The best-fit disk model for 2M04141188 indicates a close to edge-on inclination of {\it i} $\sim$ 70$\degr$ (Table~\ref{results}). An absorption+emission silicate spectrum is sometimes observed for highly inclined disks, and can be explained by silicate emission from the disk and silicate absorption from the highly extinguished wall of the disk (e.g., Luhman et al. 2007). While we do not actually see an absorption component for 2M04141188 since the full spectrum lies above the continuum (Fig~\ref{modelfits}), the lack of emission indicates larger optical depth in this small disk region, which could be due to the disk's self-absorption at the nearside of the disk. However, if disk self-absorption is evident even at such long wavelengths, one may naturally expect that much more extinction should be observed at much shorter wavelength, which is not seen. It may be that significant dust sedimentation has occurred in this disk region, resulting in the weak emission observed. 



\subsection{Prominent crystalline forsterite features: 2M04230607, CFHT-BD-Tau 20 and 2M04400067}
\label{2M0423}

The 20$\micron$ spectrum for 2M04230607 shows prominent crystalline forsterite features at $\sim$23.8, 27.5 and 34$\micron$. There is another feature observed near $\sim$16$\micron$ which is also indicative of forsterite emission (Fi.g~\ref{species}). In comparison, the 10$\micron$ spectrum for 2M04230607 shows a peak near $\sim$9.3$\micron$ which indicates enstatite emission, while a smaller forsterite peak is observed at $\sim$11.3$\micron$. The crystalline mass fraction for 2M04230607 shows an increase from $\sim$38\% in the warm inner disk region to $\sim$57\% in the outer cold region. If we look at the separate contribution from enstatite and forsterite silicates, then the enstatite mass fraction has decreased by a factor of $\sim$2 in the outer disk, while the forsterite fraction has increased by a factor of $\sim$3. Such gradients in the dust chemical composition in the inner and outer disk regions have been noted before in disks around solar-type stars and Herbig Ae/Be stars (e.g., van Boekel et al. 2004; Bouwman et al. 2008). 2M04230607 provides a similar interesting case among brown dwarf disks. We discuss the formation mechanism of the different kinds of crystalline silicates in Section \S\ref{discussion}. 



The 20$\micron$ spectrum for CFHT-BD-Tau 20 also shows a forsterite feature near 34$\micron$. There is a smaller peak observed near 27$\micron$ which is also indicative of forsterite emission. The 10$\micron$ spectrum is very weak for this object (Fig.~\ref{modelfits}b), but much stronger emission is observed in the 20$\micron$ feature. This suggests the presence of a large mass of cold dust in the disk compared to warm dust in the inner regions. We had discussed in R09 the large uncertainties in the model-fits for the 10$\micron$ spectrum of CFHT-BD-Tau 20 and the derived mass fractions. The lack of warm dust emission makes it difficult to compare the inner and outer disk mass fractions. From the 20$\micron$ model fits, we find a 37\% crystalline mass fraction and a 46\% fraction of large amorphous grains. It thus exhibits a mixed composition with signs of both grain growth and crystallization in the outer disk. The weak emission from inner surface layers could be due to significant grain growth and dust settling in the inner disk regions. This suggests that dust processing may not occur on a similar timescale in the inner and outer regions for all disks. Or there may have been some secondary collisional process that produced this new generation of cold dust in the outer disk regions. CFHT-BD-Tau 20 thus presents an interesting case where the 20$\micron$ emission strength is much higher than that observed in the 10$\micron$ feature, and where large abundances of crystalline silicates are found in the outer cold disk regions.


A similar case is of 2M04400067 that shows broad forsterite features near $\sim$27 and 34$\micron$. Some forsterite emission can also be identified near 16$\micron$. The 10$\micron$ spectrum for this disk shows a broad trapezoidal shaped feature with no prominent crystalline peaks. We find a slight increase in the crystalline fraction from $\sim$22\% in the warm component to $\sim$30\% in the cold region. The inner disk region, however, shows a high fraction of small amorphous grains, with a $\sim$60\% mass fraction, while a high 63\% mass fraction is found for the large dust grains in the outer disk region. The outer cold component of this disk thus shows stronger signs of dust processing due to both grain growth and crystallization processes. 

The disk model fits for these three disks are shown in Fig.~\ref{modelfits}b (right panel). 2M04230607 shows weak excess emission shortward of $\sim$8$\micron$. CFHT-BD-Tau 20 and 2M04400067 show stronger excess emission at these short wavelengths. The scale height at the inner disk edge is slightly higher (by a factor of $\sim$1.4) for CFHT-BD-Tau 4 than the other two disks. The disk for 2M04400067 shows slight flaring longward of $\sim$20$\micron$. For all three disks, the best-fits are obtained using an inner disk radius of 1 $R_{sub}$ ($\sim$0.0004 AU) and an intermediate inclination of {\it i} $\sim$50-60$\degr$.

\subsection{The four `outlier' cases: MHO 5, CFHT-BD-Tau 8, 2M04290068, 2M04242090}
\label{outlier}


We have four interesting `outlier' cases, where the 10$\mu$m feature shows strong silicate emission with a mixed composition of small, large and crystalline grains, but the 20$\mu$m features are very weak, and it is mainly the underlying continuum which is detected (Fig.~\ref{modelfits}c). This indicates significant dust processing as well as dust settling to the midplane in the outer disk regions, while there is still a considerable fraction of warm dust in the inner disk region. For most other disks, we find similar strengths in the 10 and 20$\micron$ features (Section \S\ref{strength}). We have compared the extent of flaring in the outer $\sim$15-35 $\mu$m part of the disk for these outliers and the other disk sources, and find similar slopes as observed for the rest. The outlier disks are thus not significantly flatter than the other disk sources that show emission in the 20$\mu$m feature (Fig.~\ref{strength}). It may be that some secondary collisional process has produced this new generation of dust in the inner disk regions, resulting in strong emission at 10$\mu$m, while grains continued to grow and settle in the outer disk. A few of such outliers have also been found among T Tauri stars (e.g., Kessler-Silacci et al. 2007), and have been explained by enhanced dust settling in the outer disk regions. Considering the variety in the disk shapes and structures observed at any given age, such outlier cases suggest that dust processing does not occur on a similar timescale in the inner and outer regions for all disks.

Results from disk modeling indicate flat disk structures for all four sources. 2M04290068 shows photospheric emission shortward of $\sim$8$\micron$, and can be classified as a transition disk. From the best-fit model, we estimate an inner disk hole of $\sim$18 $R_{sub}$ ($\sim$0.006 AU), which is a factor of $\sim$10 larger than the inner disk radius estimated for the other 3 sources. This system thus seems devoid of disk material in the inner regions of within $\sim$0.01 AU, some emission from the warm surface layers is observed at radii of $\sim$0.01-0.1 AU, while there is an absence of optically thin material at larger radii, as indicated by the flat 20$\micron$ feature. 






\subsection{CFHT-BD-Tau 9, 2M04554801 and GM Tau}

Figure~\ref{modelfits}d shows model fits for disks with comparatively weak 20$\micron$ features. The S/N for the 20$\micron$ spectra is quite low ($\sim$10). For all three sources, we find very little fraction of small dust grains ($<$5\%) from 20$\micron$ modeling, the crystalline mass fractions are between 20\% and 40\%, and large-grain mass fractions are 50-60\%. In comparison with the inner disk fractions, all three sources show an increase in the large-grain and crystalline mass fractions in the outer disk, while the small grain mass fraction has decreased by more than 30\%. These disks are thus dominated by more processed dust in the outer regions, which is consistent with the flatter shapes observed for the 20$\micron$ features. The best-fit model SED is obtained using an inclination of {\it i} $\sim$70$\degr$ for GM Tau, indicating a close to edge-on disk for this brown dwarf. For CFHT-BD-Tau 9 and 2M04554801, a face-on inclination of 30-40$\degr$ provided the best-fit. 

The 20$\micron$ spectrum for 2M04554801 indicates silica peak near 20.8$\micron$. In our simplistic model, we have only considered amorphous silica and are able to obtain a good fit to this feature. However, detailed modeling such as that presented in Sargent et al. (2009b) can correctly identify the particular polymorph of silica present in the disk. We find a 27\% mass fraction for silica from modelling the 20$\micron$ feature, which is the highest silica fraction in the whole sample. Overall, we find silica mass fractions between $\sim$7\% and 14\% in the outer regions for these brown dwarf disks, with 0\% silica fraction found for V410 X-ray 6 and 2M04141760. As discussed in Sargent et al. (2009b), thermal annealing of olivine and pyroxene produces forsterite and enstatite, along with silica. This silica will be amorphous if annealing takes place at low temperatures ($<$ 1000 K) and for short durations. That is, annealing followed by rapid quenching can produce amorphous silica, such as obsidian and tektike. Among T Tauri stars, a few silica exemplars such as ROXs 42C, have been modeled by Sargent et al. (2009b), using many different forms of silica including obsidian and amorphous silica. Sargent et al. concluded that annealed silica (a mixture of the two crystalline polymorphs stable at high temperatures and low pressures, cristobalite and tridymite) provides the best fitting of model to spectrum, of all the different forms of silica attempted in the models. Annealed silica may also be responsible for the silica feature observed in the 2M04554801 spectrum. Nevertheless, this object presents a similar silica exemplar case among brown dwarfs.

%
%
%

%
%
%

\section{Discussion}

\subsection{Strengths and Shapes of the Silicate Features}
\label{strength}

Disks dominated by pristine ISM-like dust show flared structures, while flatter structures are observed for the ones that have gone through substantial grain processing. The processing of dust into crystalline silicates as well as growth to larger sizes both affect the vertical structure of the disk, as well as the emission strength in the silicate features. Figure~\ref{strength} compares the strength and shape of the 10 and 20$\micron$ silicate features. The shape of the 10$\micron$ feature can be estimated by the parameter $S_{11.3}/S_{9.8}$, which is the ratio of the normalized fluxes at 11.3 and 9.8 $\micron$ (e.g., Bouwman et al. 2001). For the 20$\micron$ feature, the shape can be estimated by the flux ratio, $S_{23.8}/S_{19}$. The peaks at 19$\micron$ and 23.8$\micron$ are the most prominent amorphous and crystalline features, respectively, in the $\sim$15-35$\micron$ spectral range (e.g., Kessler-Silacci et al. 2006). Typical uncertainties in the flux ratios are are $\pm$0.1. We have included for comparison the data for Herbig Ae/Be  stars and T Tauri stars from van Boekel et al. (2003), Przygodda et al. (2003) and Kessler-Silacci et al. (2005; 2006). The strengths of the features are indicated by the $S_{peak}$ values, as discussed above. 

An inverse strength-shape correlation can be seen for both the 10 and 20$\micron$ features. The correlation is such that spectra with a larger value for the $S_{11.3}/S_{9.8}$ or the  $S_{23.8}/S_{19}$ flux ratio have a lower peak-over-continuum flux. This can be explained by the extent of dust processing in the disk; spectra that are dominated by more processed dust due to grain growth and/or crystallization processes are found to be flatter with a peak wavelength close to 11.3$\micron$ or 23.8$\micron$, while spectra dominated by unprocessed ISM-like dust show more peaked, narrower profiles with a peak wavelength closer to 9.8$\micron$ or 19$\micron$. The strength in the feature thus decreases or the feature becomes flatter with increasing crystallinity and/or grain growth in the disk, as indicated by the flux ratios $S_{11.3}/S_{9.8}$ and $S_{23.8}/S_{19}$. This is more prominent for disks such as V410 X-ray 6 that show weak signs of grain processing and lie in the lower right part in the plot ($S^{10}_{peak}$ and $S^{20}_{peak}$ $\sim$ 1.65). 2M04230607, on the other hand, shows strong signs of crystallinity and lies in the upper left part, near $S^{10}_{peak}$, $S^{20}_{peak}$ $\sim$ 1.35, and $S_{11.3}/S_{9.8}$, $S_{23.8}/S_{19}$ $\sim$1.05. 

For both silicate spectra, the brown dwarfs show flatter features compared to higher mass T Tauri stars and Herbig Ae/Be stars. We find a clustering for the brown dwarfs near $S^{10}_{peak}$ $\sim$ $S^{20}_{peak}$ $\sim$ 1.35. This can be explained by the differences in the location of the silicate emission zone for stars of different luminosities. The silicate feature probes smaller radii ($\leq$1 AU) in disks around brown dwarfs than in disks around T Tauri stars ($\leq$10 AU). If crystallization mainly occurs through thermal annealing of amorphous silicates at high temperatures ($\sim$800 K), then this process will be more effective in the inner warm regions. Also, due to the higher densities in the inner disk, we would expect grain growth and dust settling to be more prominent in these regions. Thus due to the smaller radii probed by the silicate features around brown dwarfs, stronger signatures of dust processing would be observed.


Increasing dust processing is thus correlated with weaker strength in the silicate emission feature. In Figs.~\ref{strength-dust} and \ref{shape-dust}, we have investigated whether the observed flattening and the shift in the peak position of the silicate features is due to an increase in the grain sizes or a higher degree of crystallinity in the disk. The strength in the 10$\micron$ silicate feature shows no dependence on the crystalline or the large-grain mass fractions (Fig.~\ref{strength-dust}; {\it left panel}). The correlation coefficient is $<$0.1 for both cases. The 20$\micron$ feature strength shows some correlation with the grain growth levels; we find a correlation coefficient of -0.3, at a confidence level of 95\%. The correlation is weaker for the 20$\micron$ crystalline mass fraction and the 13-35$\micron$ spectral slope (correlation coefficient of -0.15 at 95\% confidence level). For the feature shapes (Fig.~\ref{shape-dust}), both the 10$\micron$ large grain and crystalline mass fractions are found to increase as the 11.3 $\micron$ flux increases relative to the 9.8 $\micron$ flux. We find a correlation coefficient of 0.4 for both cases at a confidence level of 92\%. For the 20$\micron$ features, crystallization seems to be more dominant than grain growth in affecting the feature shape (Fig.~\ref{shape-dust}; {\it right panel}). The correlation coefficient is 0.34 for the crystallin case (92\% confidence level), and 0.02 for the large-grain mass fractions. The feature shape thus seems to be more strongly linked to the crystallization process, while the strength in the features shows some dependence on the grain growth levels in the disk, at least for the 20$\micron$ case. The trends we have found for brown dwarfs are consistent with the results seen among higher mass objects in Taurus, as Watson et al. (2009) report both the shape and strength to be strongly linked to the crystallinity in these disks, but no strong correlation is found for the large grain abundances. 

Figure~\ref{flaring} shows a comparison of the crystalline and large-grain mass fractions with the extent of flaring in the inner and outer disk regions. The extent of flaring in a given disk region can be estimated by the spectral index ($\alpha$) of the derived continuum from the 10 and 20$\micron$ features, using the method defined in Kessler-Silacci et al. (2006). In our definition of $\alpha$, a more positive value implies higher flaring in the disk. The 10$\micron$ crystalline mass fractions show a dependence on the 8-13$\micron$ spectral slope. We find a correlation coefficient of -0.4 at a 90\% confidence level. In comparison, the correlation coefficient is nearly zero for the large-grain mass fractions. In the outer disk regions, both crystallization and grain growth seem to be effective in causing flattening in the disk. As mentioned, with the exception of V410 X-ray 6, the brown dwarf disks in our sample have flat structures, with similar 15-35$\micron$ spectral slopes of $\sim$ -0.5. Due to the large spread in the mass fractions for similar spectral slopes, the correlation coefficients for the 20$\micron$ case are small ($\sim$ -0.1). Sargent et al. (2009a) and Watson et al. (2009) have found that the T Tauri disks with the bluest continua at 6-to-13$\micron$ and 13-to-31$\micron$ wavelengths (the ones with the lowest flaring) tend to have the largest crystalline and large grain mass fractions. A number of T Tauri stars in the Watson et al and Sargent et al samples are found to have flared disks, resulting in a large range in the spectral indices, which makes it easier to identify the possible trends between the disk geometry and the mass fractions. Despite the lack of any strong correlations for the brown dwarfs, we find the strengths/shapes and the extent of flattening in these disks to be affected by both the crystallization and grain growth processes.

The dust settling and sedimentation process resulting in flattening of a disk may also be affected by the level of turbulence in the disk. In disks with higher accretion rates, turbulent mixing may occur alongside grain growth and prolong the timescales over which large grains settle to the optically thick disk midplane (e.g., Dullemond \& Dominik 2004). In Fig.~\ref{accretion}, there may be some dependence of the grain growth level in the inner disk region on the accretion rate (correlation coefficient = -0.3); MHO 5 and 2M04414825 are among the weakly accreting systems, but show high grain growth levels with $\sim$70\% mass fractions. 2M04141760 is the most actively accreting system in the sample and also has the most least processed disk. CFHT-BD-Tau 6, on the other hand, is also a weakly accreting system but has a large-grain fraction similar to 2M04141760. The accretion rates have been obtained from Muzerolle et al. (2003; 2005). For the case of the 10$\micron$ crystalline mass fraction, 2M04141760 is again the only outskirt in the figure, and ignoring this point would result in a flat distribution or a zero correlation coefficient. A flattened disk due to either grain growth and/or crystallization does not necessarily imply a weakly accreting system. However, as noted earlier, nearly all disks in our sample have flattened disks, irrespective of the accretion rate or the mass fractions. In the outer disk region, we find similar levels of grain growth and crystallinity for $log\dot{M}$ between $\sim$-8 and -10.5  $M_{\sun}~  yr^{-1}$. The number of known Taurus brown dwarfs with accretion rate measurements is small. A larger number of data points would be valuable to investigate any dependence on the mass accretion rate.

\subsection{Inner vs. Outer Disk Evolution: Dependence on Stellar Mass}
\label{inout}

In Figure~\ref{10vs20}, we have compared the 10 and 20$\micron$ mass fractions for the small and large amorphous grains, and the crystalline silicates. Also included for comparison is data for 65 T Tauri disks in Taurus from Sargent et al. (2009a). The T Tauri sample covers spectral types (SpT) between K5 and M5, which would imply stellar masses between $\sim$1 and 0.1$M_{\sun}$ (using Baraffe et al. 2003 evolutionary models for an age of $\sim$1 Myr). Our brown dwarf sample consists of a narrow range in SpT between M5 and M7, implying masses of $\sim$0.1-0.04$M_{\sun}$. There is thus no age effect involved in making such a comparison as all sources are presumably at the age of Taurus ($\sim$1 Myr). For the Taurus brown dwarfs, more than half of the disks show negligible mass fractions for small amorphous grains in the outer disk regions. We find just two sources that show $>$50\% small grain mass fractions in both the warm and cold components. These are the two least processed disks discussed in Section \S\ref{least}. In comparison, a majority of T Tauri disks have small amorphous grain fractions of $>$50\% in the cold component, and in some cases these are much higher than the fractions obtained from the inner disk. In comparing the grain growth levels, most brown dwarfs show higher large-grain mass fractions in the outer disk. We have six sources that have 10$\micron$ large-grain fractions of $<$20\%, while the outer disk fractions have increased to $>$20\%. Three other brown dwarfs show similar large-grain mass fractions of $\sim$40\% in the warm and cold components. Among the T Tauri sample, nearly all disks show an absence of grain growth in the outer disk region, with a 20$\micron$ large-grain mass fraction of $<$1\%. 

If we compare the crystalline levels (Fig.~\ref{10vs20}; {\it bottom panel}), then the T Tauri stars show a large clustering near $\sim$10\% fraction, and almost the full sample has a crystalline mass fraction of $\leq$20\% both in the inner and outer disk regions. The T Tauri distribution also shows two `tails' of sources with negligible outer disk crystallinity but significant inner disk crystalline mass fractions, and vice versa. Among the brown dwarfs, we find some correlation in the warm and cold crystalline mass fractions, with more than half of the sample showing similar fractions of $\sim$20-60\% at both 10 and 20$\micron$. The two brown dwarfs with lower crystalline fractions ($<$20\%) are the two least processed disks of V410 X-ray 6 and 2M04141760. The solid line in Fig.~\ref{10vs20}c is a straight line fit to the brown dwarfs and has a correlation coefficient of 0.5. {\it There is thus (weak) evidence that crystallinity increases linearly from the inner to the outer disk regions for brown dwarfs. }

The dependence of the 20$\micron$ mass fractions on the stellar mass can be seen in Fig.~\ref{spt-frac}. There is a lack of later type stars with disks that have abundant small grains, and there is a lack of later type stars with negligible crystalline and large grain mass fractions. We find a mean 20$\micron$ small, large and crystalline mass fraction of 17.2\%, 43.1\% and 30.3\%, respectively, for the brown dwarfs. In comparison, the higher mass star sample has a mean 20$\micron$ small, large and crystalline mass fraction of 62.2\%, 14.7\% and 16.1\%, respectively. There is thus a factor of $\sim$2-3 increase in the grain growth and crystallinity levels in the cold disk regions for later type stars, while the fraction of small amorphous grains has decreased by a factor of $\sim$4.


We thus find two possible trends from our analysis: {\it brown dwarfs show stronger signs of dust processing, in terms of the large-grain and crystalline mass fractions, in the outer disk compared to the inner disk regions}, and, {\it the extent of dust processing in the cold component is more significant for the sub-stellar disks compared to the T Tauri stars}. The trend is more prominent if we consider the four outlier cases discussed in Section \S\ref{outlier}, that show significant grain growth and dust settling in the outer disk regions. Watson et al (2009) find a trend in the Taurus T Tauri sample of more flattened disks having greater crystallinity. Sargent et al (2009a) also find a possible weak trend in a similar sample of flattened disks tending to have higher large grain mass fractions. For nearly all of our targets, we find a flat disk structure, and so stronger signs of dust processing in terms of higher crystalline and large-grain mass fractions can be expected. V410 X-ray 6 shows a comparatively flared disk structure and has the least processed disk, so the correlation noted by Watson et al. and Sargent et al. also holds true for the sub-stellar cases. Increasing grain growth in the outer disk also suggests that we are looking at a denser region at larger radii for the brown dwarfs, as discussed in Section \S\ref{strength}. If the 20$\micron$ feature probes lower scale heights or a deeper layer in brown dwarf disks than the surface layers or larger scale heights in T Tauri disks, then we would expect to see grains of larger sizes due to larger density and dust settling for the case of the brown dwarfs. This difference in the disk scale height could also explain the exceptionally high fractions of small dust grains in the cold component of the T Tauri disks (Fig.~\ref{spt-frac}), which are nearly negligible for the case of the brown dwarfs. Small dust grains should be more abundant at larger radii due to decreased densities, but this abundance would decrease with decreasing scale height in the disk, at a given disk radius. We can expect stronger signs of crystallinity in brown dwarf disks compared to T Tauri disks, primarily because the silicate emitting region would lie very close to the central source in brown dwarfs. Using a simple estimation, if the stellar mass drops by an order of magnitude (1 $M_{\sun}$ --$>$ 0.1 $M_{\sun}$), the disk radius at which annealing temperatures can be reached will be located 10 times closer to the star (e.g. 1AU --$>$ 0.1AU) (Section \S\ref{silicate}). Also, using Kepler's Law, M$_{*}$P$^{2}$ = A$^{3}$, the Keplerian time will be $\sim$10 times smaller if the mass and radius are reduced by a factor of 10. This would lead to very rapid dust evolution around brown dwarfs. Thus due to the whole 10 and 20$\micron$ tracing region lying close to the central source in a brown dwarf disk, stronger signatures of crystallinity would be observed.

\subsection{Dependence on X-ray emission strength}
\label{X-ray}



Figure~\ref{xray}a compares the crystalline mass fractions with the X-ray luminosity for brown dwarfs and T Tauri stars in Taurus. For the two brown dwarfs 2M04414825 and CFHT-BD-Tau 9, the X-ray luminosities are from our new {\it Chandra} observations computed in the 0.3-8 keV bands. For the rest of the sources, the luminosities are from Grosso et al. (2007) XMM-Newton observations in the 0.5-8 keV range. There is a $\sim$10\% systematic difference between {\it Chandra} and the XMM data ($\sim$ -0.1 in log$L_{X}$). The 10$\micron$ crystalline mass fractions for brown dwarfs show an inverse correlation with the X-ray strength (Fig.~\ref{xray}a; top left panel). We find a correlation coefficient of -0.6 at a 90\% confidence level, obtained from a straight line fit to the plotted points (excluding the upper limits). The dependence is more evident if we compare the `extreme' cases such as V410 X-ray 6 and 2M04414825, that show weak crystallinity in the disk but are among the stronger X-ray emitters in the sample. No correlation, however, is observed for the 20$\micron$ crystalline fractions (correlation coefficient = -0.02 at a confidence level of 99.99\%), and we find similar mass fractions of $\leq$20\% for a range in X-ray luminosities. Among the T Tauri stars, both the 10 and 20$\micron$ crystalline mass fractions show a nearly flat distribution, and there is no clear dependence observed on the X-ray luminosity. The correlation coefficients are -0.01 -- -0.02, at a confidence level of 99.99\%. There are, however, five T Tauri sources that show weak X-ray strength ($L_{X}<$ 10$^{30}$ergs/s) but high 20$\micron$ crystalline mass fractions of $\geq$30\% (Fig.~\ref{xray}a; bottom right panel). Though these make up a small $\sim$7\% fraction of the T Tauri sample, there may be a possible inverse trend observed among the weaker X-ray emitters for T Tauri stars. 


Figure~\ref{xray}b compares the grain growth levels in the disks with X-ray luminosities. For the case of the brown dwarfs, the 10$\micron$ fractions again show a dependence on the X-ray strength; sources with stronger X-ray emission of $L_{X}$$\geq$10$^{29}$ergs/s have higher grain growth levels of $>$40\%. We find a correlation coefficient of 0.42 at a 92\% confidence level obtained from a straight line fit to the plotted points. The correlation is again more evident when comparing extreme cases such as 2M04414825 and MHO 5 with V410 Anon 13, that either show very high or negligible fractions of large grains in the disk. A weak anti-correlation may be present for the 20$\micron$ large-grain fractions (correlation coefficient of -0.3 at 94\% confidence level). However, the cold component large-grain fractions have higher uncertainties and we have fewer data points at 20$\micron$, making it difficult to confirm any possible trend. Among the T Tauri stars, a large scatter in the 10$\micron$ large-grain mass fractions between 0-80\% is observed for a range in X-ray luminosities. There are three T Tauri sources with strong X-ray emission ($L_{X}$$\geq$10$^{30}$ergs/s) and comparatively higher 20$\micron$ large-grain mass fractions of $>$40\%. But it is unlikely to confirm any X-ray dependence (correlation coefficients $<$0.05), considering that the cold component large-grain mass fraction is less than 5\% for more than 90\% of the T Tauri sources. 

We have two brown dwarfs in our sample, V410 X-ray 6 and 2M04141760, which have been classified as X-ray flaring sources (Grosso et al. 2007). V410 X-ray 6 is a stronger X-ray emitter, and shows variability in $L_{X}$ between 8E28 and 6.3E29 ergs/s. 2M04141760 shows variability in $L_{X}$ between 1E28 and 8.5E28 ergs/s. In Fig.~\ref{xray}a (top left panel), 2M04141760 is a clear outlier, and the inverse correlation between crystallinity and $L_{X}$ will be stronger if we ignore this variable source. Similar is the case for the large-grain mass fractions (Fig.~\ref{xray}b; top left panel), where V410 X-ray 6 is an outlier and ignoring this variable source increases the correlation coefficient between large-grain fractions and $L_{X}$ to 0.6. 

There is thus evidence of a possible anti-correlation between the X-ray emission strength and the crystallinity in the warm component of the disks. Sudden X-ray activity may be responsible for varying the crystalline structure of dust grains in the disk surface layers. If X-ray amorphization happens quickly, then perhaps the maximum X-ray luminosity, as that observed for V410 X-ray 6, 2M04141760 and 2M04414825, is what directly determines the inner disk crystallinity, rather than some average X-ray luminosity. If we only consider the rightmost points for these objects in the plot in question, then there is a pretty reasonable correlation in the upper left plots of Fig.~\ref{xray}a and b. Processes such as amorphization by ion irradiation have been discussed to explain the absence of crystalline silicates in the ISM (e.g., Bringa et al. 2007; J\"{a}ger et al. 2003; Kemper et al. 2004). Some laboratory experiments have shown low-energy (keV) ions to be efficient in amorphizing silicate dust (e.g., Demyk et al. 2001; J\"{a}ger et al. 2003). However, the amophization timescales estimated by some of these studies are quite long. In the ISM, Kemper et al. (2004) suggest that the amorphization process occurs on a timescale much shorter than the grain destruction timescale, and it would take $\sim$9 Myr to achieve the low crystallinity level observed for the ISM dust ($\sim$0.4\%). Bringa et al. (2007) estimate a $\sim$70 Myr timescale for amorphization of ISM silicates by heavy-ion cosmic rays at GeV energies. In a study of variability in T Tauri disks, Bary et al. (2009) found the 10$\micron$ spectra to vary over month- and year-long timescales, but shorter timescales on day- to week-long periods were not detected. On the other hand, for the young eruptive star EX Lupi, \'{A}brah\'{a}m et al. (2009) obtained a 10$\micron$ spectrum just 2 months after an eruption in 2008, and found significant variations when compared with an earlier 2005 silicate spectrum. \'{A}brah\'{a}m et al. have noted the fast removal of crystalline silicates to be possibly due to amorphization by X-rays. It is thus difficult to estimate the expected timescale for the crystalline structure to reform, or otherwise appear in the spectrum. Coronal X-ray emission may correlate with other high energetic radiation in effectively destroying the crystalline silicates, and this may reduce the timescale over which such processes take place.




An opposite trend is observed for the large dust grains in the warm disk regions for brown dwarfs. Ilgner \& Nelson (2006) have considered X-ray irradiation from the central star to be the main source of ionisation, and have studied the vertical diffusion of chemical species, that mimics the effects of turbulent mixing in the disk. These authors have shown that X-ray ionisation induces turbulence that results in a decrease or a complete removal of the ``dead'' zones in the disk. Such zones are defined to be the regions that are too neutral and decoupled from the disk magnetic field for MHD turbulence to be maintained, as opposed to the ``active'' zones that are sufficiently ionised for the gas to be well-coupled to the magnetic field, and thus able to maintain turbulence. M-type pre-main sequence stars should be fully convective, so they can have strong magnetic fields that could give rise to X-rays.  X-rays would therefore be suggestive of a strong stellar magnetic field.  The stronger magnetic field then drives a more vigorous Balbus-Hawley instability (Balbus \& Hawley 1991) in the disk, which gives rise to turbulence in the disk. Increasing X-ray irradiation could thus be an indirect indicator of the presence of a strong magnetic field, which could result in active turbulent mixing in the disk. With increasing turbulence, the level of sedimentation would decrease and a higher fraction of large-grains would be detected in the upper disk layers, as probed by the silicate emission features. 



The outer disk crystalline and large-grain mass fractions however are unaffected, which suggests that X-rays cannot penetrate into the surface layers at larger radii in the disk. For the case of T Tauri stars, no clear trend is observed for any of the two processes with the strength in X-ray emission. In Fig.~\ref{xray}, there is a significant overlap in the X-ray luminosities (-1.4 $<$ log(Lx) $<$ -0.2) between the brown dwarf and T Tauri star samples. From the work of Reiners et al. (2009), the magnetic field strengths range between 2-3 kG for M3-M6 stars. For later types objects at spectral type of M7-M9, the field strengths are 1-2 kG (Reiners \& Basri 2010). The field strengths for the T Tauri stars and sub-stellar objects are found to be quite similar ($\sim$2kG) at a young age of $\sim$5Myr (Reiners et al. 2010). The similarity in the magnetic field strengths between T Tauri stars and brown dwarfs implies that any possible correlation found for the brown dwarfs mass fractions with X-ray luminosities must also hold true for the T Tauri stars. The lack of any correlation for the T Tauri stars suggests that there may be more decoupled zones in these disks, compared to the sub-stellar cases. Or, perhaps the amorphization process resulting in low crystallinity at high X-ray irradiation (or the opposite process for the grain growth) may occur too quickly or at a much shorter timescale in T Tauri disks. Such processes may proceed at a slower pace in brown dwarfs, making it possible to study them.

\subsection{The presence of crystalline silicates in the cold outer disk regions}
\label{discussion}

A fundamental finding of our work is the almost constant crystallinity observed regardless of the disk radius being probed. Crystalline silicates are certainly not less abundant in the outer cold regions of circum(sub)stellar disks. Two main mechanisms are considered to dominate the crystalline formation: (1)- gas-phase condensation at high temperature above $\sim$1200 K; (2)- thermal annealing at temperature above $\sim$800 K. 

In the gas-phase condensation model, the pristine interstellar dust of olivine and pyroxene composition is transported inwards to the warm inner regions, where these grains are evaporated. The elements of Mg, Si, Fe and O liberated due to evaporation then recondense when the gas cools down to form the forsterite and enstatite condensates. In the model calculations of Gail (2004), the evaporation rate reaches a peak at a disk radius of $\sim$0.6 AU, and the pristine dust mixture is completely evaporated within this radius. The inner regions of $<$0.6 AU are mainly occupied by the equilibrated components of forsterite and enstatite, although a small fraction of these silicates is also transported in the outer cold regions due to turbulent diffusion and large-scale circulation that mixes the grains inwards and outwards. The abundance of crystalline forsterite and enstatite at 10 AU is a factor of $\sim$4 less than the abundance at 1 AU (Gail 2004). This difference in abundances with disk radii could explain the trend noted above of higher crystalline mass fraction for the sub-stellar objects compared to the higher mass stars, as the silicate feature probes smaller radii of $\leq$1 AU for brown dwarfs compared to 10-20 AU for T Tauri disks. Gail (2004) estimate a timescale of $\sim$1 Myr for the transport, diffusion, annealing, evaporation and recondensation processes and to eventually reach chemical equilibrium, which is consistent with observations. 

Figure~\ref{ens-fors} compares the enstatite and forsterite mass fractions in the inner and outer disk regions for brown dwarf and T Tauri disks. The 10 and 20$\micron$ mass fractions for these dust species in brown dwarfs show a similar range between 0 and $\sim$30\%. However, the brown dwarf 2M04230607 is a prominent outlier, and shows a higher abundance of forsterite in the outer disk region. Among the T Tauri disks, there are about 10\% sources that show a similar high forsterite and enstatite fraction in the cold component of the disk. We consider the case where these different kinds of crystals form through the gas-phase condensation mechanism (e.g., Gail 2004). At high temperatures, forsterite condensates are formed, along with silica (due to dissociation of silicate minerals). This is followed by enstatite formation at a slightly lower temperature due to gas-solid reactions between forsterite and silica. One explanation for the higher forsterite fraction in the outer disk could be that the forsterite at larger radii is not converted to enstatite through solid-gas reactions with silica. This could happen if the surrounding SiO$_{2}$ gas is cooled rapidly or if the silica grains coagulate, thus preventing the reactions. We could also consider non-equilibrium conditions in the inner disk, such that only forsterite is formed, and then there is a rapid outward transport of forsterite to prevent reactions with silica to form enstatite. Thereafter, equilibrium conditions may prevail in the inner disk regions (e.g., Bouwman et al. 2008). 

Some cases of compositional differences among comets are also known. More specifically, a contrast has been noted between comets with Mg-rich crystals, e.g., Hale-Bopp, which comes from the Oort cloud and probably formed near Jupiter in the giant planet region, and comets with Fe-rich crystals, such as, 9P/Tempel 1, which came from the transneptunian region, i.e., further from the Sun (e.g., Lisse et al. 2007; Keller \& Gail 2004). This may be indicative of the degree of hydration (or other conditions) in the nebular region where they formed, since Fe-rich silicates hydrate more easily than Mg-rich ones (e.g., Brownlee et al. 2006). Alternatively, the Fe-rich crystals may have formed close to the young Sun (within the present orbit of Mercury), since their annealing requires a high temperature of $\sim$1400 K (e.g., Nuth \& Johnson 2006), and then transported outwards to the comet formation zone. The transport, however, must have occurred early on, some 5 to 10 Myr after nebula formation, after which the formation of giant planet cores would have inhibited further radial transport across their orbits (Lisse et al. 2006). Thus these objects may have formed at different times, as the properties of the dust in the outer nebula evolved from the pre-solar nebula properties (e.g., Lisse et al. 2006; Ciesla 2009). A difference in the location or the times of formation could explain the compositional differences observed for 2M04230607. There may have been a rapid outward transport of crystalline forsterite to prevent reactions with the surrounding silica to form enstatite, resulting in a low enstatite fraction in the outer disk, or forsterite may have formed later in the outer disk due to some in-situ formation mechanism, resulting in a high forsterite fraction in the cold component. 

There may be some local crystalline formation mechanism occurring in the cold component, such as low-temperature crystallization due to physical or chemical processes that could temporarily raise the local temperature in a turbulent disk environment and cause crystallization, followed by slow cooling over a few seconds timescale (e.g., Molster et al. 1999). A plausible mechanism could be flash heating of dust by shock waves, which has been proposed as a formation mechanism for chondrules found in meteorites (e.g., Harker \& Desch 2002). Tanaka et al. (2010) have recently formulated a model in which low-temperature crystallization can be induced by the latent heat of chemical reaction with the ambient gas, and may be triggered at temperatures of a few hundred kelvin and at the low gas densities typically observed in the outer regions of a protoplanetary disk. Such non-thermal crystallization could proceed if the heat of reaction is of the order of 10$^{26}$ K cm$^{-3}$, and the heat is transferred to the grains on a short timescale of a millisecond or less. Tanaka et al. have argued that this amount of stored energy density is less than that required for chondrule formation, and could be triggered by certain mechanisms likely common in protoplanetary disks such as grain-grain collisions or heating by shock waves due to gravitational instabilities. Interestingly, they find that a higher stored energy density is required to crystallize amorphous silicate of enstatite composition than forsterite composition, suggesting a higher abundance of enstatite in the warm inner region. Their model could provide an alternative to large-scale radial mixing for cases such as 2M04230607, that show a higher fraction of forsterite in the cold component. If both forms of crystalline silicates are formed in the hot inner disk and then transported outwards, similar abundances should be expected throughout the entire disk. A gradient in the dust chemical composition suggests some local crystalline formation mechanism, which could be low-temperature crystallization.

Alternatively, amorphous dust can be thermally annealed into crystalline dust at temperatures above $\sim$1000 K (e.g., Nuth \& Johnson 2006), followed by an efficient radial mixing mechanism to transport crystalline material from the hot inner disk to the cold outer parts (e.g., Nuth et al. 2000; Keller \& Gail 2004; Okamoto et al. 2004). The very first detection of crystalline silicates in comets, or in any celestial source (except meteorites), was reported by Campins \& Ryan (1989). Since then, several comets have been noted to have a high abundance of crystalline dust, for e.g., Hale-Bopp has a silicate composition which is $\sim$67\% crystalline, while the transneptunian comet Temple 1 has an even higher $\sim$80\% crystalline mass fraction (e.g., Lisse et al. 2007). Since these comets are formed in regions much colder than the annealing temperatures in excess of 1000 K, several mechanisms have been proposed to explain the transport of high-temperature silicates from the inner regions of the solar nebula to the cold outer regions where comets formed. It is interesting to discuss the possible analogies between comets and brown dwarf disks, in terms of the presence of high-temperature material in the cold outer regions, and the mechanisms responsible for their outward transport. Ciesla (2009) has considered the ``jet model'' to explain the high abundance of crystalline grains in comet Wild 2 collected by the Stardust spacecraft. Their jet model is similar to the X-wind model from Shu et al. (1996). The X-wind model was proposed to launch ``proto-chondrules'' as projectiles from the ``X-region'' close to the star. These proto-chondrules would be heated to various degrees close to the young Sun and then launched in ballistic trajectories above the disk as a result of winds driven through the interaction of the magnetic field of the Sun with the inner edge of the solar nebula. The proto-chondrules would land back in the outer regions of the disk, which would ordinarily be too cold to melt and crystallize the chondrules. 


In the Ciesla (2009) model, jets launched from the inner solar nebula deliver high-temperature materials to the outer comet formation region. The dust grains are injected into the upper disk layers and then vertically diffuse to the disk midplane. The abundance of crystalline grains in a comet would be dependent on the delivery timescale to the outer nebula, or the rate of injection. If the delivery of the high-temperature grains occurs over a short timescale ($\leq$10$^{4}$ years), i.e., shorter than the timescale for radial redistribution, then the crystalline grains can mix with the native amorphous grains, and high crystalline fractions can be achieved in the comets or cometesimals thus formed. If the delivery occurs over a longer timescale (10$^{5}$-10$^{6}$ years), then cometesimal growth would have begun in an environment where high-temperature grains were absent, leading to formation of comets with low abundance of high-temperature silicate grains from the inner solar nebula. Such cometesimals could have an outer layer of high-temperature grains that slowly settled to the midplane. Thus if the delivery/injection to the outer cold regions occurs over a short timescale, then a homogenous mixture would result, while slower delivery would form a heterogenous comet (Ciesla 2009). 

The brown dwarf disks we have studied in this work are at young ages of $\sim$1 Myr. The high crystalline abundance found in the outer cold regions of 2M04230607 or CFHT-BD-Tau 20 suggests delivery of high-temperature crystalline silicates to the cold regions at an early stage of $<$10$^{4}$ years. Both of these disks also show a $\sim$20-40\% fraction of large amorphous grains, indicating a homogenous composition which can be expected in the case of early mixing. None of these two sources, however, are known to drive outflows, and so the jet model may not be an applicable mechanism. Nearly a dozen brown dwarf outflow sources have been discovered to date (e.g., Whelan et al. 2009). Two of these outflow sources are found to be rich in crystalline silicates in the inner disk regions (e.g., Phan-Bao et al. 2008; R09), but the outer disk silicate spectrum is either flat or shows amorphous features. Typical outflow mass loss rates for brown dwarfs are between $10^{-10}$ and $10^{-9}$ $M_{\sun}$ yr$^{-1}$ (e.g., Whelan et al. 2009). Perhaps the outflow emission strengths are too weak to efficiently transport high-temperature material to the cold regions. A detailed study of silicate features for the outflow sources will be valuable in determining how effective outflows can be in the radial transport of crystalline silicates. 

More than one mechanism may be responsible for the transport of high-temperature silicates. An alternative scenario has recently been presented by Vinkovi\'{c} (2009), wherein a non-radial radiation pressure force could loft a crystalline grain several scale heights above the disk midplane in the warm inner regions, followed by gliding over the disk surface towards colder disk regions. The grain then re-enters and settles towards lower scale heights when the peak wavelength is larger than the grain size and the radiation force keeping the dust afloat ceases. In the Vinkovic (2009) model, dust particles of sizes $\leq$ 1mm in optically thick protoplanetary disks are well coupled with the gas. The dynamics of the dust particles are dominated by the gas drag, and dust motion is very similar to the gas orbital, which is almost Keplerian motion. The radiation pressure force serves as the slow perturbation that leads to the rearrangement of dust orbits. The net radiation pressure force is directed exactly parallel to the disk surface irrespective of its curvature. The large grains are thus able to migrate along any disk curvature. Such surface flows might be possible, however, Vinkovi\'{c} model assumes a laminar disk without turbulence. It is not clear if such a gliding process or radiation-pressure mixing is indeed more effective than other process such as turbulent diffusion, X-winds, or other mechanisms that could transport crystalline material to the outer region. Nevertheless, the high crystallinity level found in the outer regions for a few of these $\sim$1 Myr old brown dwarf disks indicates rapid outward transport as well as vertical diffusion during the early stages of formation of these disks.

\section{Summary}

We present a compositional analysis of the 10 and 20$\micron$ silicate emission features for brown dwarf disks in the Taurus-Auriga star-forming region. The spectra show a variety in the emission features characteristic of silicate grains, both amorphous and crystalline in composition. A comparison of the crystalline and large-grain mass fractions in the warm and cold components of the disk, as well as with the higher mass T Tauri stars in Taurus, indicates the following:

\begin{enumerate}

\item Brown dwarfs show stronger signs of dust processing in the cold component compared to T Tauri stars. The grain growth and crystallinity levels in the cold disk regions for later type stars are a factor of $\sim$2-3 higher compared to the warm component, while the fraction of small amorphous grains is smaller by a factor of $\sim$4. 

\item The brown dwarf 2M04230607 provides an interesting case where a difference in the crystalline silicate composition is observed in the inner and outer disk regions; the forsterite mass fraction in the outer disk is a factor of $\sim$3 higher than the inner disk fraction. Simple large-scale radial mixing cannot account for the gradient in the dust chemical composition and some local crystalline formation mechanism may be effective in this disk. 

\item We have four `outlier' cases among the brown dwarfs that show strong 10$\micron$ features but very weak 20$\micron$ features, implying significant grain growth/dust settling has occurred at larger radii in these disks. Another outlier case is of CFHT-BD-Tau 20 that shows a weak 10$\micron$ feature but strong emission at 20$\micron$. Such outliers suggest that dust processing mechanisms may not proceed simultaneously in the inner and outer regions for all brown dwarf disks. 

\item {\it Chandra} X-ray observations have been presented for two Taurus brown dwarfs 2M04414825 and CFHT-BD-Tau 9. These observations were obtained to probe any dependence of crystallization on X-ray emission strength. We find 2M04414825, which has a $\sim$12\% crystalline fraction, to be an order of magnitude brighter in X-ray than CFHT-BD-Tau 9, which has a $\sim$35\% crystalline mass fraction. Combining with previously published X-ray data for brown dwarfs in Taurus, we find the inner disk crystalline mass fraction to be anti-correlated with the X-ray emission strength. 




\end{enumerate}

\section*{Acknowledgments}
Support for this work was provided by the National Aeronautics and Space Administration through Chandra Award Number 12200172 issued by the Chandra X-ray Observatory Center, which is operated by the Smithsonian Astrophysical Observatory for and on behalf of the National Aeronautics Space Administration under contract NAS8-03060. HC acknowledges support from NASA and the National Science Foundation. This work is based in part on observations made with the Spitzer Space Telescope, which is operated by the Jet Propulsion Laboratory, California Institute of Technology under a contract with NASA. This work has made use of the SIMBAD database.

\begin{onecolumn}
\begin{landscape}
\begin{table*}
\begin{minipage}{\linewidth}
\tiny
\caption{X-ray properties of 2M04414825 and CFHT-BD-Tau 9}
\label{resume}
\begin{tabular}{lccccccccccccc@{}c@{}c}
\hline
Source  & RA (J2000) & Dec (J2000) & $\Theta$\footnote{$\Theta$ is the off-axis angle.} (arcmin) & net counts & $\sigma_{net counts}$ & bkg counts\footnote{$bkg$ is for ``background''.} & PB\footnote{$PB$ is the probability associated with the hypothesis that the detected source is compatible with a background fluctuaction.} & $E_{25}$\footnote{$E_{25}$, $E_{50}$, and $E_{75}$, are the 25\%, 50\% and 75\% quartiles of the photons median energy distribution.} & $E_{50}$$^{d}$ & $E_{75}$$^{d}$ & nH ($10^{22} cm^{-2}$)  & kT (keV) & $log\left(L_X \right)$ \\ \hline

2M04414825   & 70.451042& 25.575151& 0.3& 373& 20& 18& 0              & 1.03& 1.26& 1.76& 0.36& 1.5& 29.315 \\
CFHT-BD-TAU 9& 66.110222& 26.830584& 0.3& 7&   2&   6& $4\times10^{-8}$& 0.82& 0.88& 1.34& 0.36& 0.98& 27.68\\

\hline
\end{tabular}
\end{minipage}
\end{table*}
\end{landscape}
\end{onecolumn}

\begin{table*}
\begin{minipage}{150mm}
\caption{Characteristics of the grain species used for the fitting procedure.}
\label{density}
\begin{tabular}{ccccccc}
\hline
Species  & Composition & Size [$\micron$] & Shape & Reference & $\rho_{b}$ [$g ~cm^{-3}$] \\ \hline

Amorphous olivine & MgFeSiO$_{4}$ & 0.1 \& 2.0 & Homogeneous spheres & Dorschner et al. (1995) & 3.71\\
Amorphous pyroxene & Mg$_{0.8}$Fe$_{0.2}$SiO$_{3}$ &  0.1 \& 2.0 & Homogeneous spheres & Dorschner et al. (1995) & 3.71 \\
Crystalline forsterite & Mg$_{1.9}$Fe$_{0.1}$SiO$_{4}$ & 0.1 & Irregular (CDE) & Fabian et al. (2001) & 3.33\\
Crystalline enstatite & MgSiO$_{3}$ & 0.1 & Irregular (CDE) & J\"{a}ger et al. (1998) & 2.80\\
Amorphous silica & SiO$_{2}$ & 0.1 & Irregular (CDE) & Fabian et al. (2000) & 2.21\\

\hline
\end{tabular}
\end{minipage}
\end{table*}

\begin{figure*}         
      \includegraphics[width=55mm]{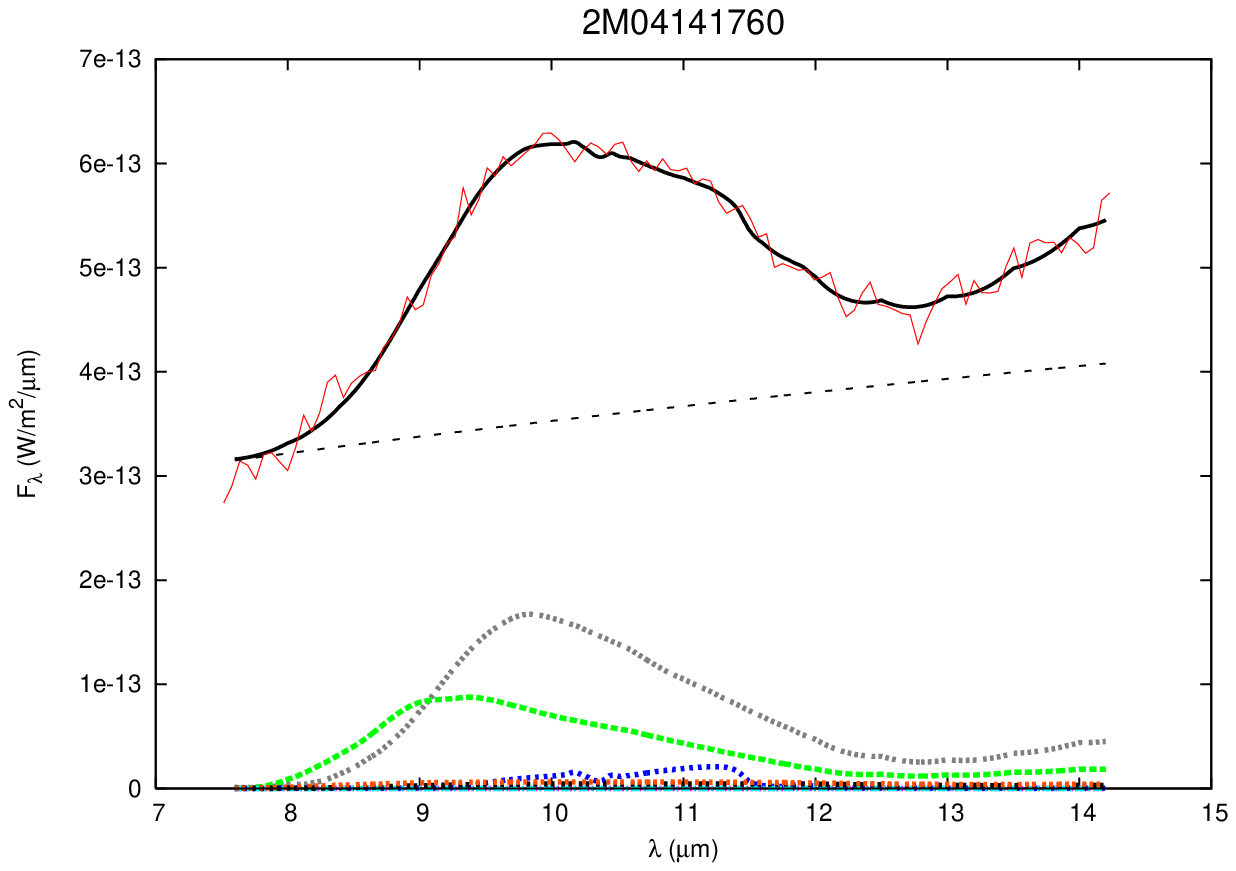} 
      \includegraphics[width=55mm]{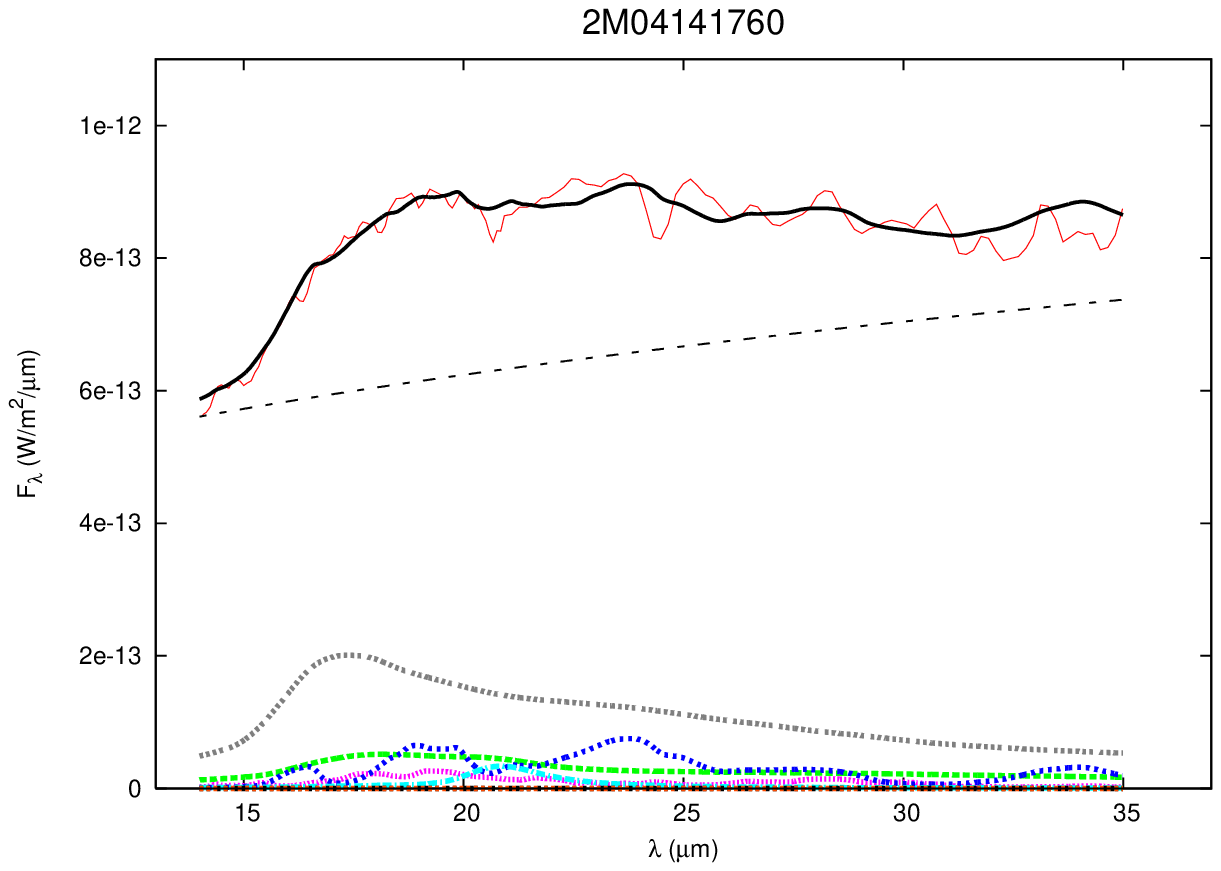} 
      \includegraphics[width=55mm]{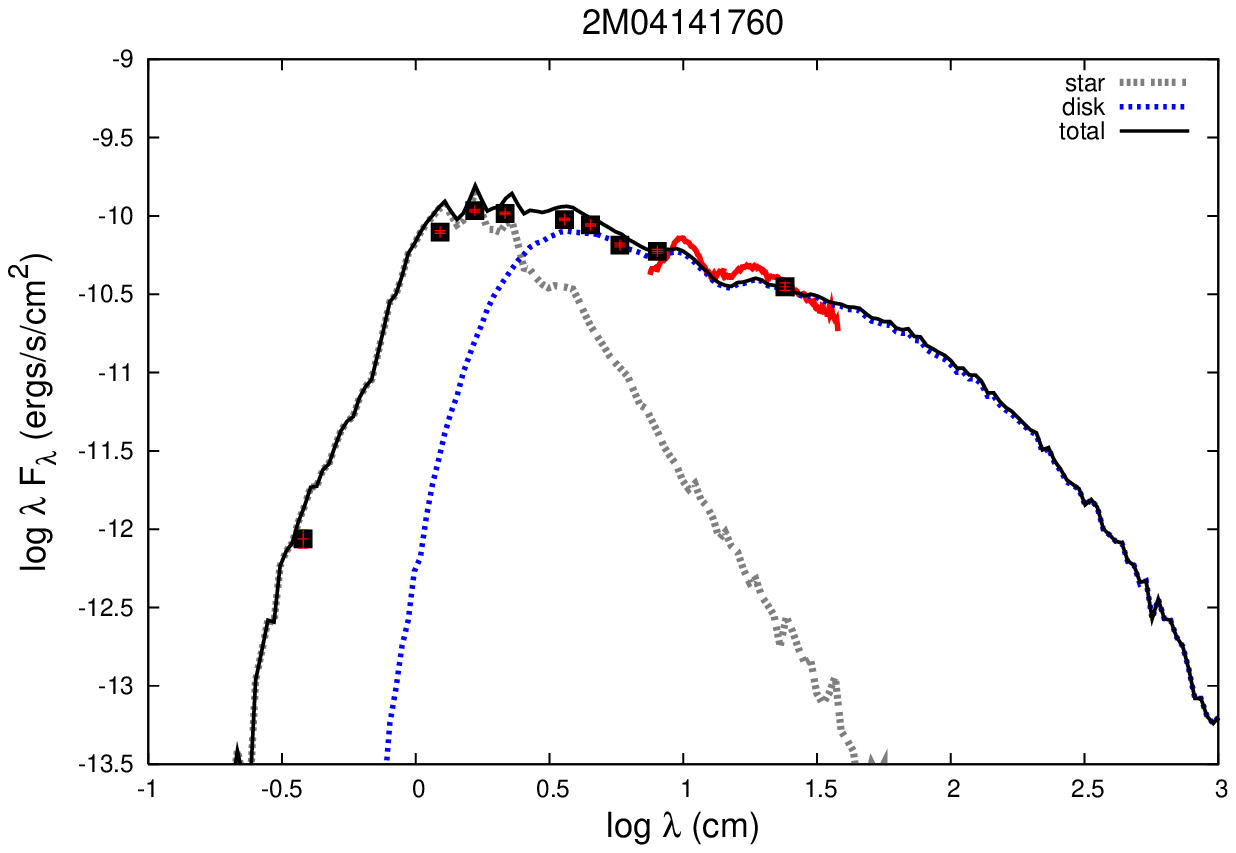} \\       
      \includegraphics[width=55mm]{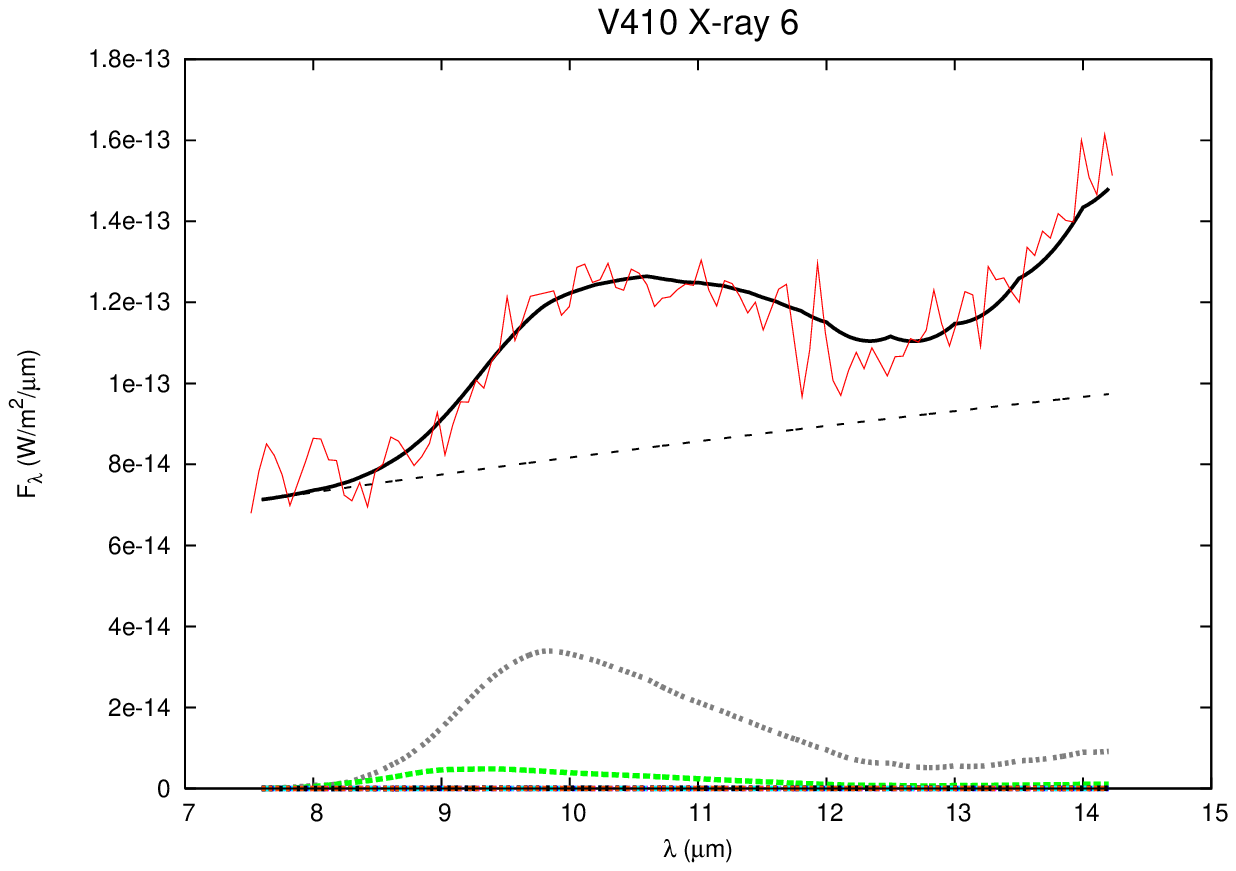}    
      \includegraphics[width=55mm]{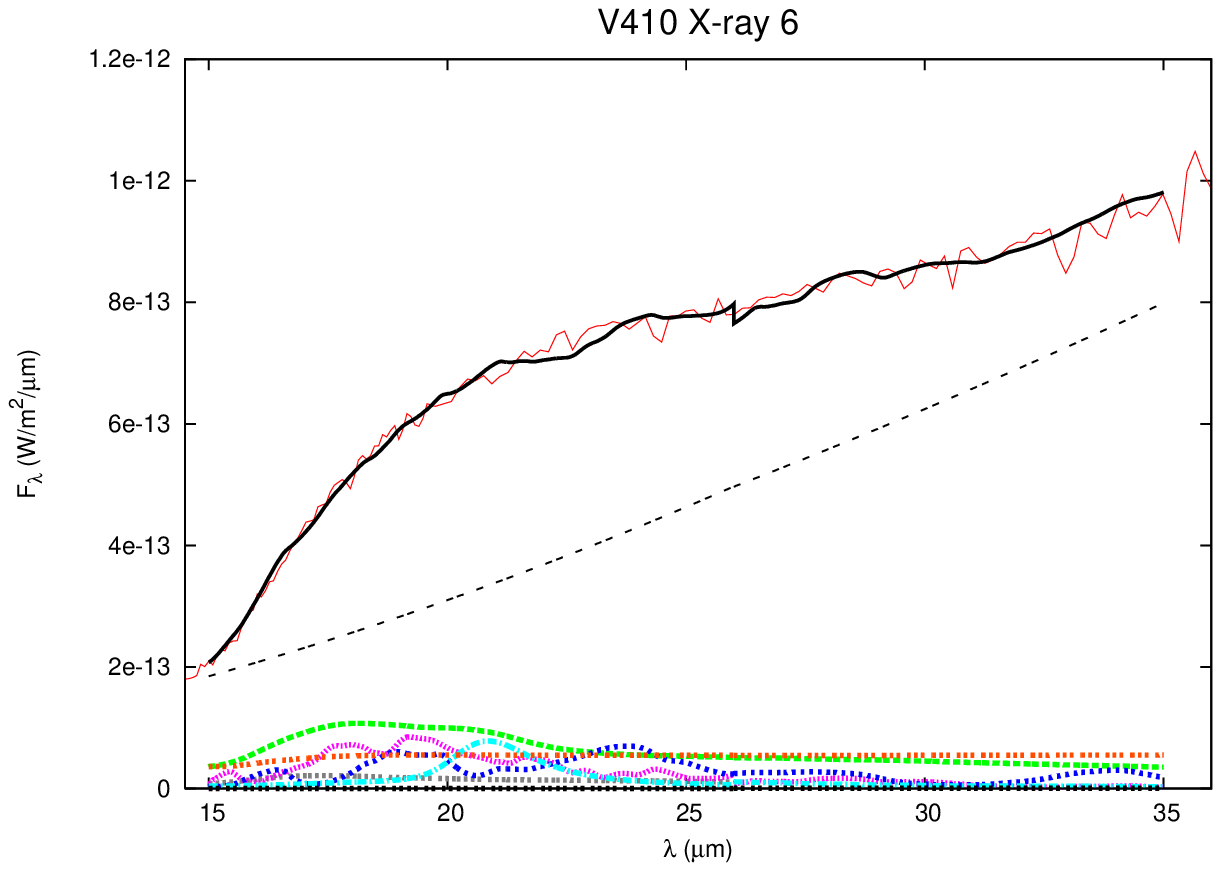}  
      \includegraphics[width=55mm]{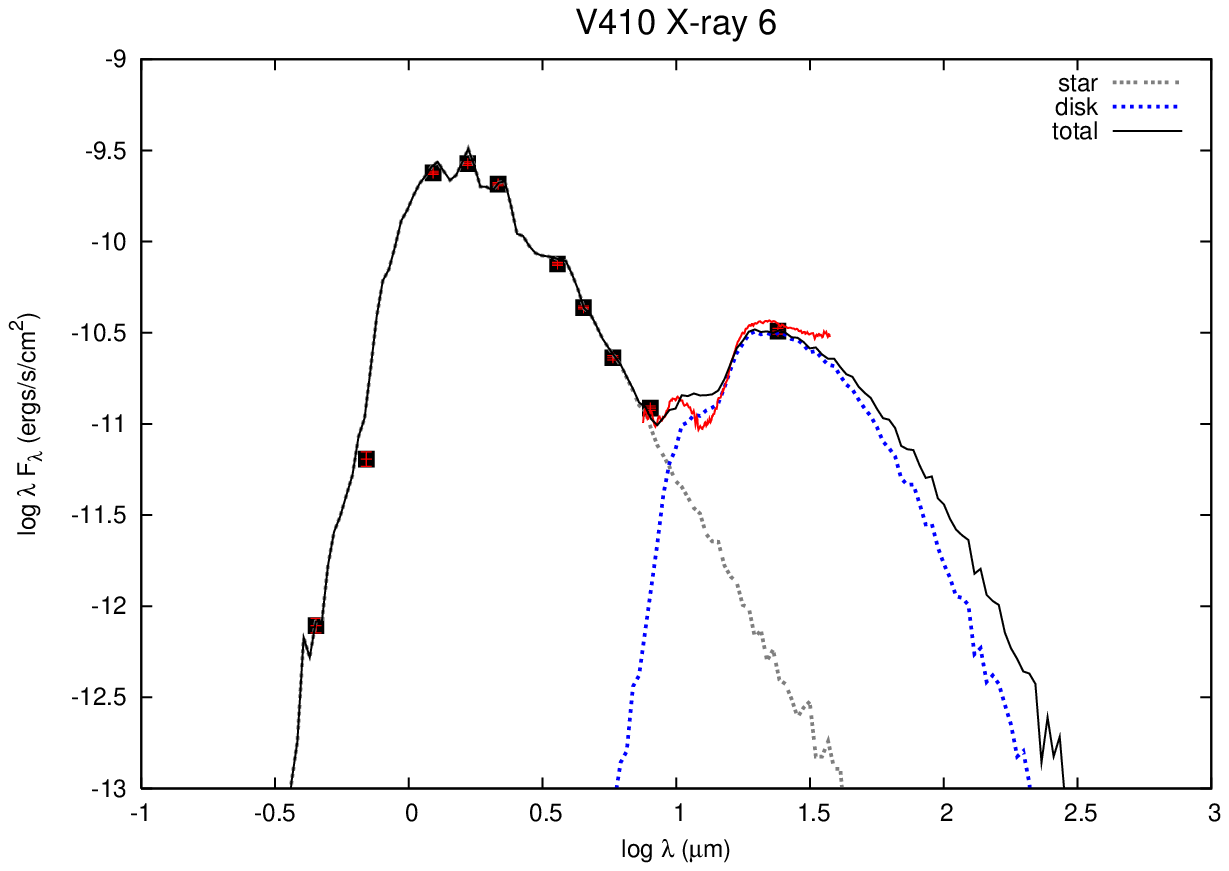} \\      
      \includegraphics[width=55mm]{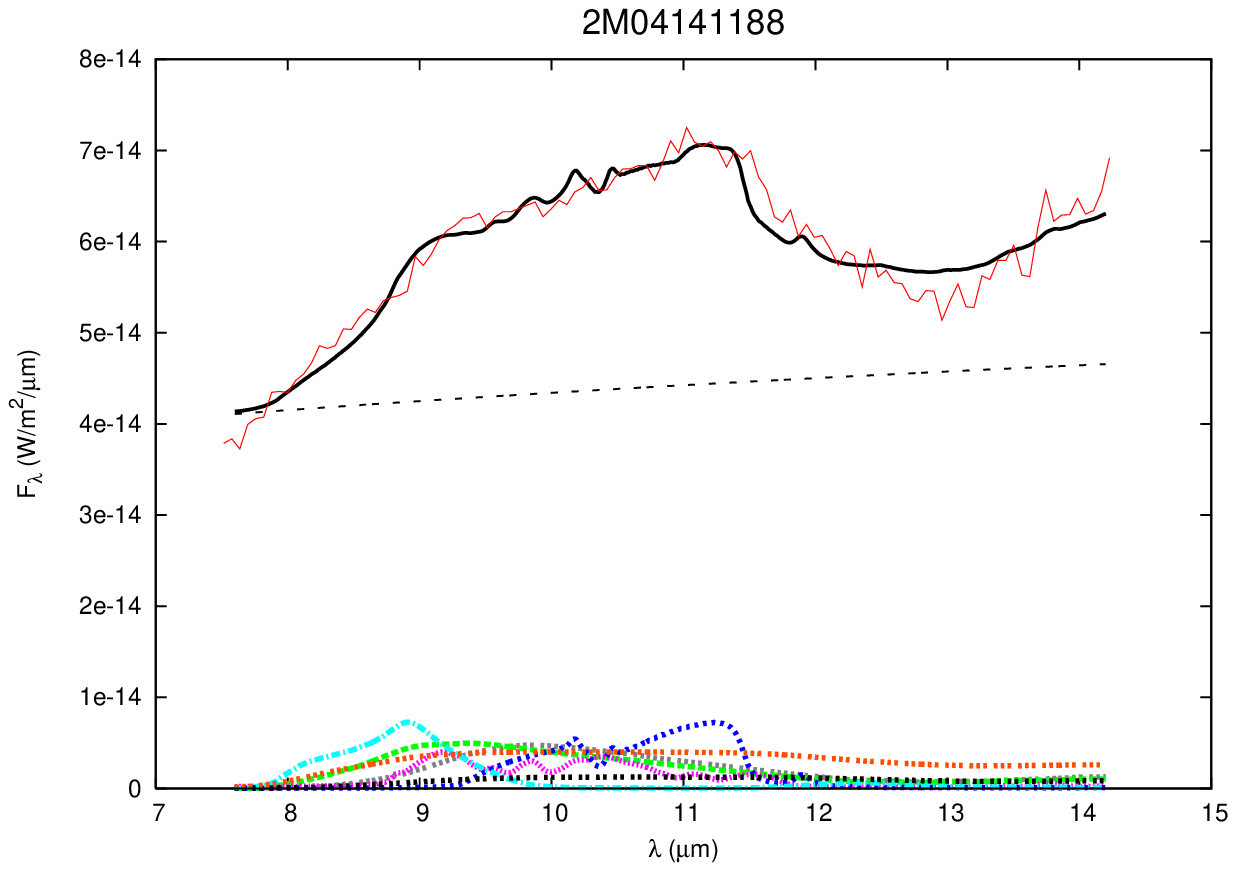} 
      \includegraphics[width=55mm]{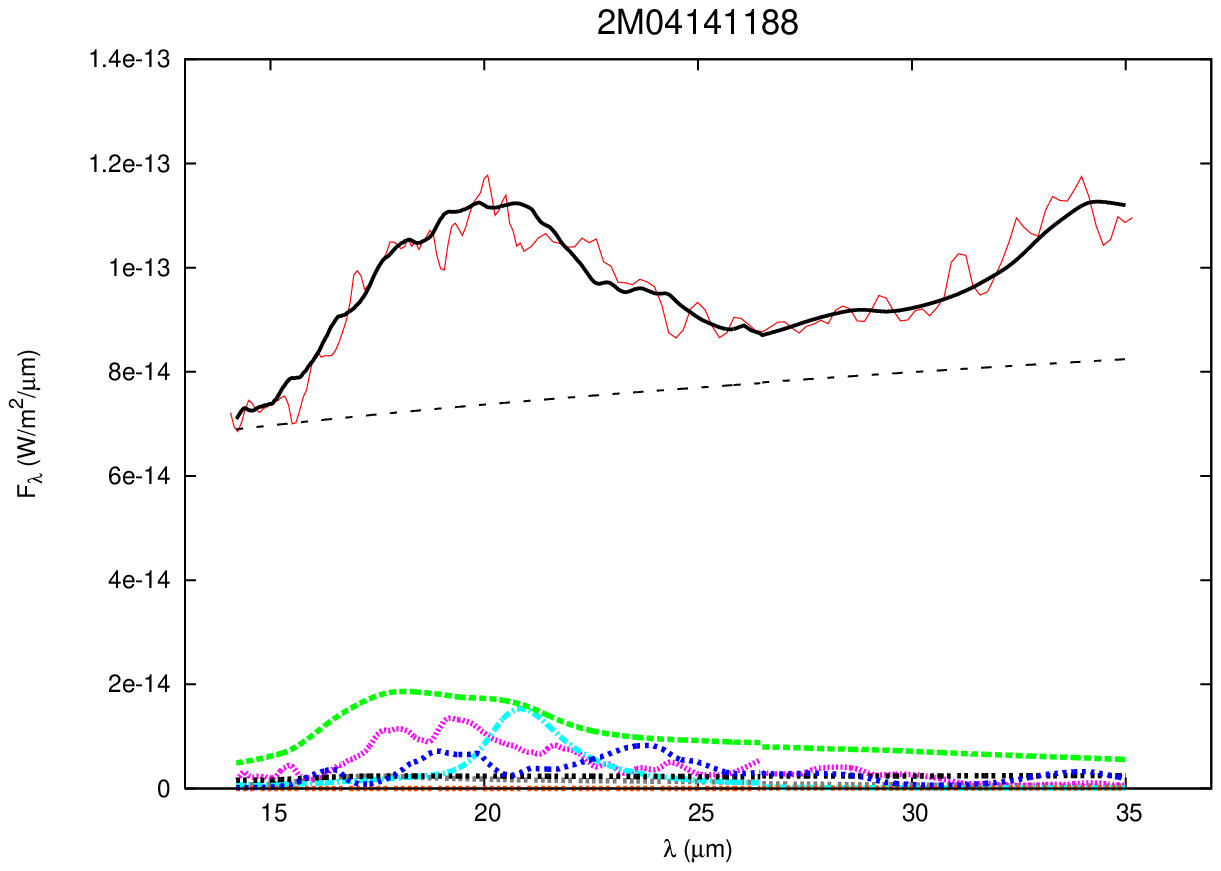} 
      \includegraphics[width=55mm]{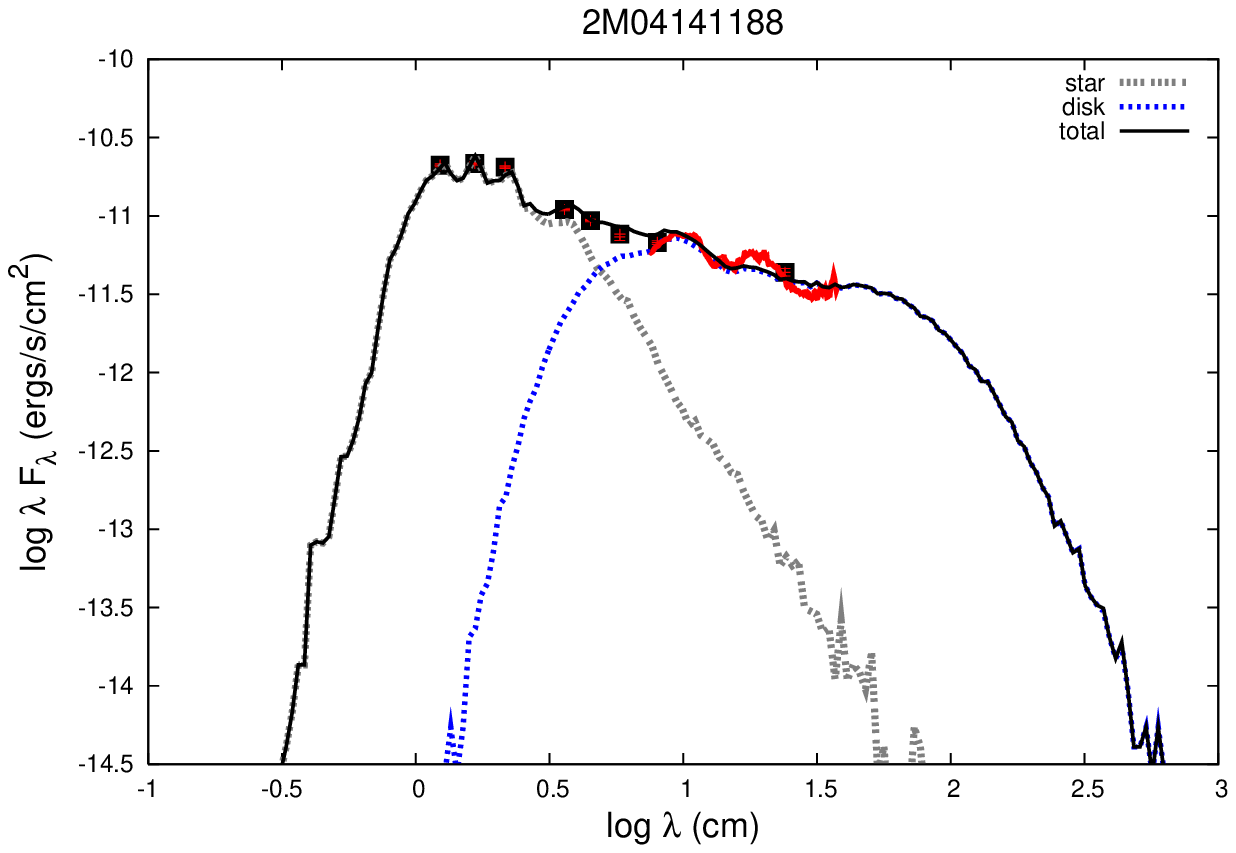} \\                
    \caption{(a): {\it Left} and {\it middle} panels show the model fits to the 10$\micron$ and 20$\micron$ silicate features, respectively. Colors represent the following: red--observed spectrum; black--model fit, grey--small amorphous olivine; green--small amorphous pyroxene; cyan--silica; blue--forsterite; pink--enstatite; black dashed--large amorphous olivine; orange--large amorphous pyroxene. Thin dashed line represents the continuum. {\it Right} panel shows the model SED fit. Grey line represents the stellar photosphere contribution, disk emission is indicated by blue line.  } 
    \label{modelfits}  
 \end{figure*}

\begin{figure*}
\setcounter{figure}{1}           
      \includegraphics[width=55mm]{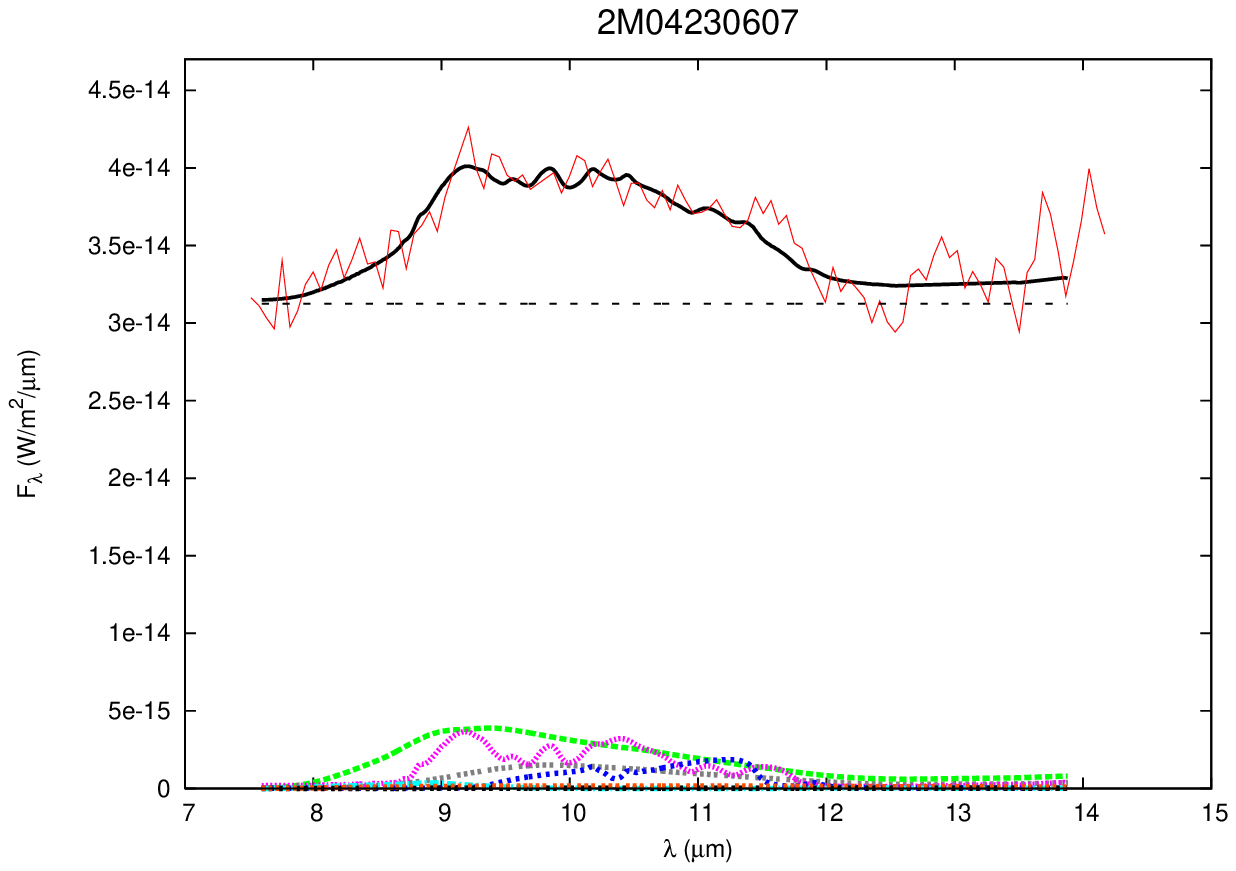} 
      \includegraphics[width=55mm]{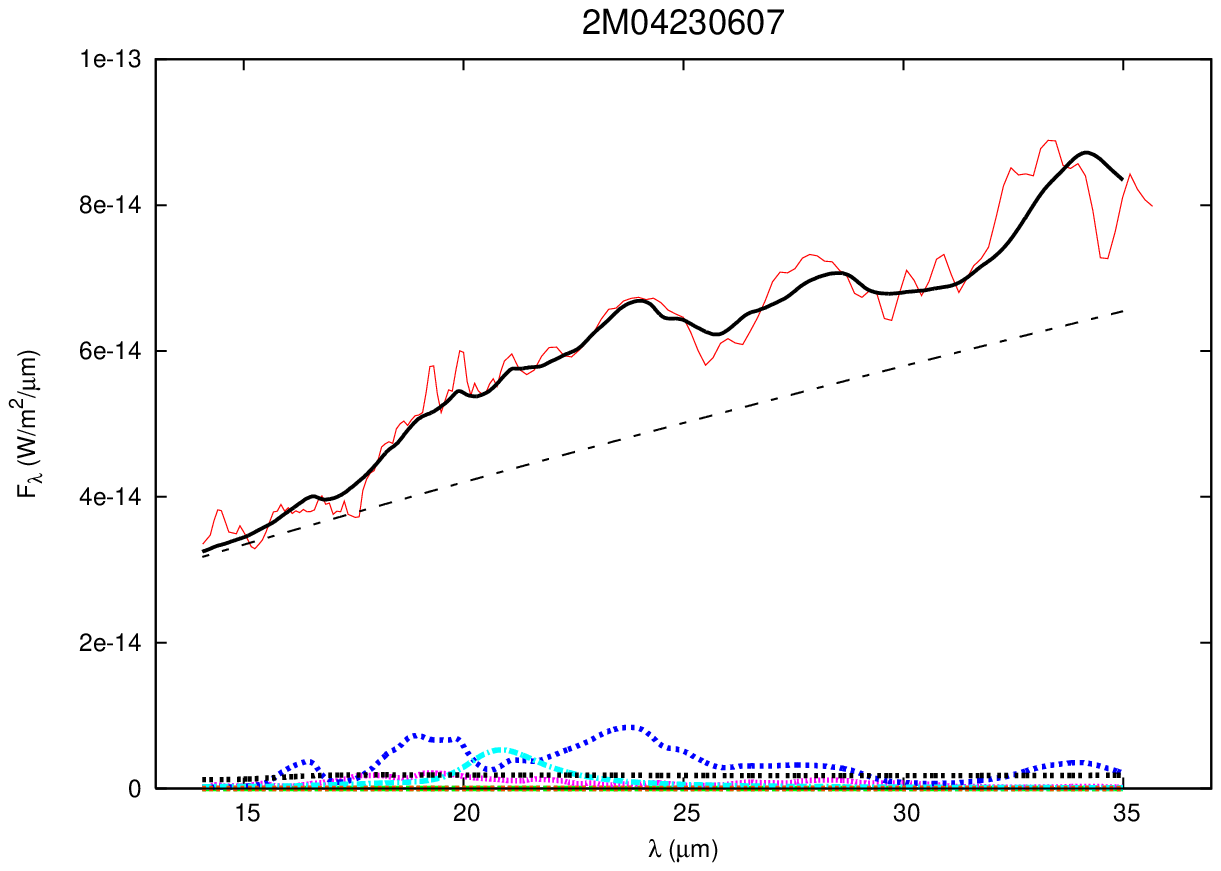} 
      \includegraphics[width=55mm]{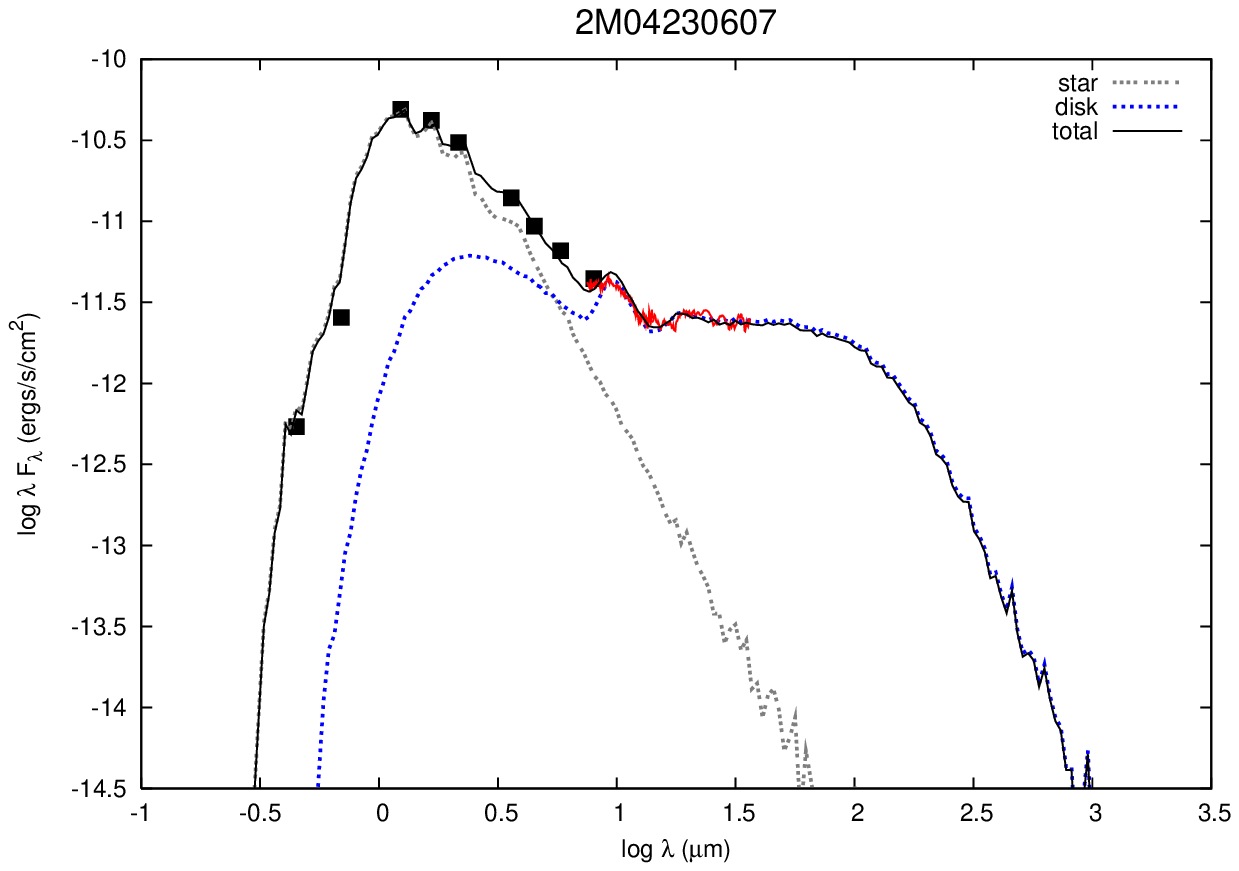} \\      
      \includegraphics[width=55mm]{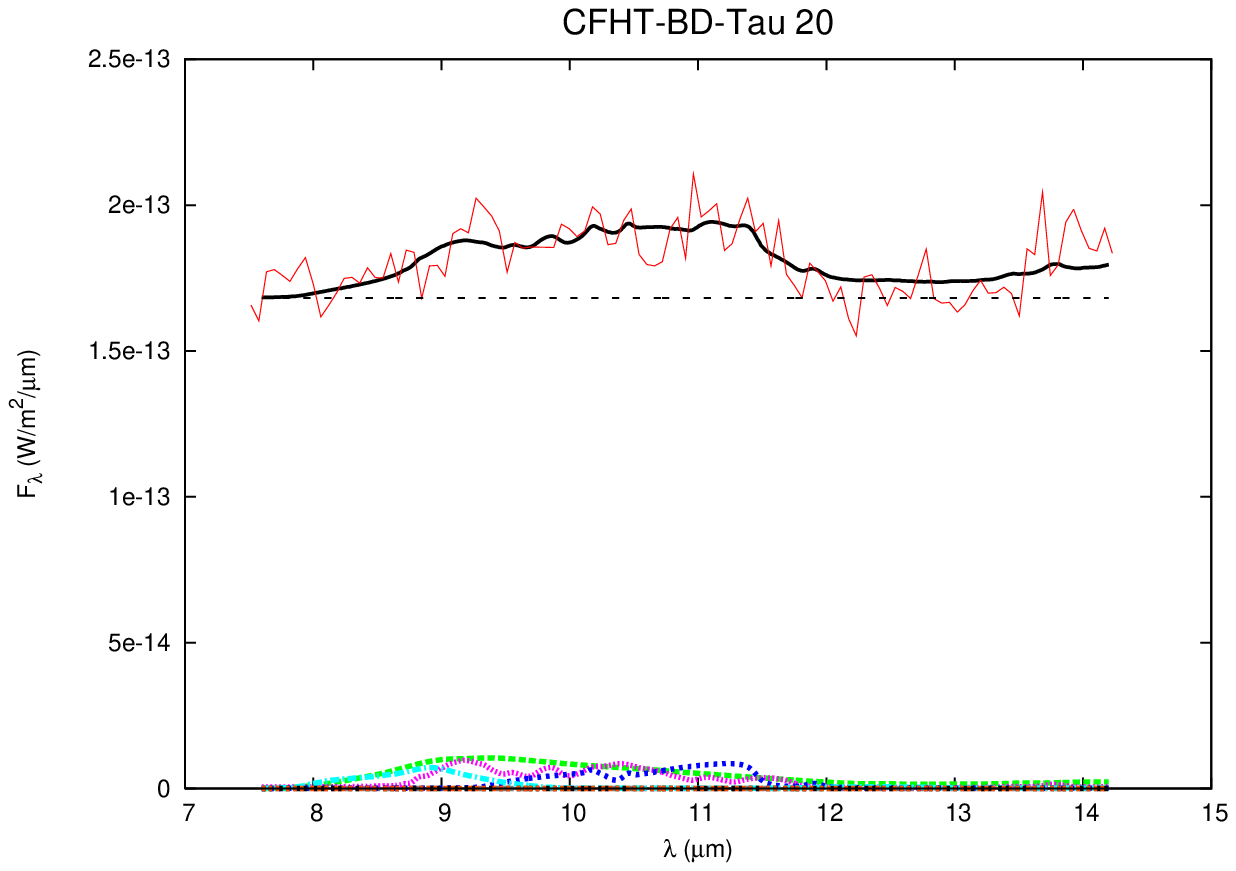} 
      \includegraphics[width=55mm]{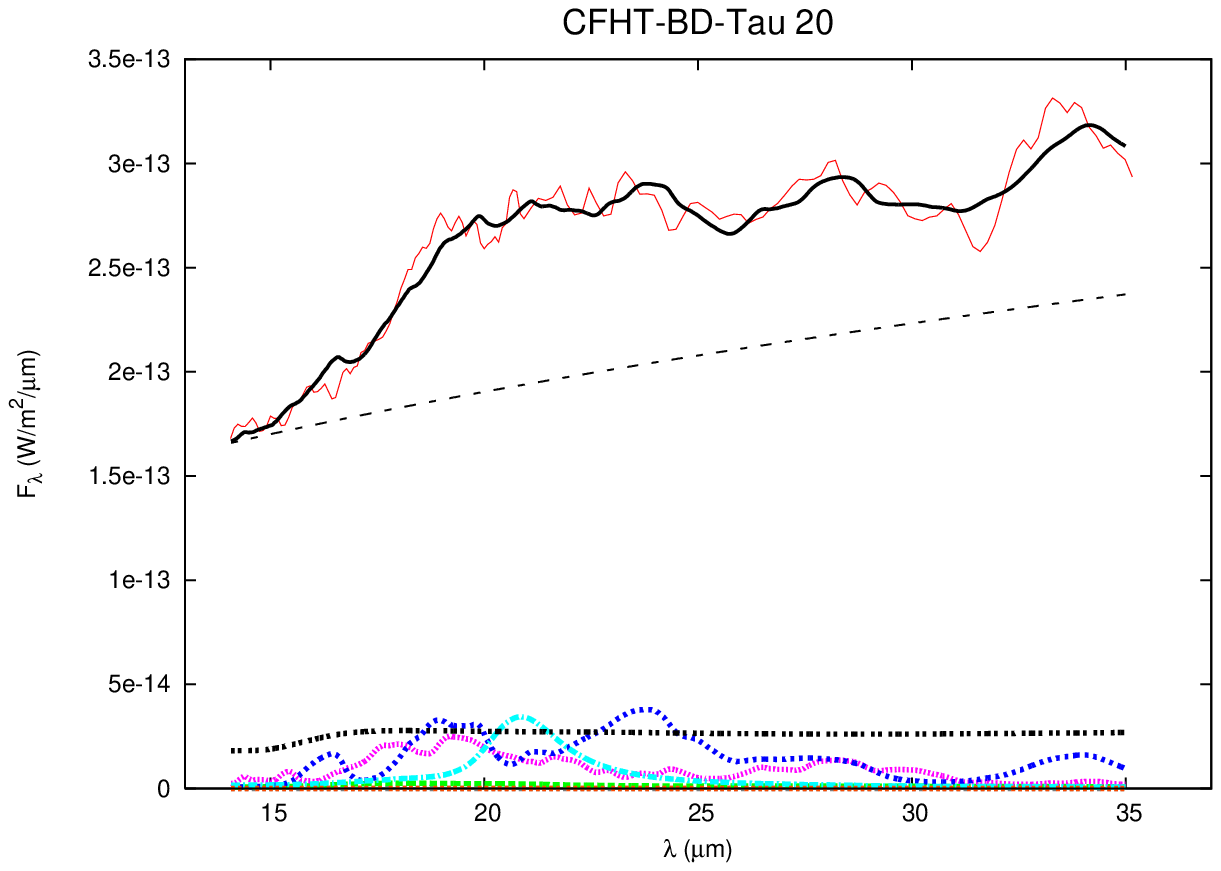} 
      \includegraphics[width=55mm]{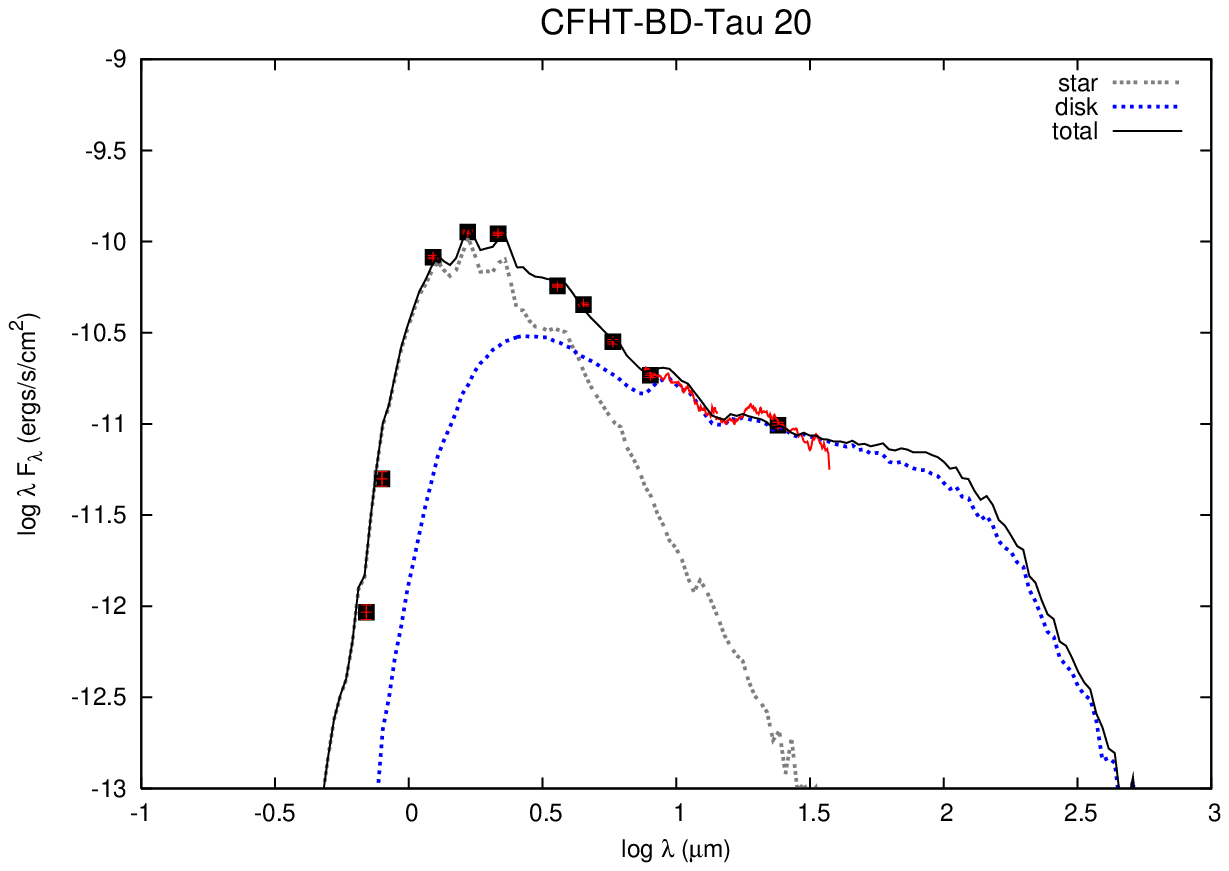} \\  
      \includegraphics[width=55mm]{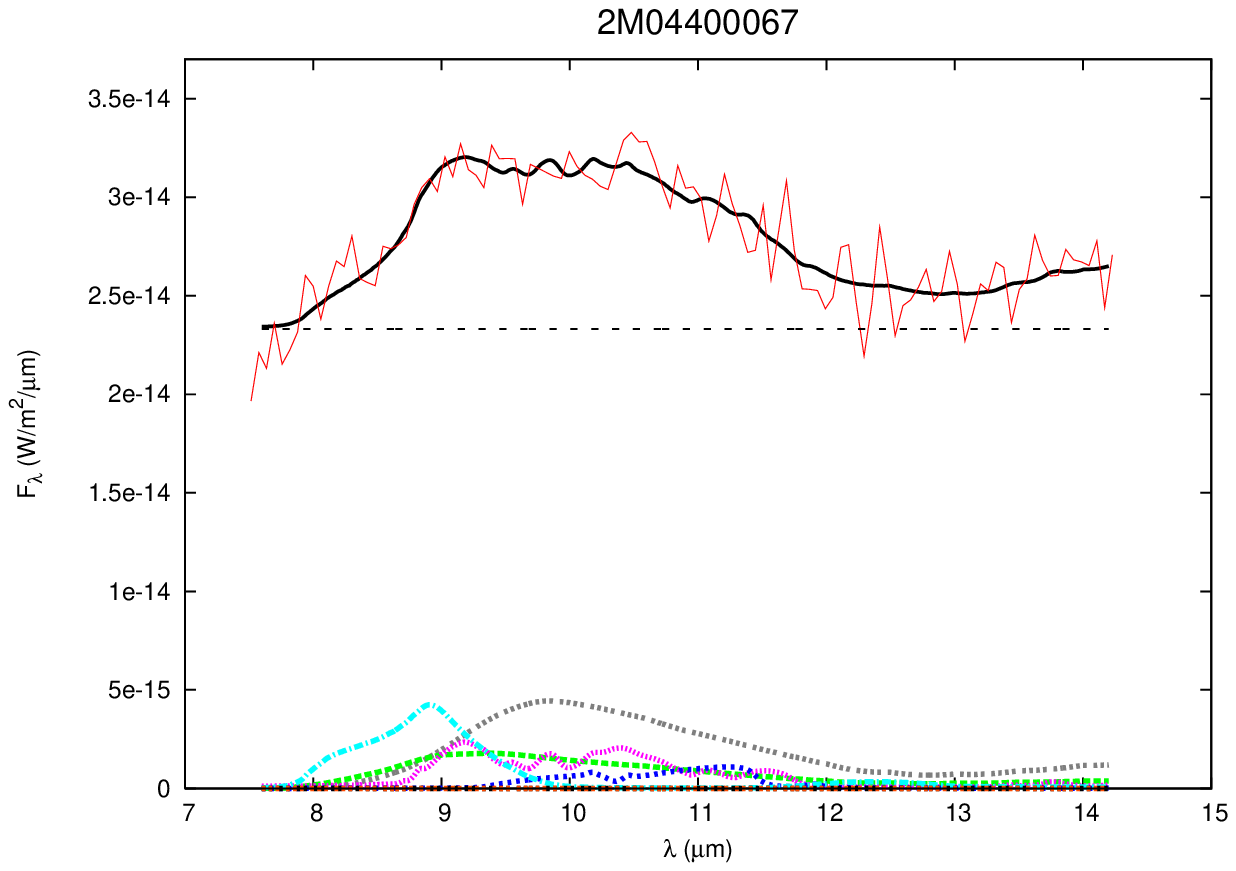} 
      \includegraphics[width=55mm]{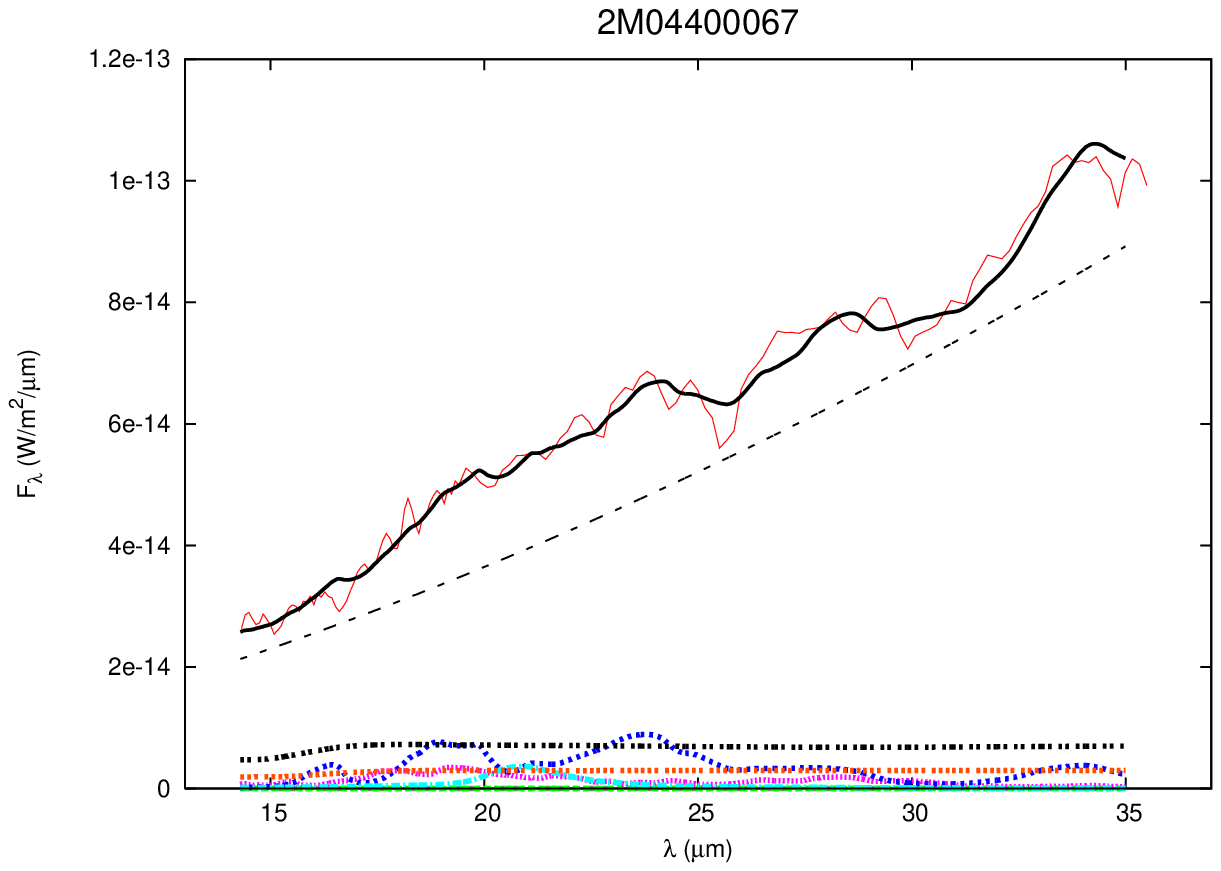} 
      \includegraphics[width=55mm]{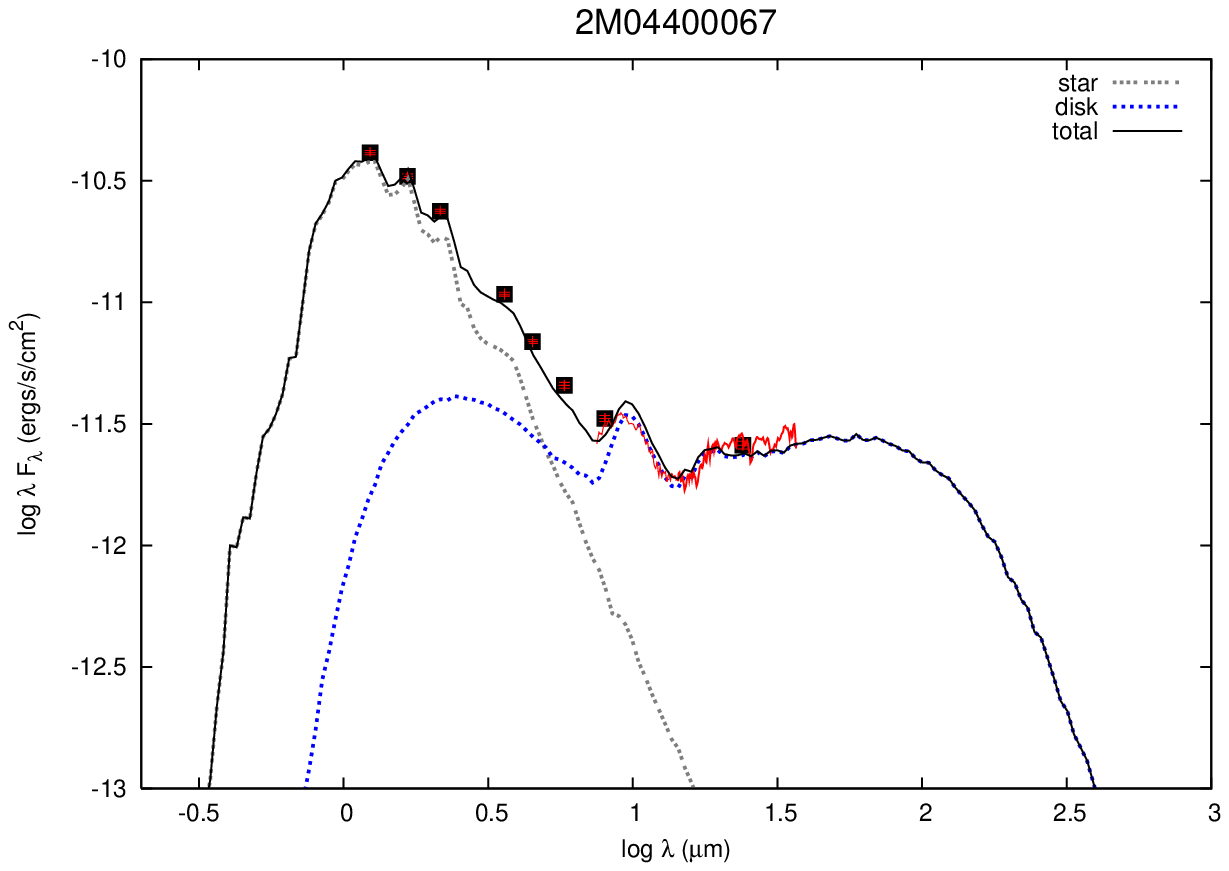} \\                        
    \caption{(b): Model fits for disks with prominent forsterite features at 20$\micron$. Symbol are the same as in Fig. 2a. } 
 \end{figure*}

\begin{figure*}
\setcounter{figure}{1}    
      \includegraphics[width=55mm]{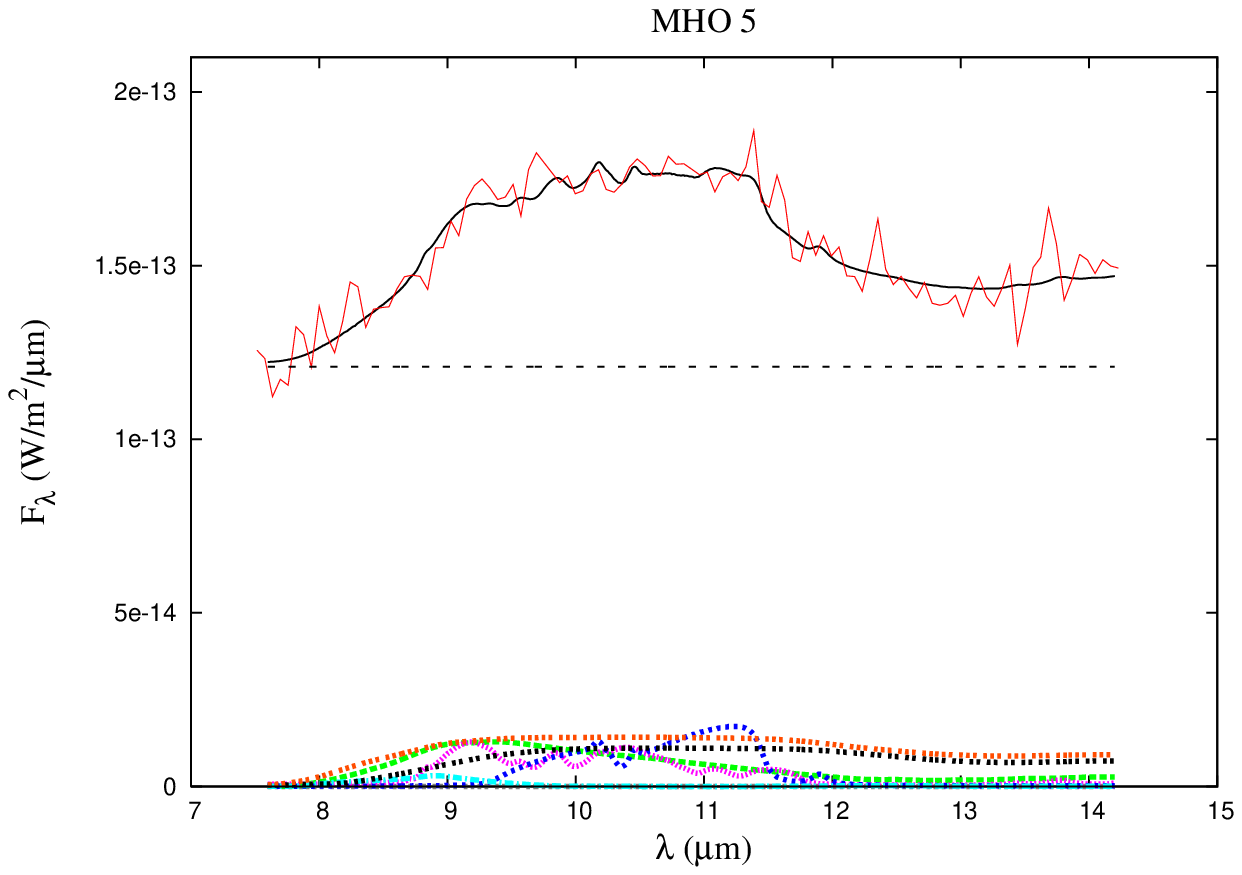} 
      \includegraphics[width=55mm]{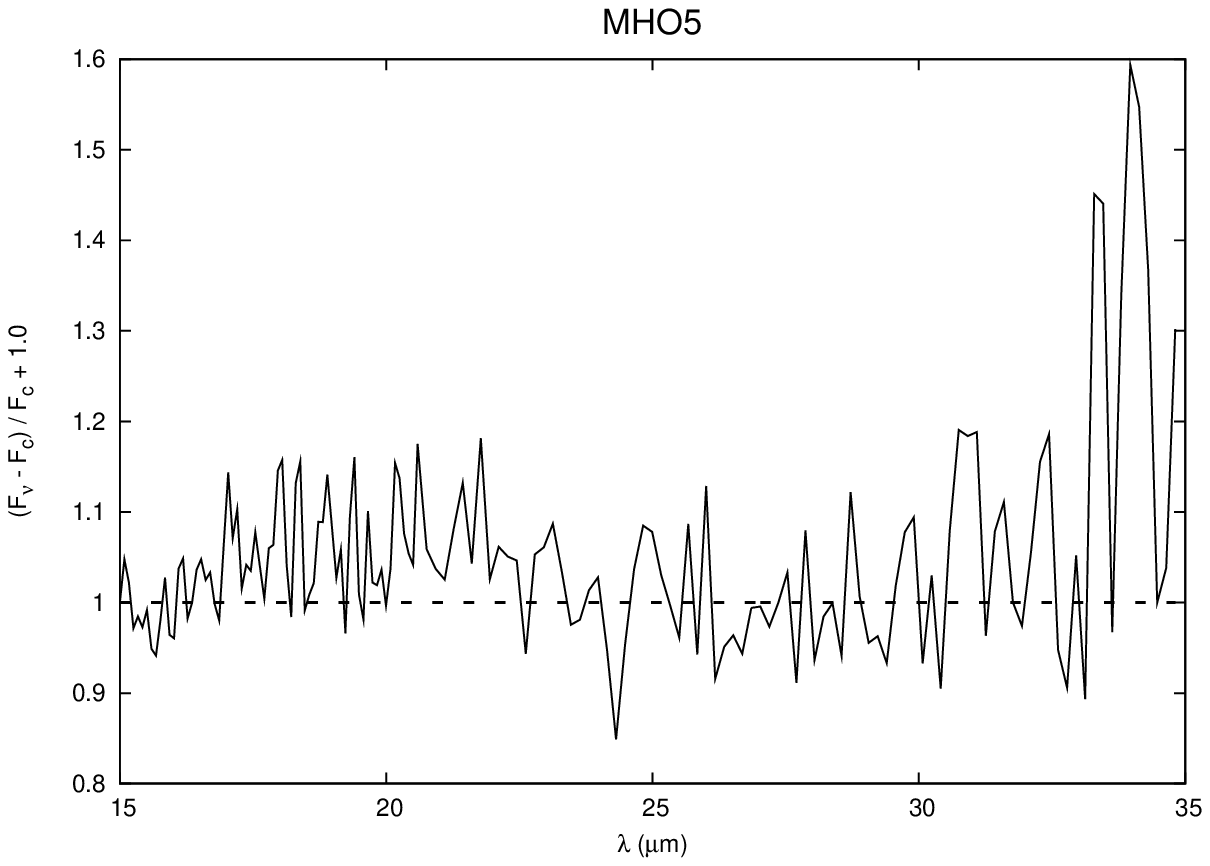} 
      \includegraphics[width=55mm]{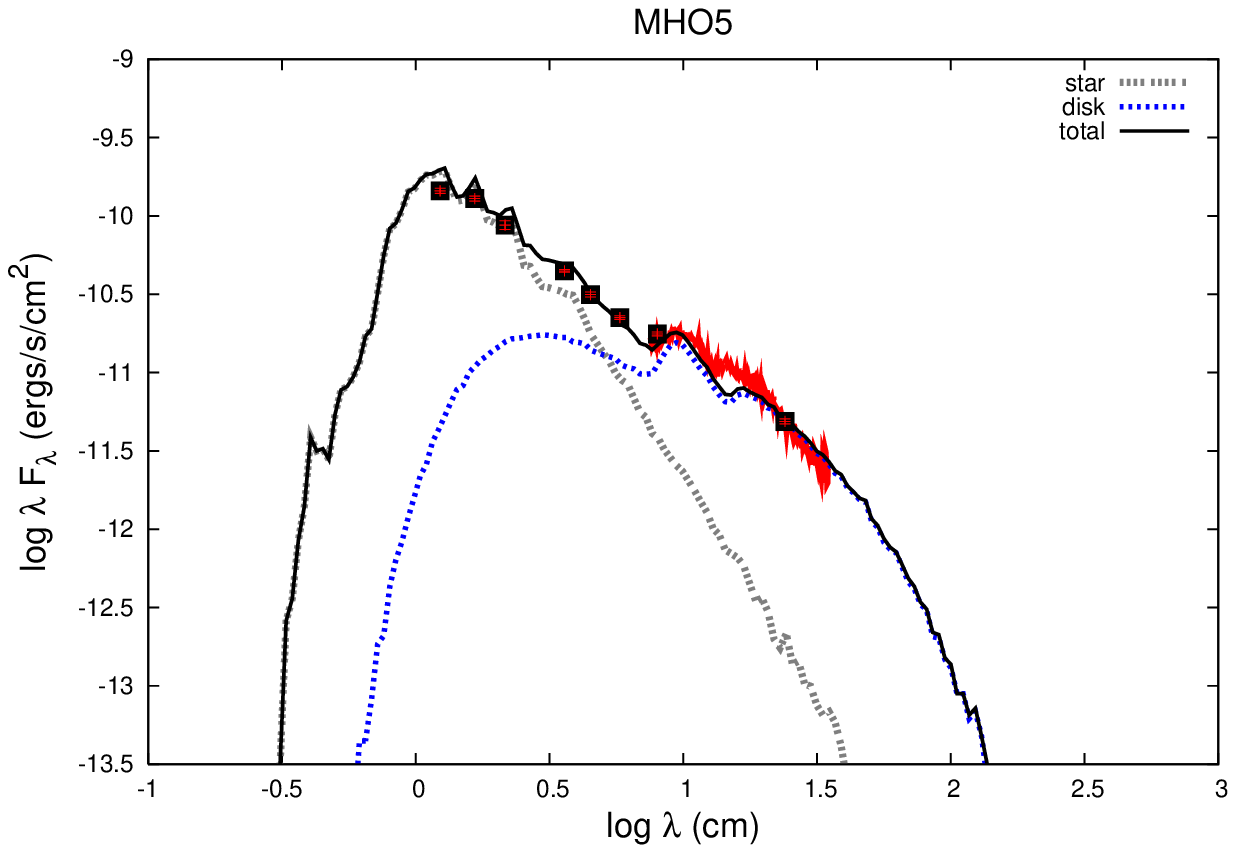} \\      
      \includegraphics[width=55mm]{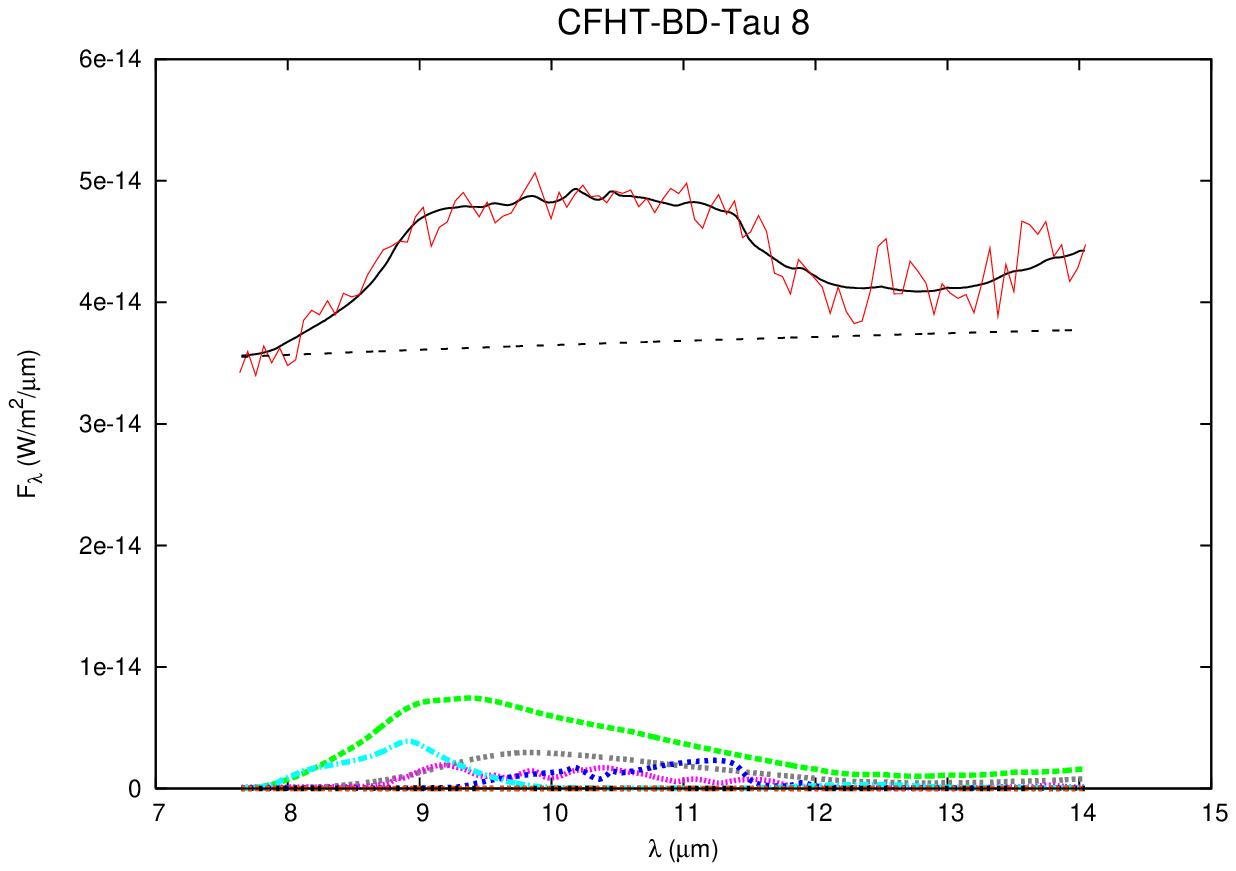} 
      \includegraphics[width=55mm]{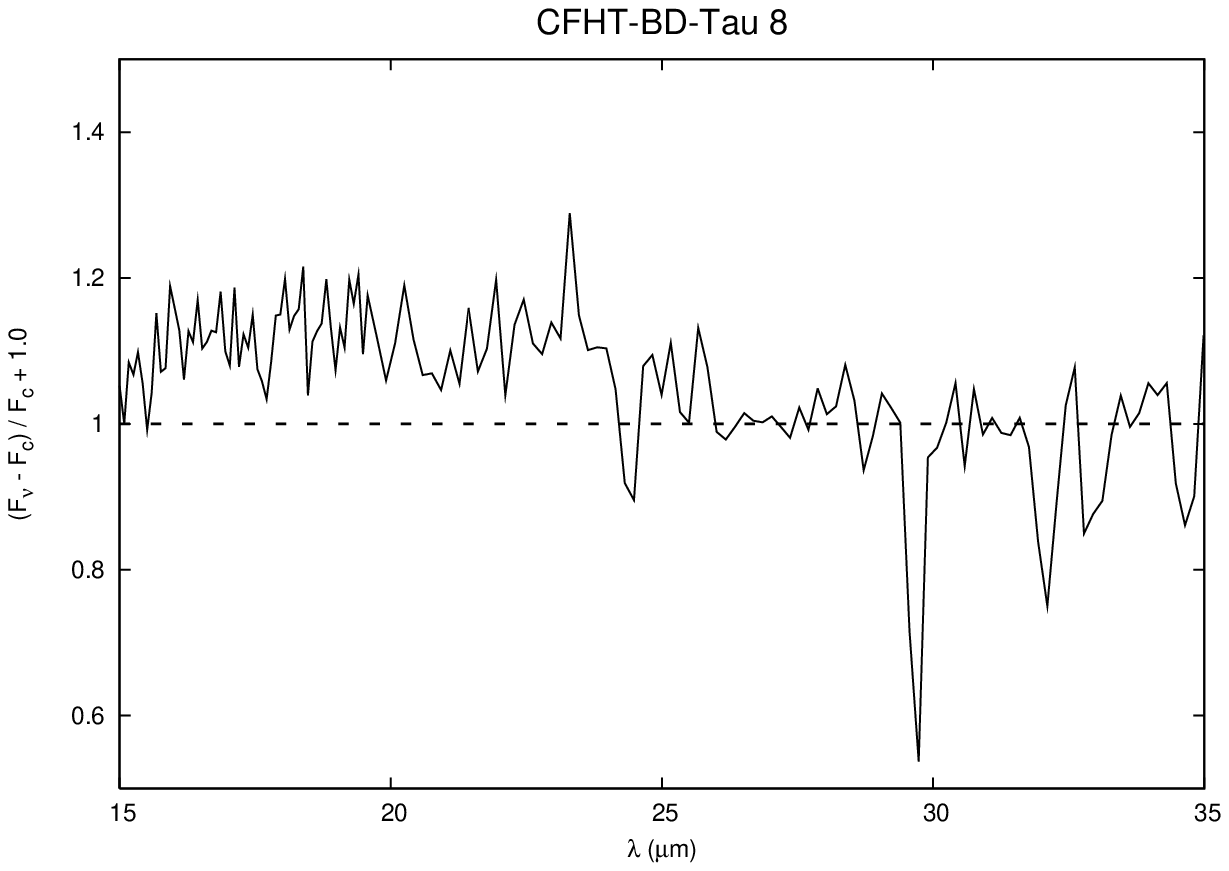} 
      \includegraphics[width=55mm]{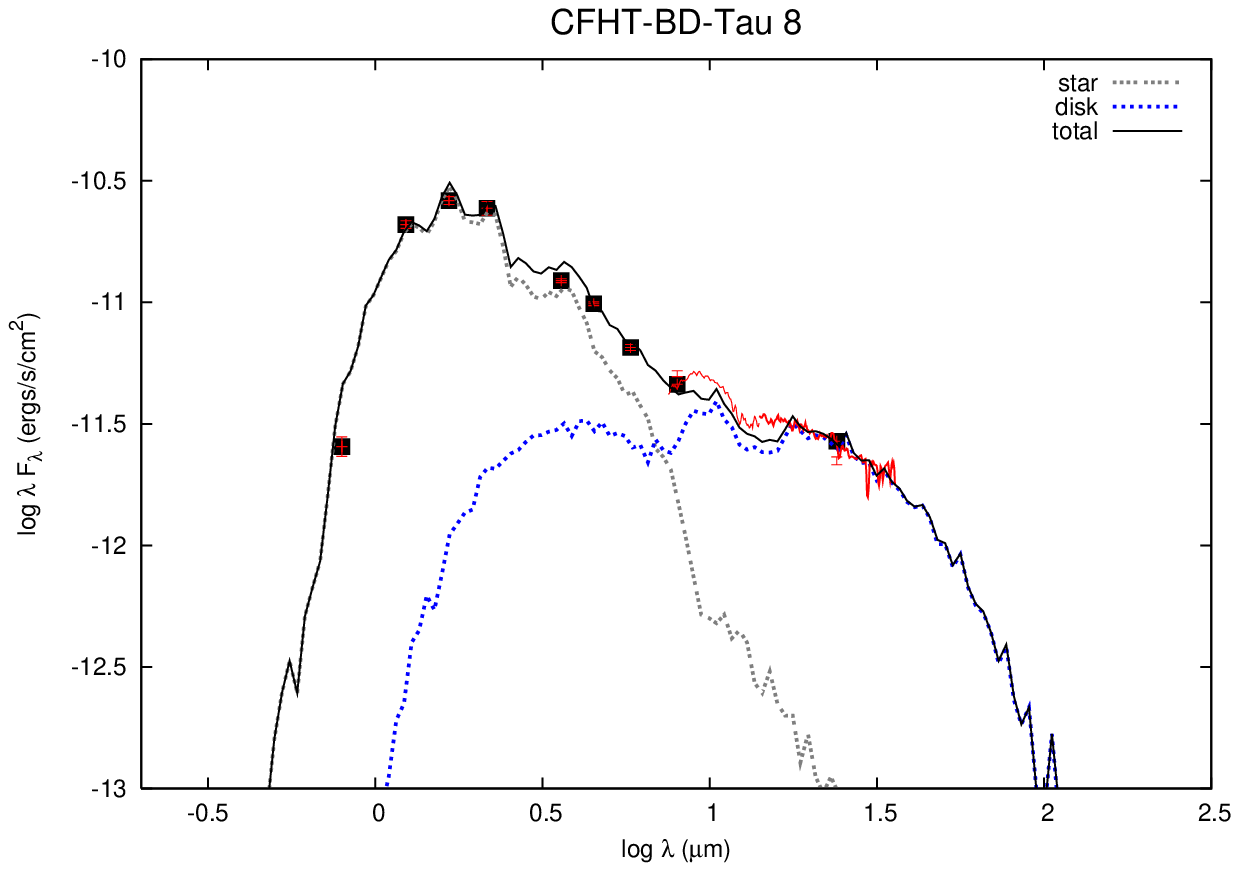} \\      
      \includegraphics[width=55mm]{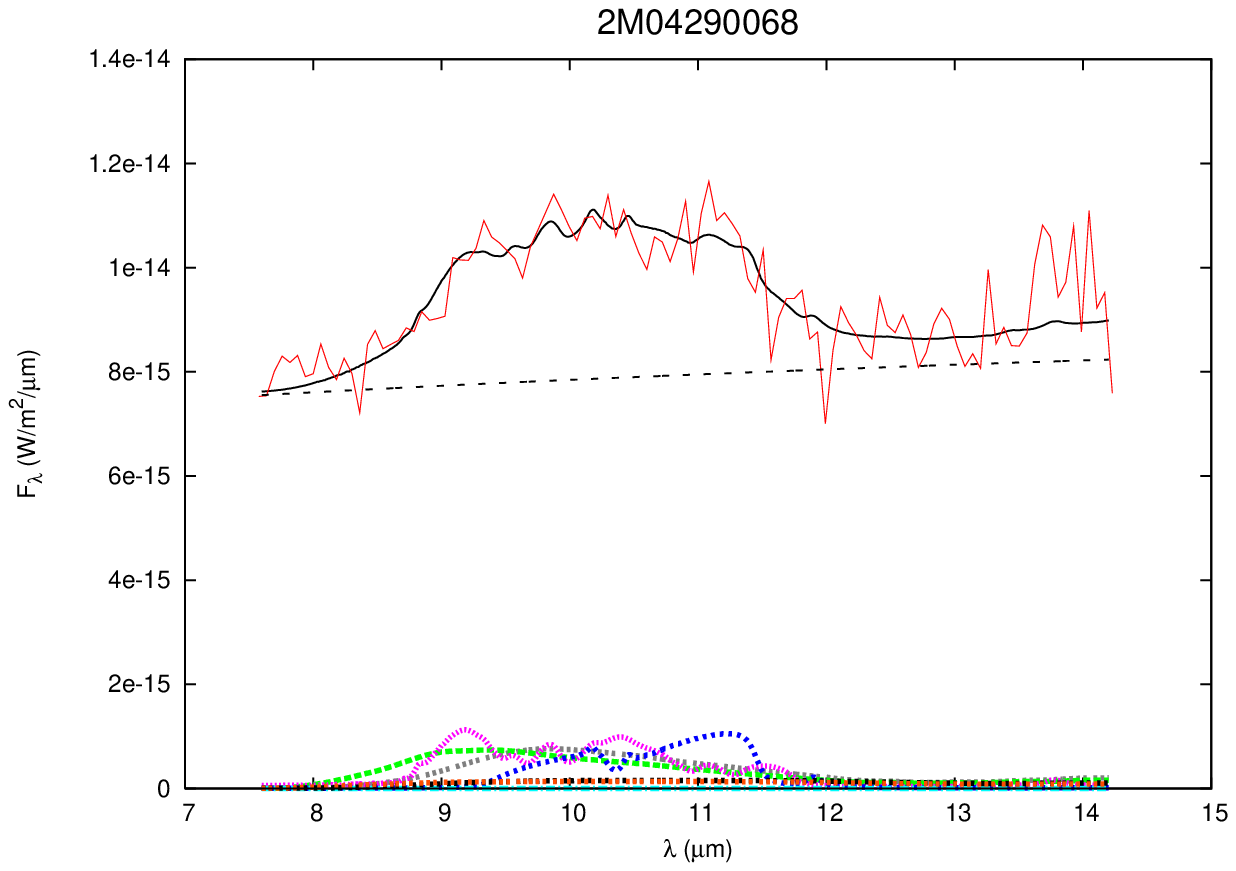} 
      \includegraphics[width=55mm]{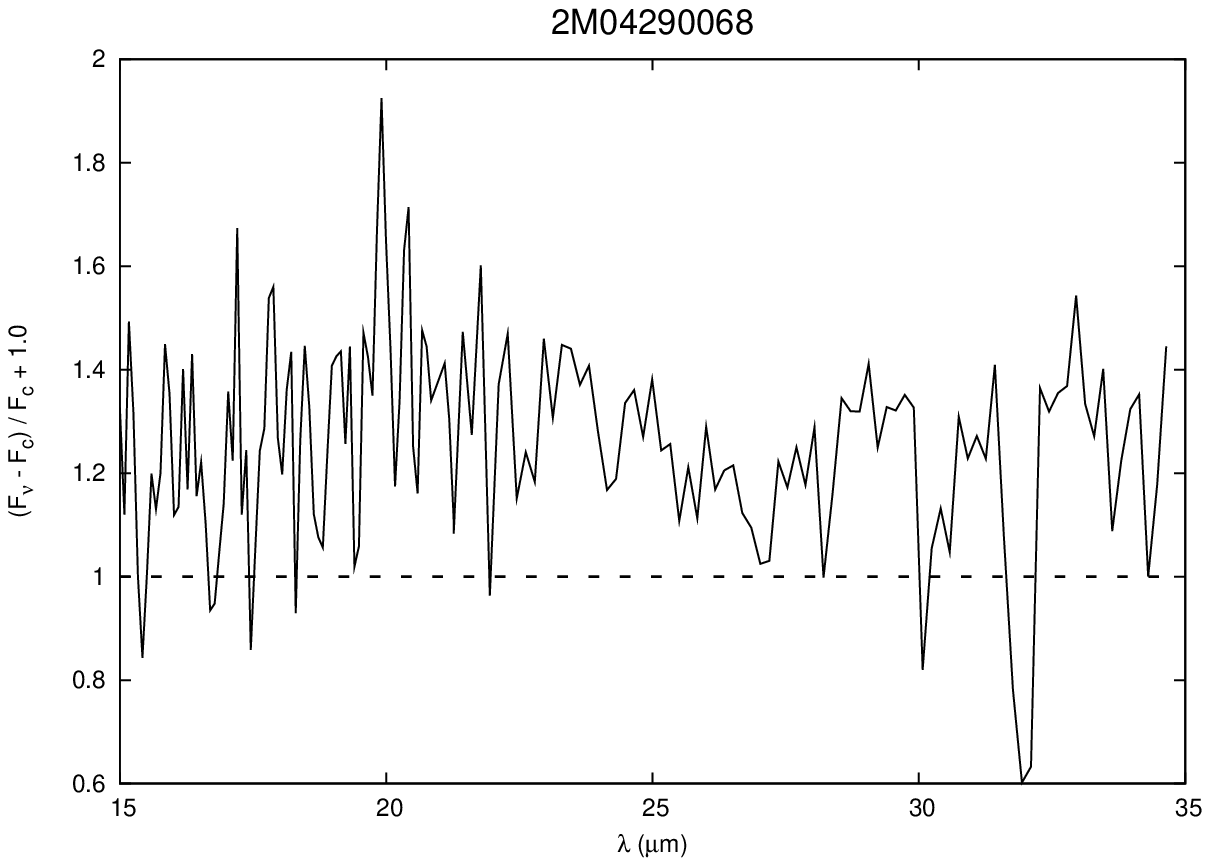} 
      \includegraphics[width=55mm]{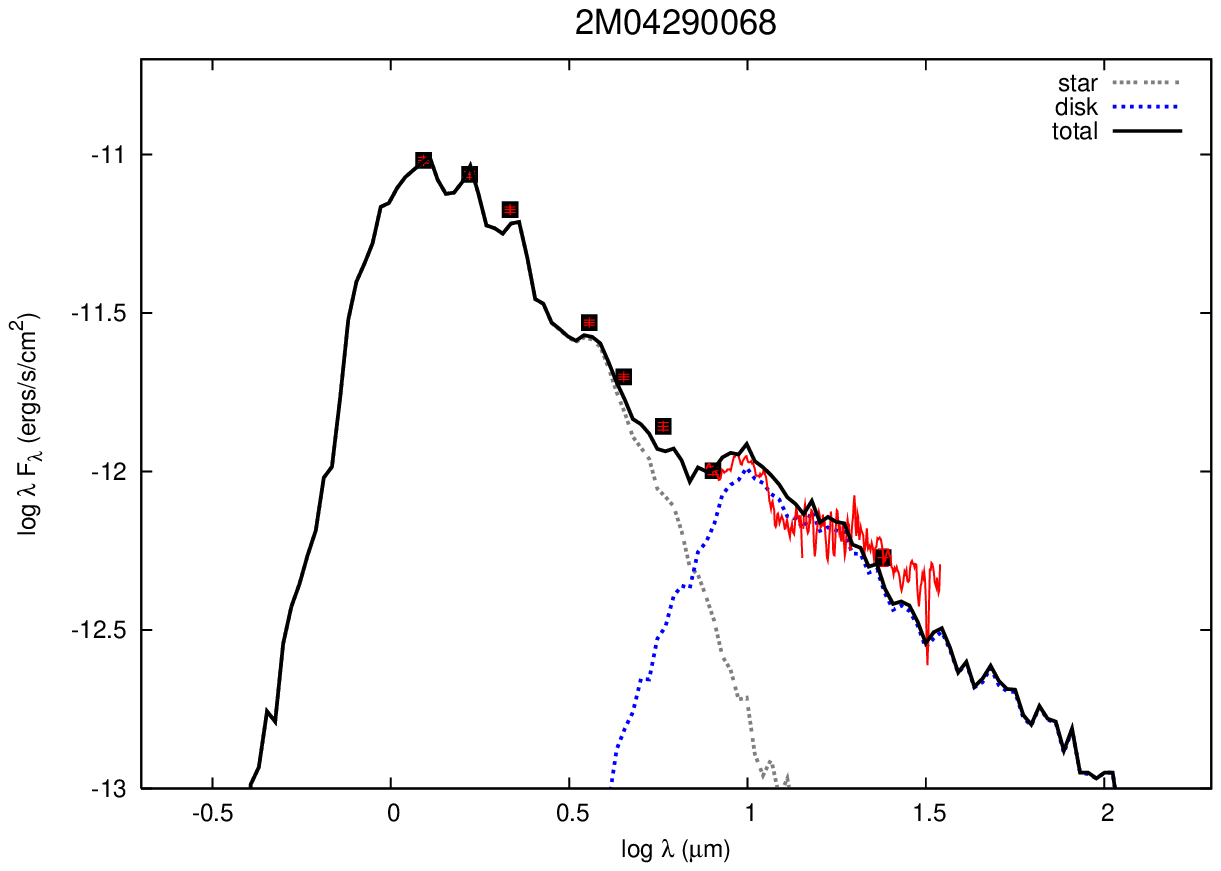} \\      
      \includegraphics[width=55mm]{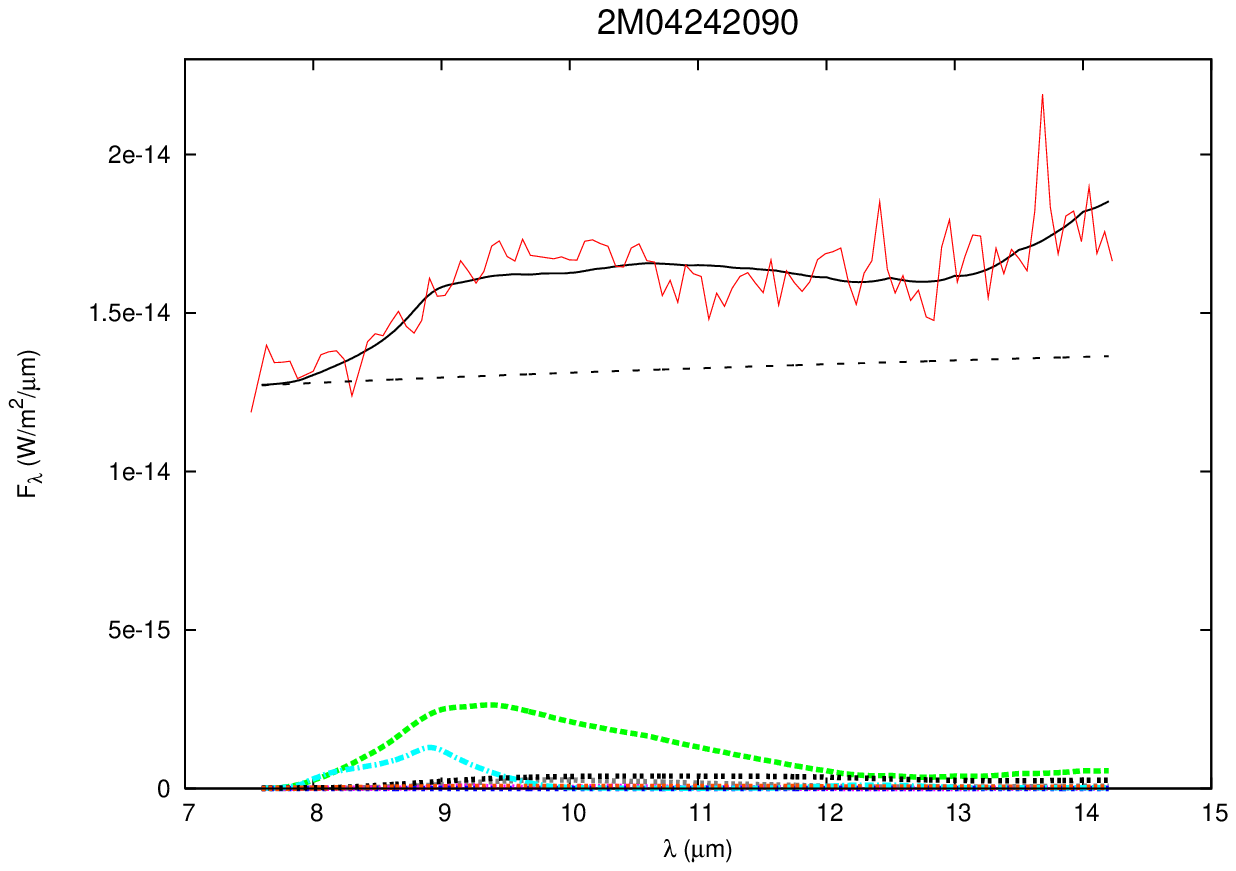} 
      \includegraphics[width=55mm]{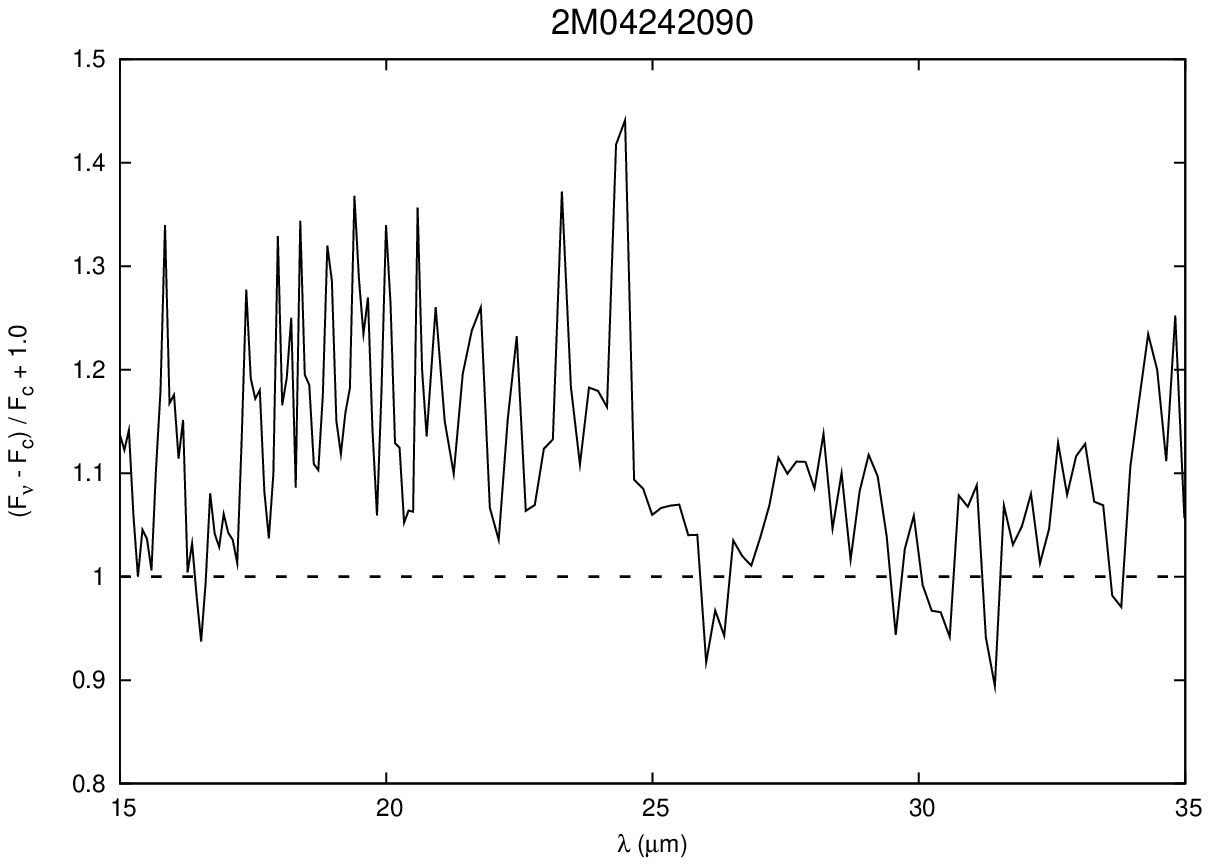} 
      \includegraphics[width=55mm]{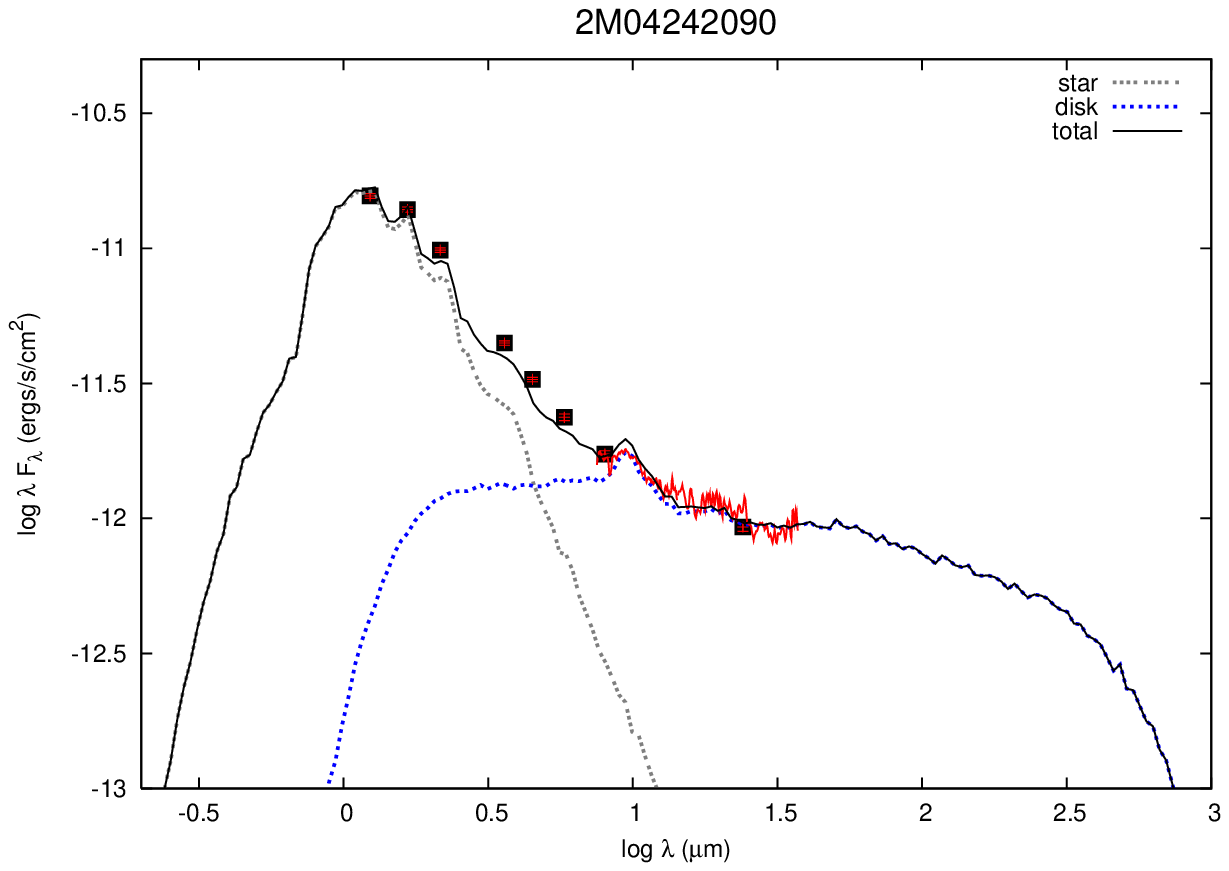} \\                
    \caption{(c): Model fits for the four `outliers' with negligible 20$\micron$ emission. The {\it middle} panel here shows the normalized continuum-subtracted 20$\micron$ spectra in units of ($F_{\nu} - F_{c})/F_{c}$. The dashed horizontal line represents the continuum. For the {\it left} and {\it right} panels, symbols are the same as in Fig. 2a.  } 
 \end{figure*}

\begin{figure*}
\setcounter{figure}{1}     
      \includegraphics[width=55mm]{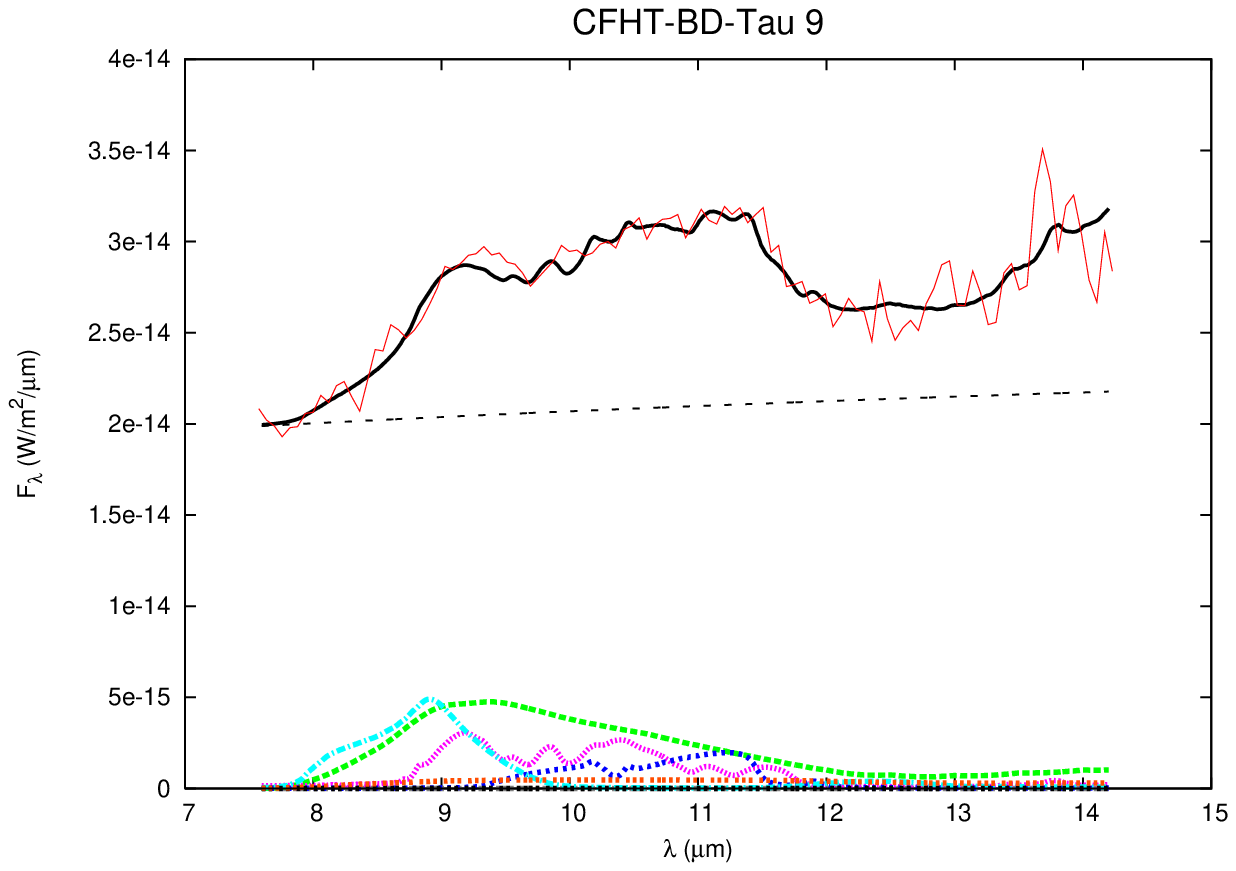} 
      \includegraphics[width=55mm]{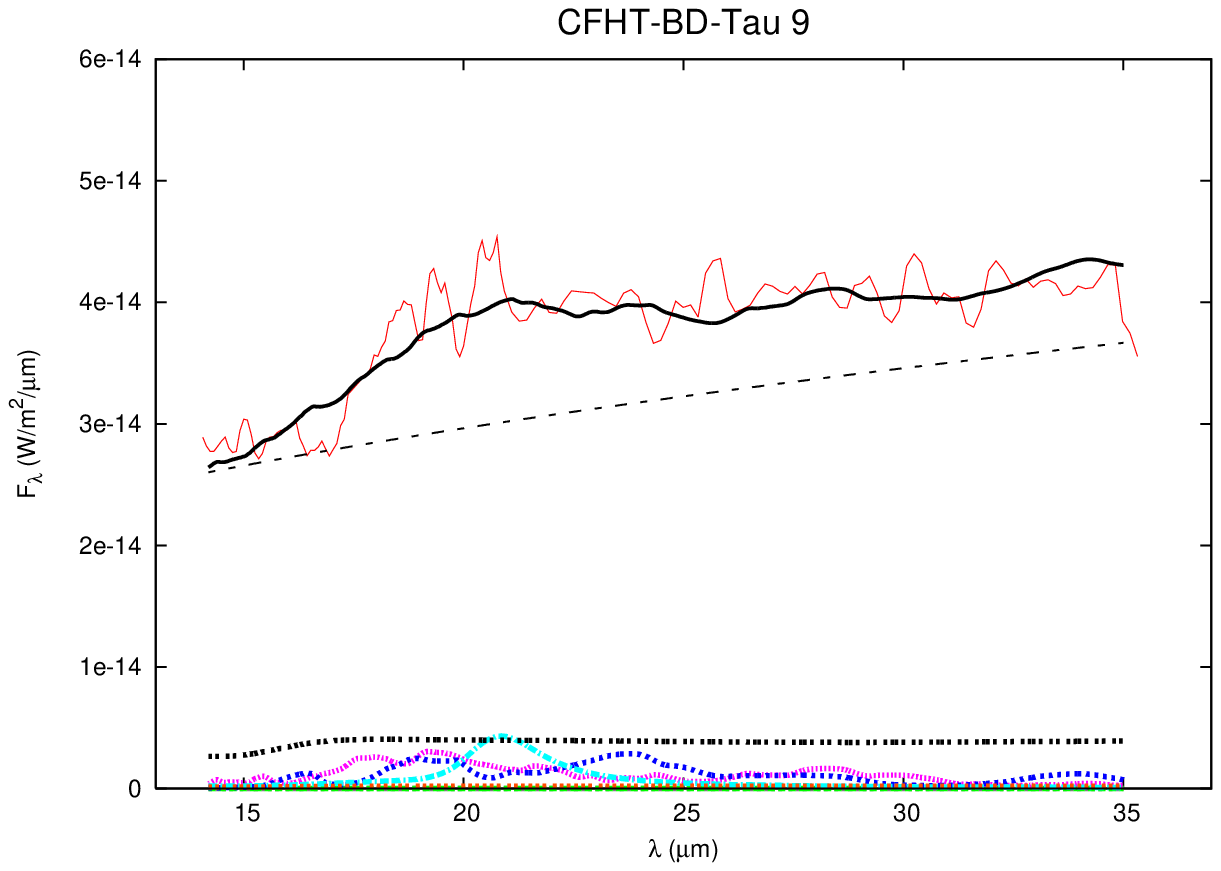} 
      \includegraphics[width=55mm]{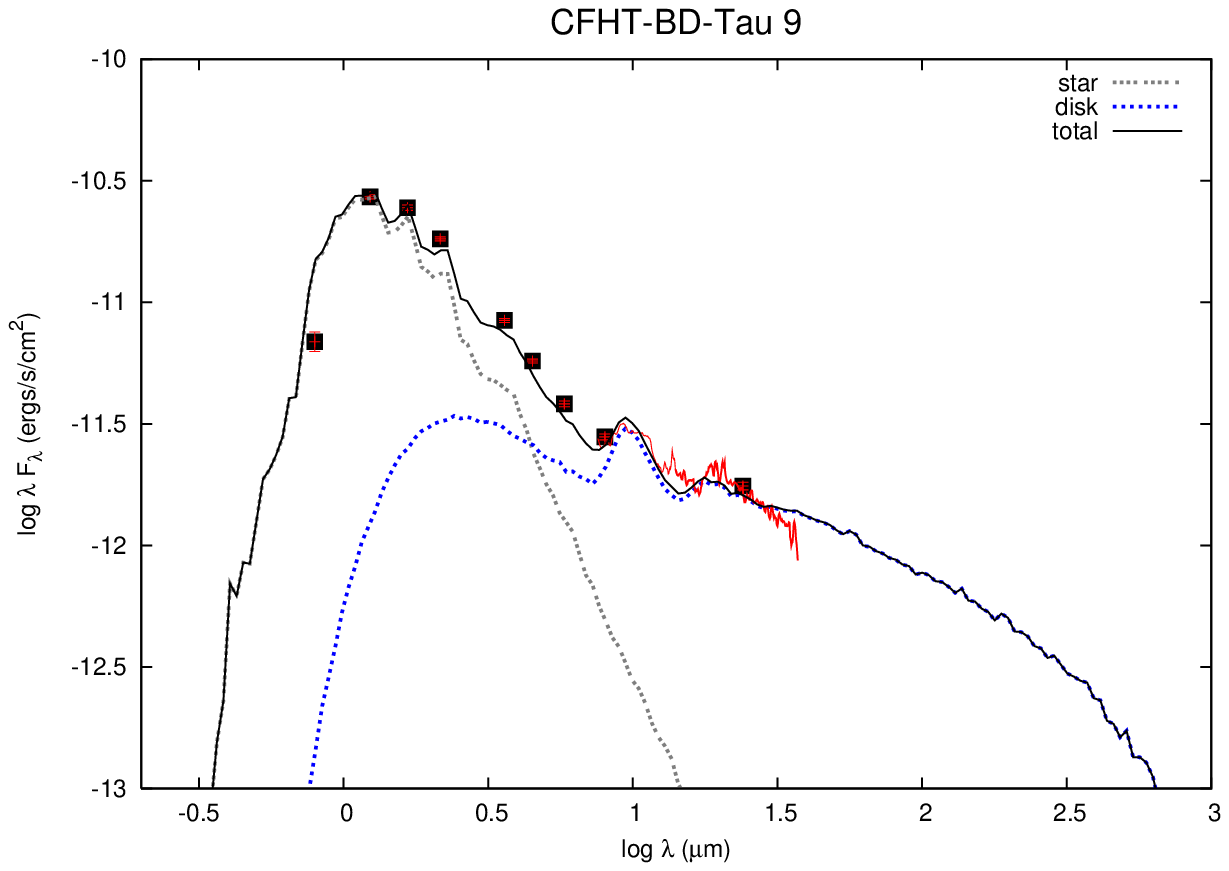} \\     
      \includegraphics[width=55mm]{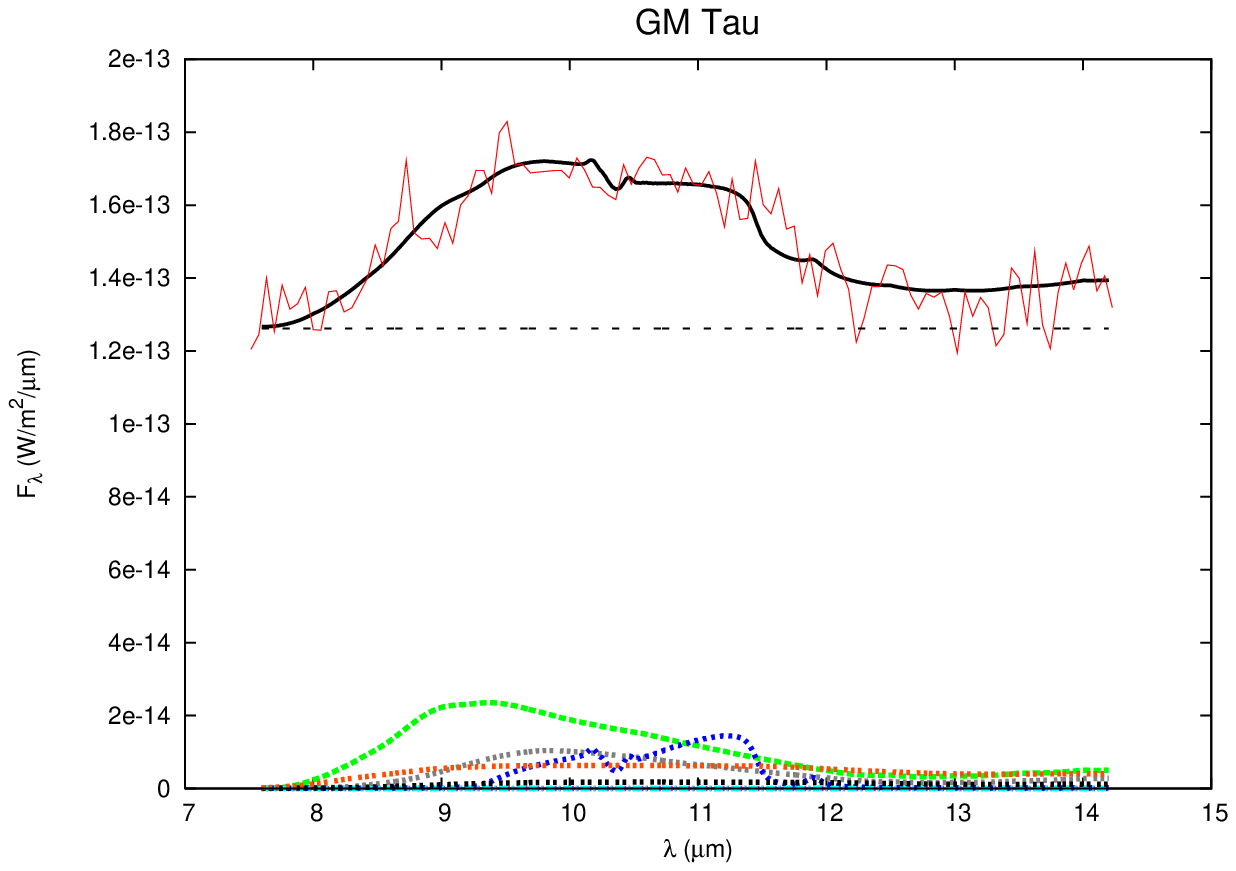}  
      \includegraphics[width=55mm]{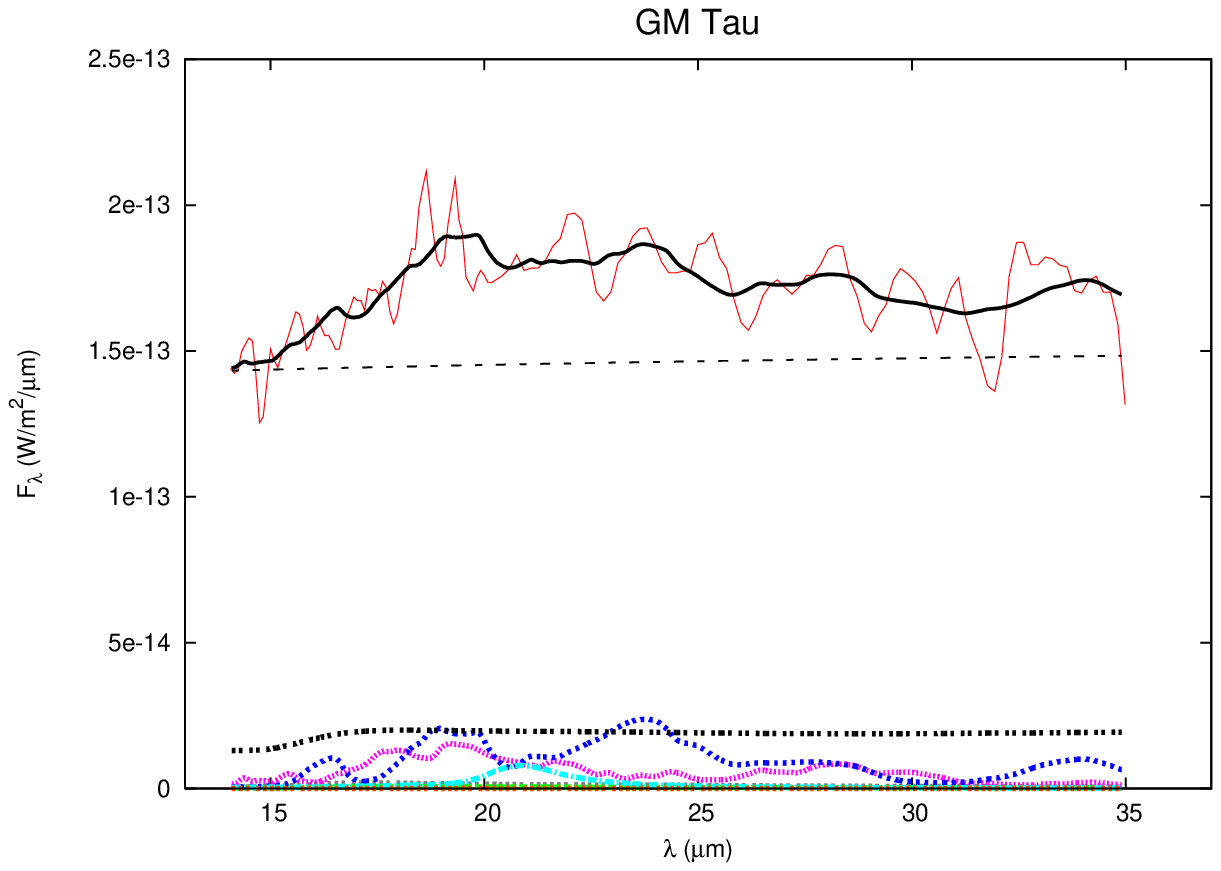}        
       \includegraphics[width=55mm]{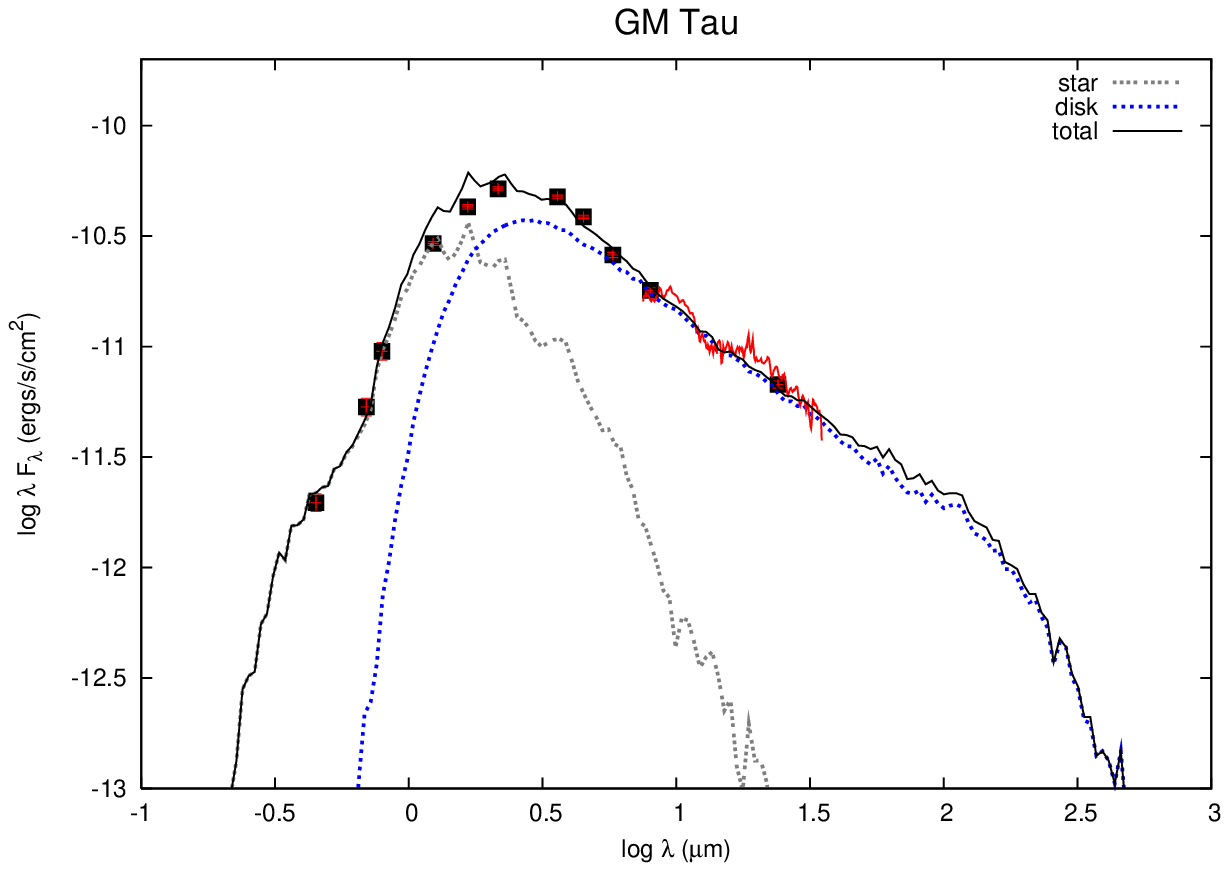} \\  
      \includegraphics[width=55mm]{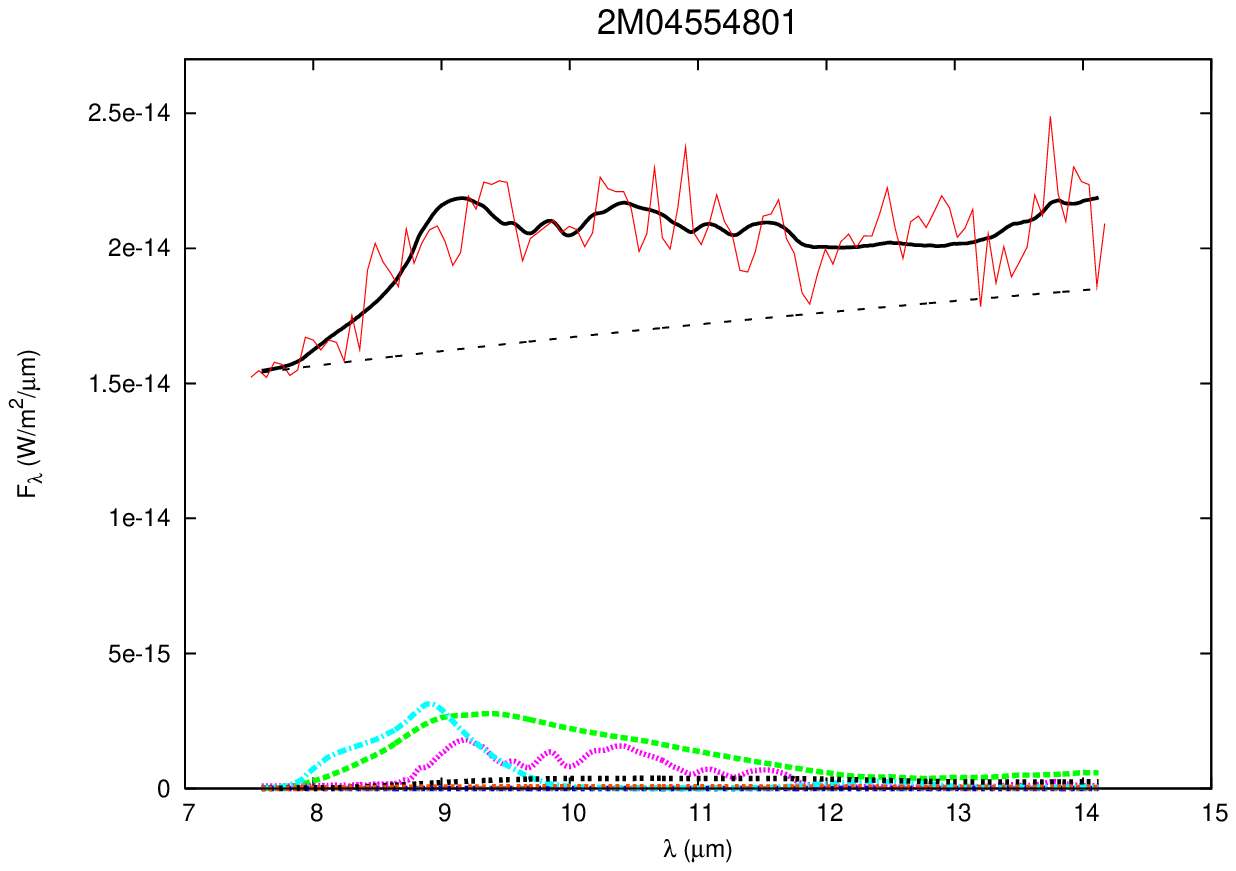}  
      \includegraphics[width=55mm]{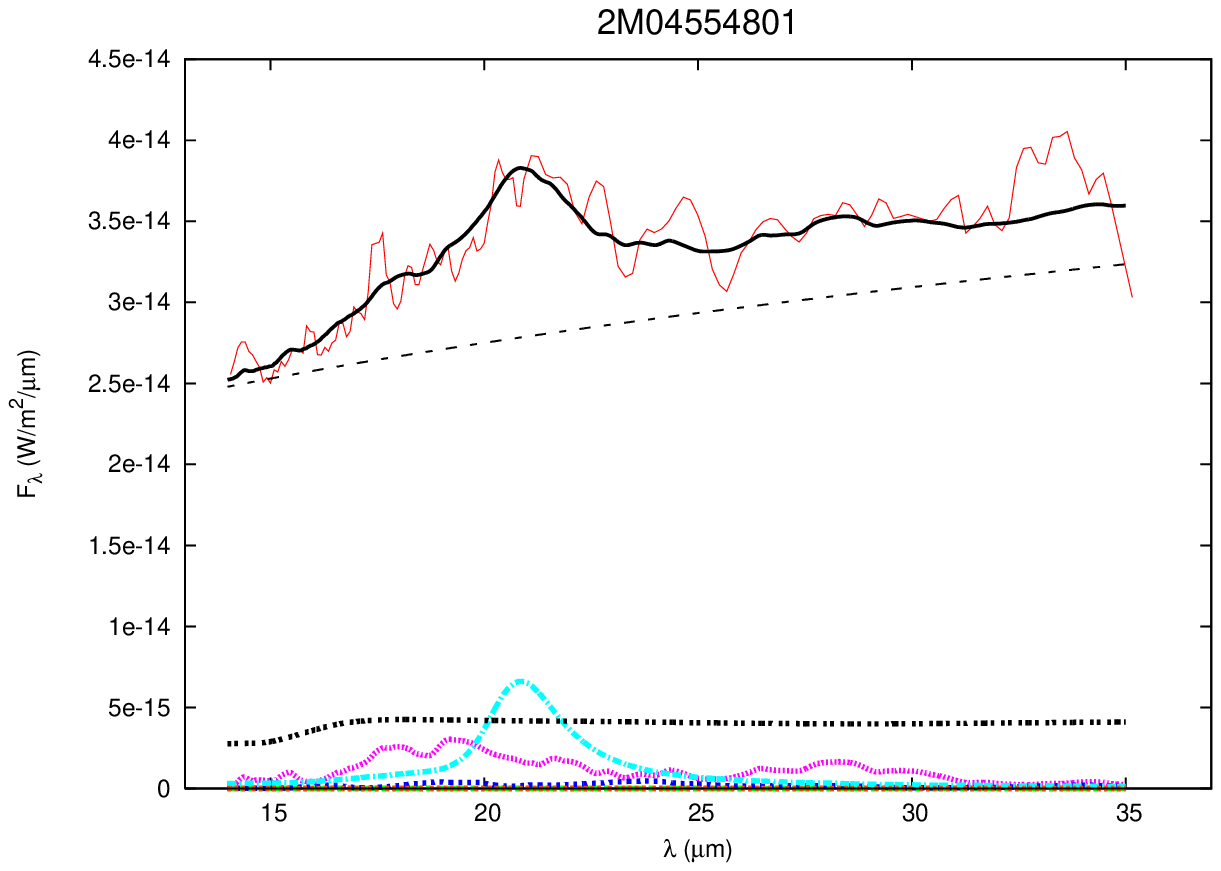}       
      \includegraphics[width=55mm]{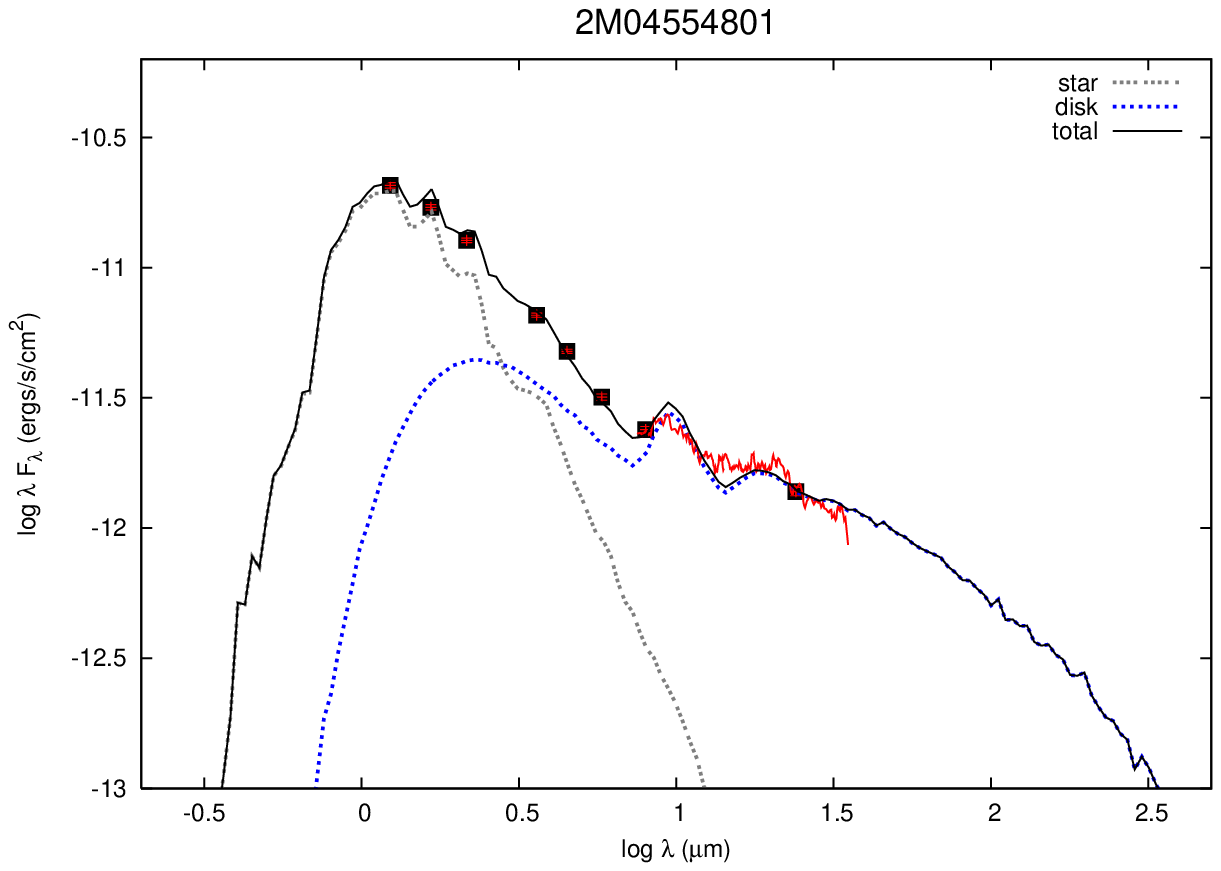} \\         
    \caption{(d): Model fits for the disks that show weaker features at 20$\micron$. Symbol are the same as in Fig. 2a.  } 
 \end{figure*}

\begin{onecolumn}
\begin{landscape}
\begin{table*}
\begin{minipage}{\linewidth}
\tiny
\caption{Dust Properties in the Inner and Outer Disk}
\label{results}
\begin{tabular}{lccccccccccccc@{}c@{}c}
\hline
Name  & SpT\footnote{References for spectral type: Luhman 2004, Luhman et al. (2006), Luhman (2006), Mart\'{i}n et al. (2004). Uncertainty in spectral type is $\pm$0.25.} & & Inner Disk (10$\micron$) & & & Outer Disk (20$\micron$) & & $R_{in}$ [$R_{sub}$]\footnote{Results for inner disk radius, $R_{in}$, the disk inclination, and the flaring power, $\beta$, form the best-fit disk models. The unit for $R_{in}$ is $R_{sub}$, the dust sublimation radius; 1$R_{sub}$ $\sim$ 0.0004 AU.} & {\it inclination}$^{b}$ & $\beta$$^{b}$ \\ \hline
  &    & Small\footnote{Percentage of small amorphous olivine and pyroxene silicates compared to all other silicates.} & Large\footnote{Percentage of large amorphous olivine and pyroxene silicates compared to all other silicates.} & Crystalline [ens; fors]\footnote{Percentage of sub-micron crystalline silicates (enstatite and forsterite) compared to all other silicates, including silica. The separate enstatite and forsterite fractions are listed in square brackets. } & Small & Large & Crystalline [ens; fors] &&& \\
  &    & (\%)  &  (\%) &  (\%) &  (\%) &  (\%) &  (\%) &&& \\ 

\hline

2MASS J04141188+2811535	&	M6.25	& 33.2$\pm$10 & 39.5$\pm$5 & 22.7$\pm$7 [8.2$\pm$4; 14.5$\pm$6] & 27.2$\pm$8 & 40.5$\pm$14 & 28.5$\pm$8 [9.7$\pm$5; 18.8$\pm$6] & 8.8 & 70-80$\degr$ & 1.172 \\
2MASS J04141760+2806096	&	M5.5	 &  81.4$\pm$7 & 10.5$\pm$6 & 4.7$\pm$3 [0.0001$\pm$0.5; 4.7$\pm$3] & 58$\pm$8 & 18.4$\pm$11 & 20.1$\pm$8 [7.4$\pm$4; 12.7$\pm$7] & 2.95 & 50-60$\degr$ & 1.144 \\
V410 X-ray 6	&	M6	& 85.6$\pm$8 & 14.3$\pm$5 & 0.009$\pm$0.6 [0.004$\pm$0.6; 0.005$\pm$0.6] & 60.6$\pm$9 & 19.7$\pm$7 & 14.3$\pm$5 [9.2$\pm$4; 5.1$\pm$3] & 118.5 & 70-80$\degr$ & 1.06 \\
2MASS J04230607+2801194	&	M6.5	 & 56.2$\pm$8 & 4.2$\pm$5 & 38.5$\pm$8 [24.7$\pm$7; 13.8$\pm$4] & 0.0004$\pm$11 & 21.9$\pm$18 & 56.9$\pm$15 [11.6$\pm$8; 45.4$\pm$13] & 1.0 & 50-60$\degr$ & 1.16 \\
CFHT-BD-Tau 20	&	M5.5	 & 41.7$\pm$10 & 0.013$\pm$9 & 49.9$\pm$13 [25.2$\pm$9; 24.7$\pm$10] & 1.6$\pm$9 & 46.2$\pm$16 & 37.2$\pm$12 [14.8$\pm$10; 22.4$\pm$7] & 1.0 & 50-60$\degr$ & 1.103 \\
CFHT-BD-Tau 9	&	M6.25	& 44.6$\pm$5 & 18.2$\pm$6 & 29.6$\pm$4 [17.2$\pm$2; 12.4$\pm$4] & 0.0001$\pm$9 & 59.8$\pm$14 & 26.1$\pm$8 [13.5$\pm$8; 12.6$\pm$8] & 1.0 & 30-40$\degr$ & 1.1 \\
2MASS J04400067+2358211	&	M6.5	  & 59.7$\pm$6 & 5$\pm$3 & 24.8$\pm$6 [15.3$\pm$4; 9.4$\pm$5] & 0.0006$\pm$10 & 63.4$\pm$12 & 28.7$\pm$7 [10.1$\pm$4; 18.5$\pm$6] & 1.0 & 50-60$\degr$ & 1.185 \\
GM Tau & M6.5 & 51.1$\pm$8 & 30.2$\pm$4 & 15.7$\pm$5 [0.001$\pm$2; 15.7$\pm$5] & 2.6$\pm$11 & 47.7$\pm$19 & 43.3$\pm$10 [17.1$\pm$10; 26.2$\pm$10] & 1.0 & 70-80$\degr$ & 1.03 \\
2MASS J04554801+3028050	&	M5.6	 & 39.2$\pm$5 & 31.2$\pm$7 & 16.7$\pm$10 [16.7$\pm$10; 0.0005$\pm$2] &  0.0002$\pm$11 & 52.4$\pm$18 & 19.9$\pm$14 [17.3$\pm$12; 2.6$\pm$7] & 1.0 & 30-40$\degr$ & 1.1 \\
MHO 5	&	M6.2	  & 13.5$\pm$4 & 63.7$\pm$11 & 22.5$\pm$5 [10.5$\pm$3; 12$\pm$4] & -- & -- & -- & 1.0 & 30-40$\degr$ & 1.13 \\ 
CFHT-BD-Tau 8	&	M6.5	  &  72.2$\pm$7 & 0.004$\pm$5 & 20.4$\pm$6 [8.2$\pm$4; 12.2$\pm$5] &  -- & -- & -- & 1.0 & 70-80$\degr$ & 1.12 \\
2MASS J04290068+2755033	&	M8.25	& 41.3$\pm$3 & 18.6$\pm$7 & 36.3$\pm$7 [17.6$\pm$4; 18.7$\pm$6] &  -- & -- & -- & 17.9 & 70-80$\degr$ & 1.06 \\
2MASS J04242090+2630511	&	M7	&  66.5$\pm$11 & 24.4$\pm$15 & 1.1$\pm$2 [0.55$\pm$2; 0.55$\pm$2] &  -- & -- & -- & 1.0 & 30-40$\degr$ & 1.16 \\

\hline
\end{tabular}
\end{minipage}
\end{table*}
\end{landscape}
\end{onecolumn}

\begin{figure*}     
           \includegraphics[width=85mm]{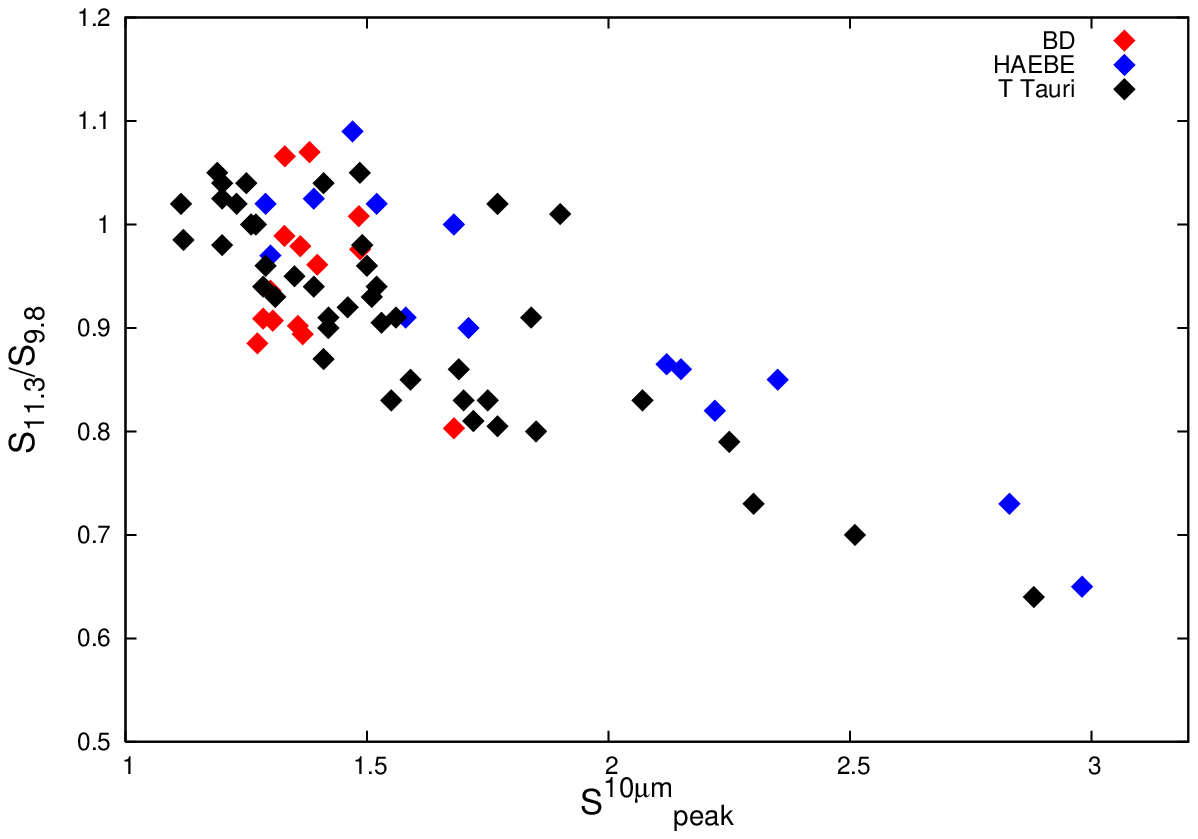} 
     \includegraphics[width=85mm]{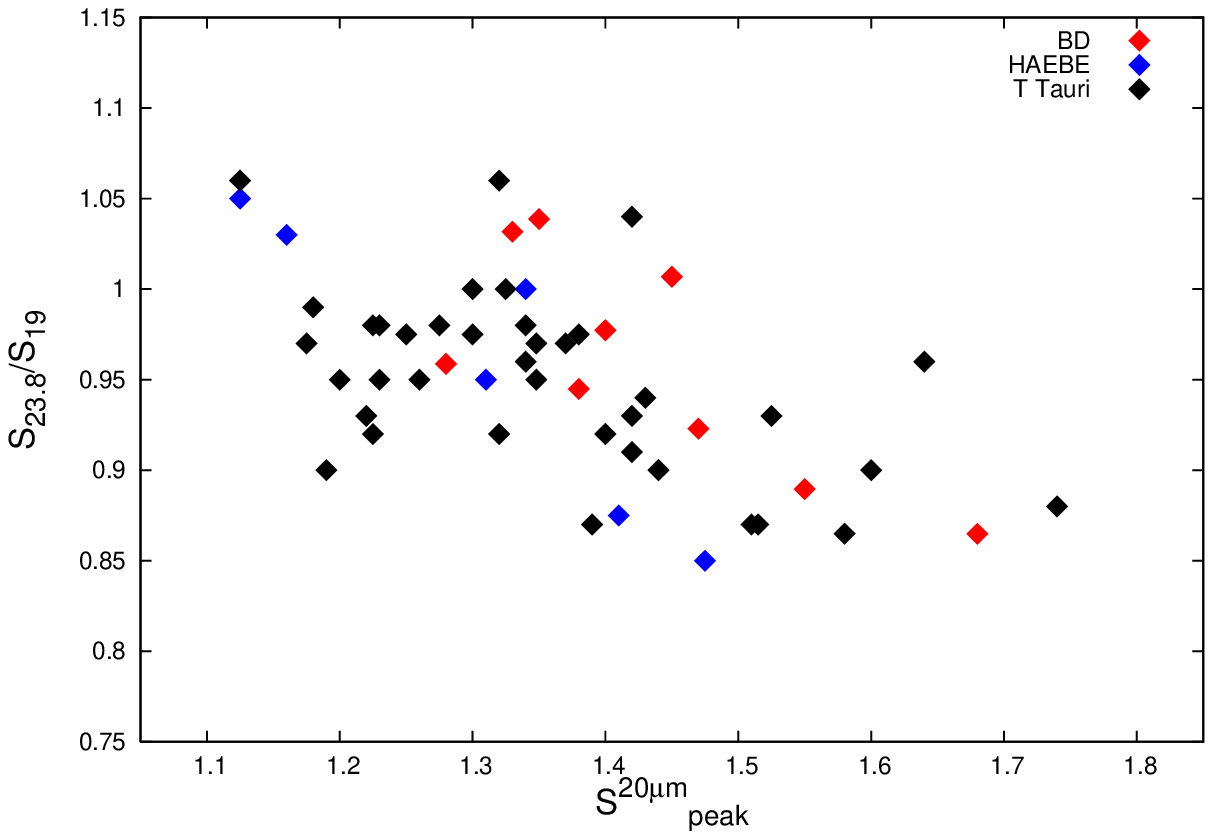} \\               
    \caption{The shape vs. strength of the 10$\micron$ ({\it left}) and 20$\micron$ ({\it right}) silicate emission features. Colors represent the following: red--Taurus brown dwarfs; blue--Herbig Ae/Be stars; black--T Tauri stars. Data for Herbig Ae/Be stars and T Tauri stars is from van Boekel et al. (2003), Przygodda et al. (2003) and Kessler-Silacci et al. (2005; 2006).  } 
    \label{strength}  
 \end{figure*}
 
\begin{figure*}     
     \includegraphics[width=85mm]{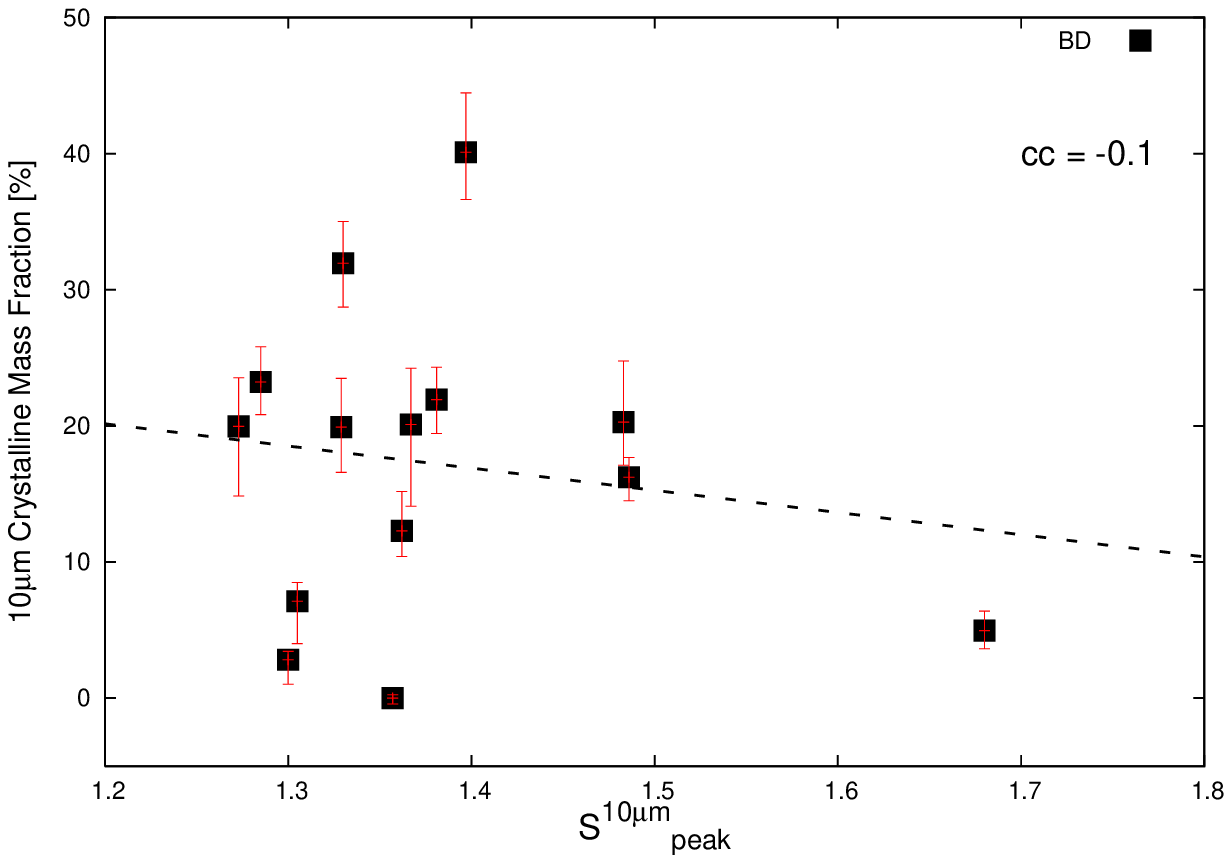}  
     \includegraphics[width=85mm]{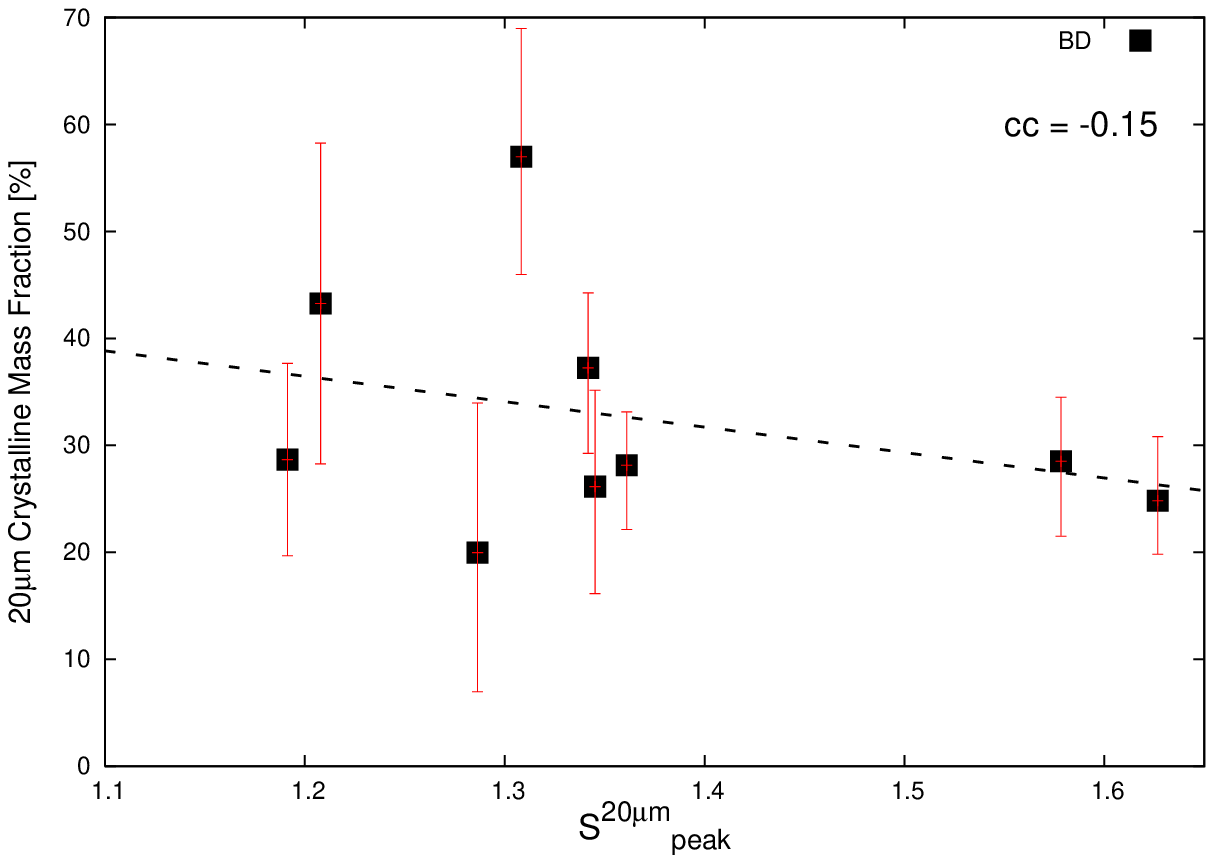} \\      
           \includegraphics[width=85mm]{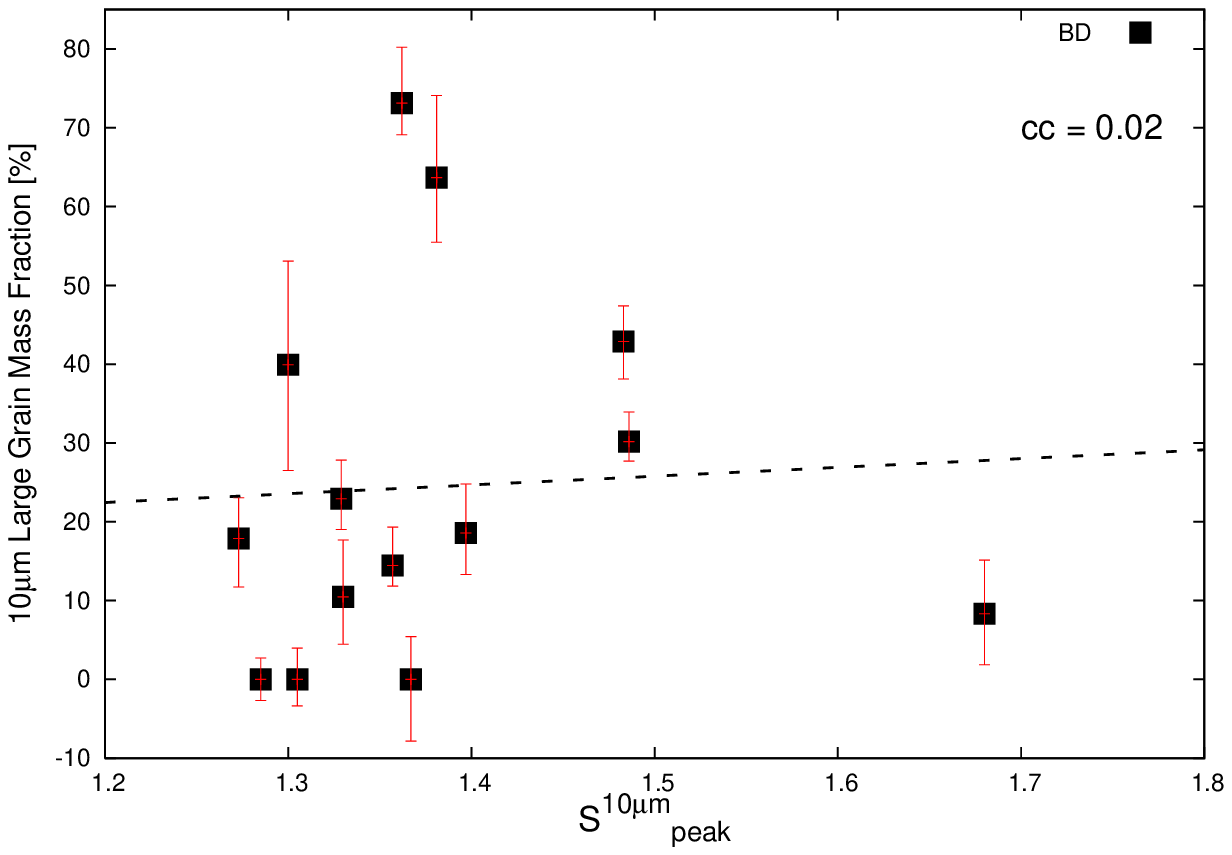} 
     \includegraphics[width=85mm]{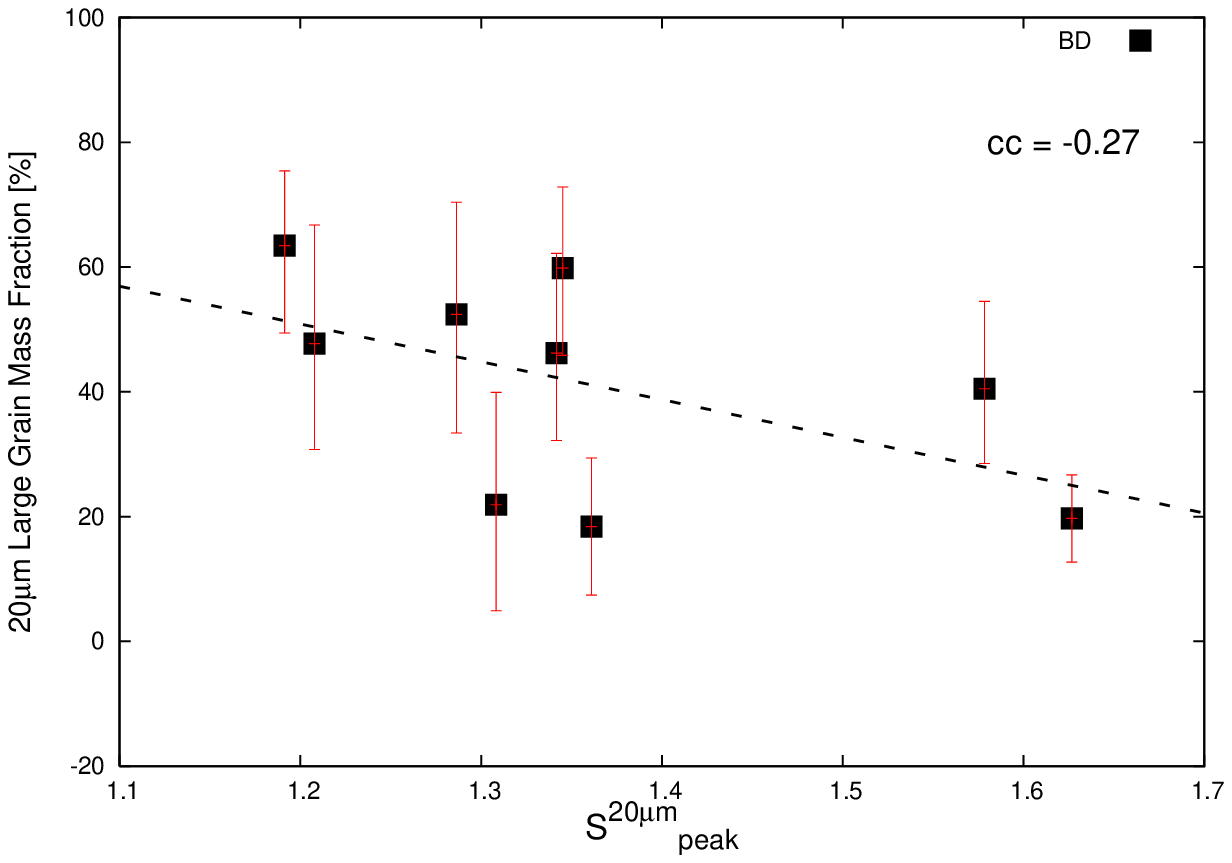} \\               
    \caption{The strength in the 10$\micron$ ({\it left}) and 20$\micron$ ({\it right}) features for brown dwarfs versus the crystalline and large-grain mass fractions. The correlation coefficients (cc) are noted in the top right corner.  } 
    \label{strength-dust}  
 \end{figure*} 
 
 \begin{figure*}     
     \includegraphics[width=85mm]{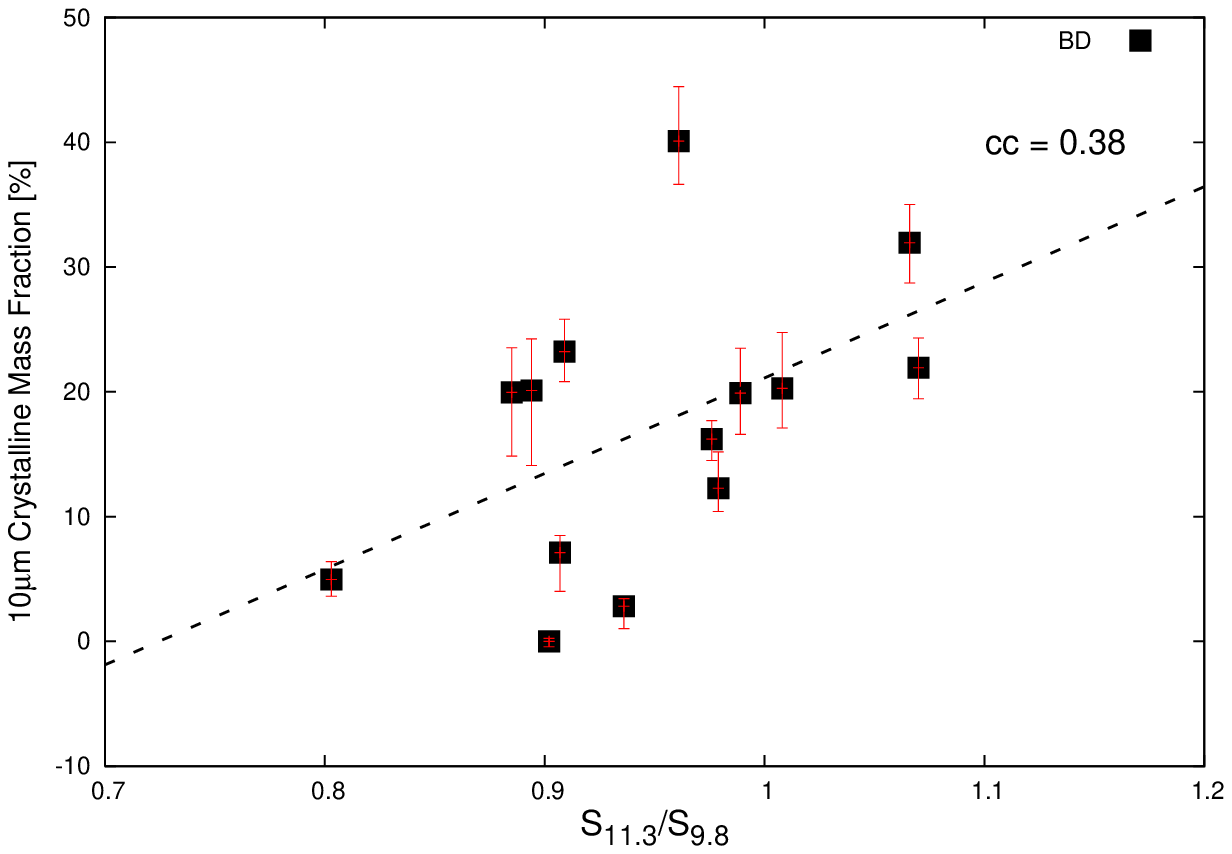}  
     \includegraphics[width=85mm]{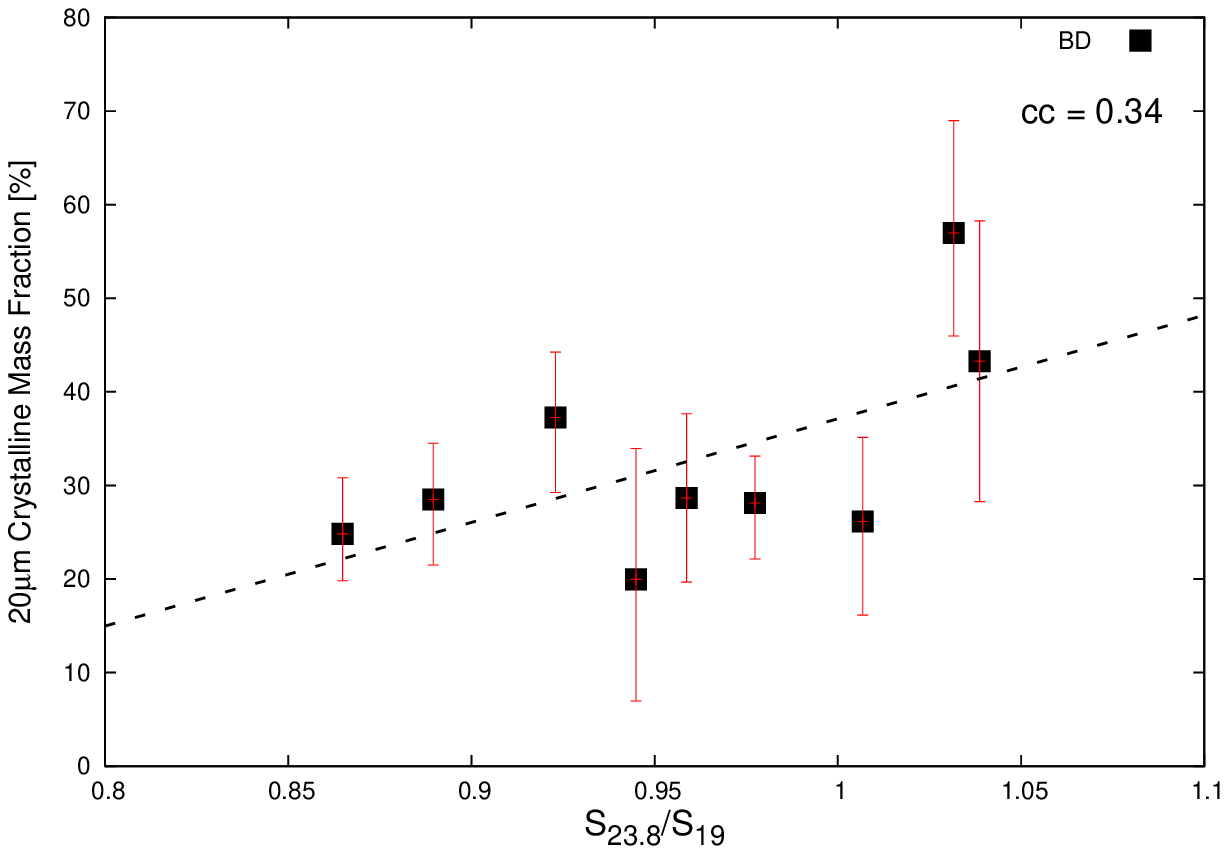} \\      
           \includegraphics[width=85mm]{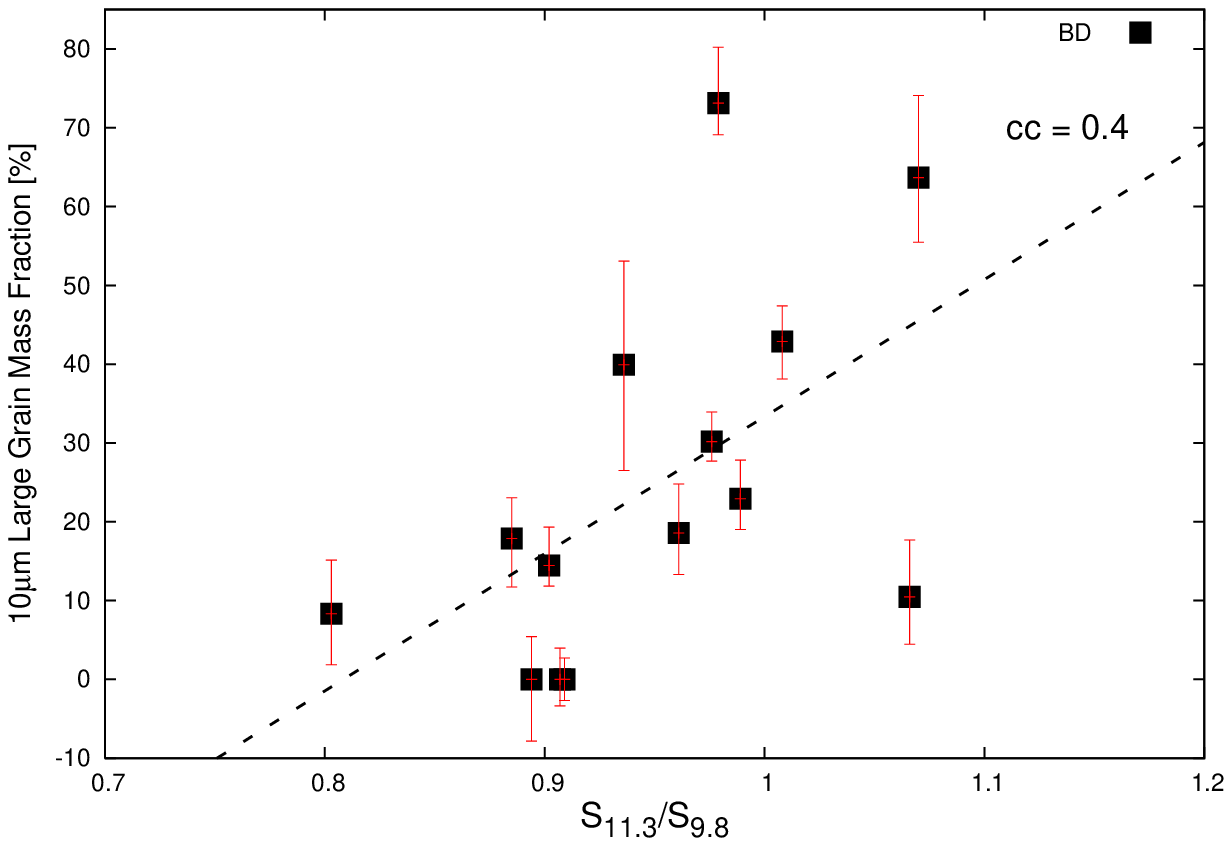} 
     \includegraphics[width=85mm]{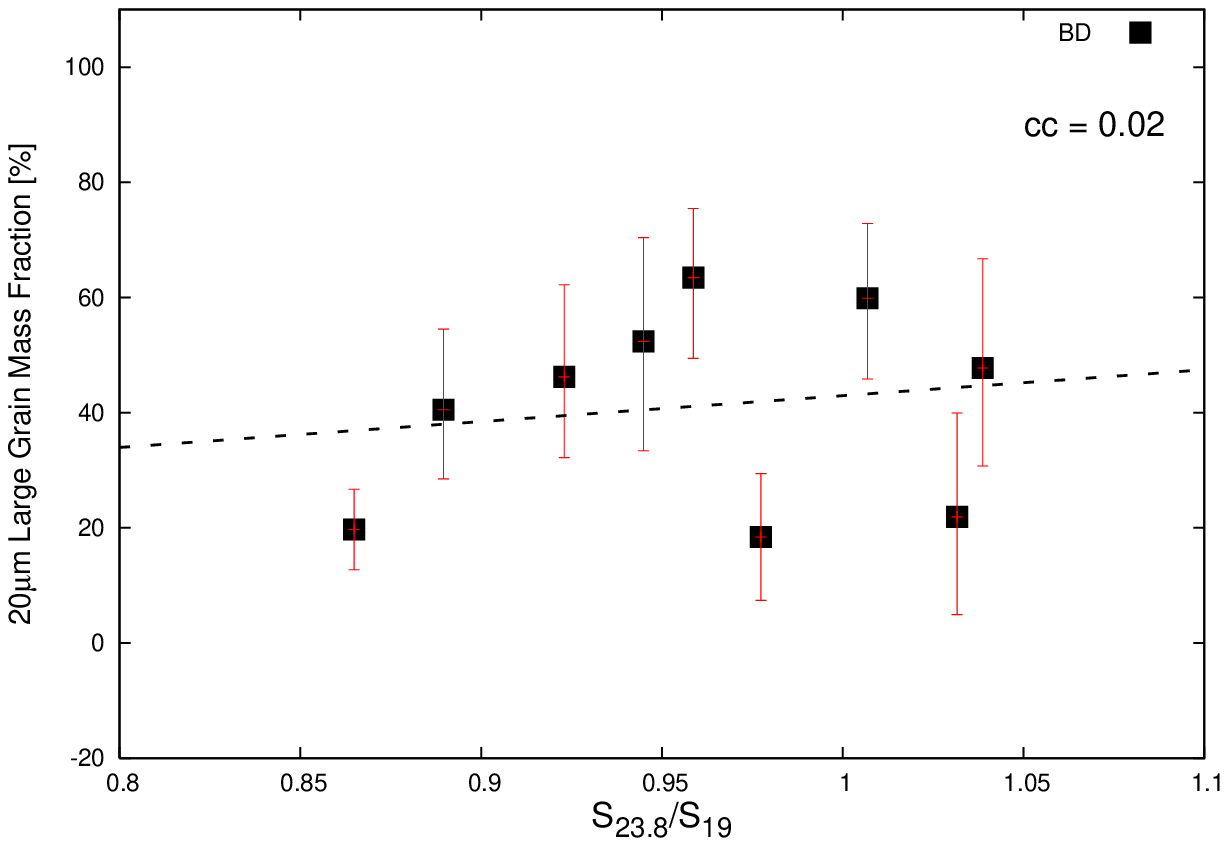} \\               
    \caption{The shape of the 10$\micron$ ({\it left}) and 20$\micron$ ({\it right}) features for brown dwarfs versus the crystalline and large-grain mass fractions.  } 
    \label{shape-dust}  
 \end{figure*} 
 
  \begin{figure*}     
     \includegraphics[width=85mm]{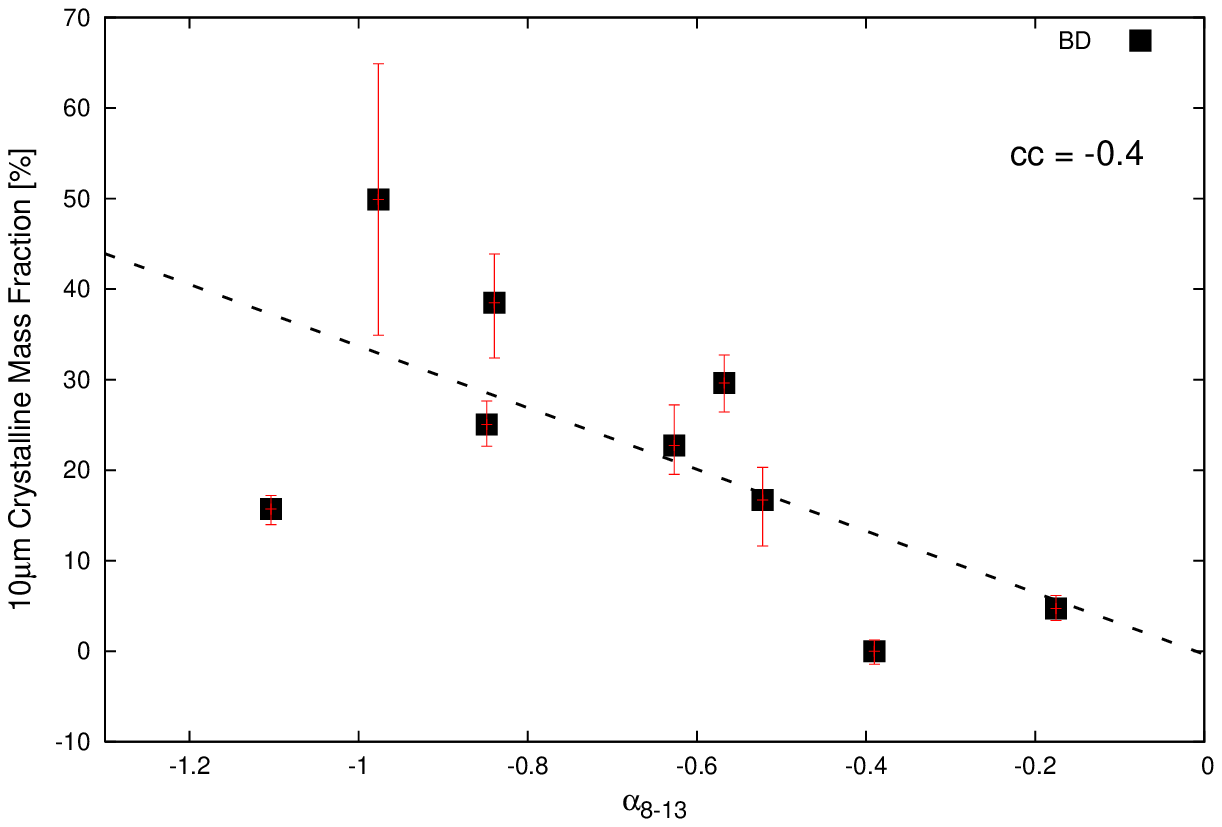}  
     \includegraphics[width=85mm]{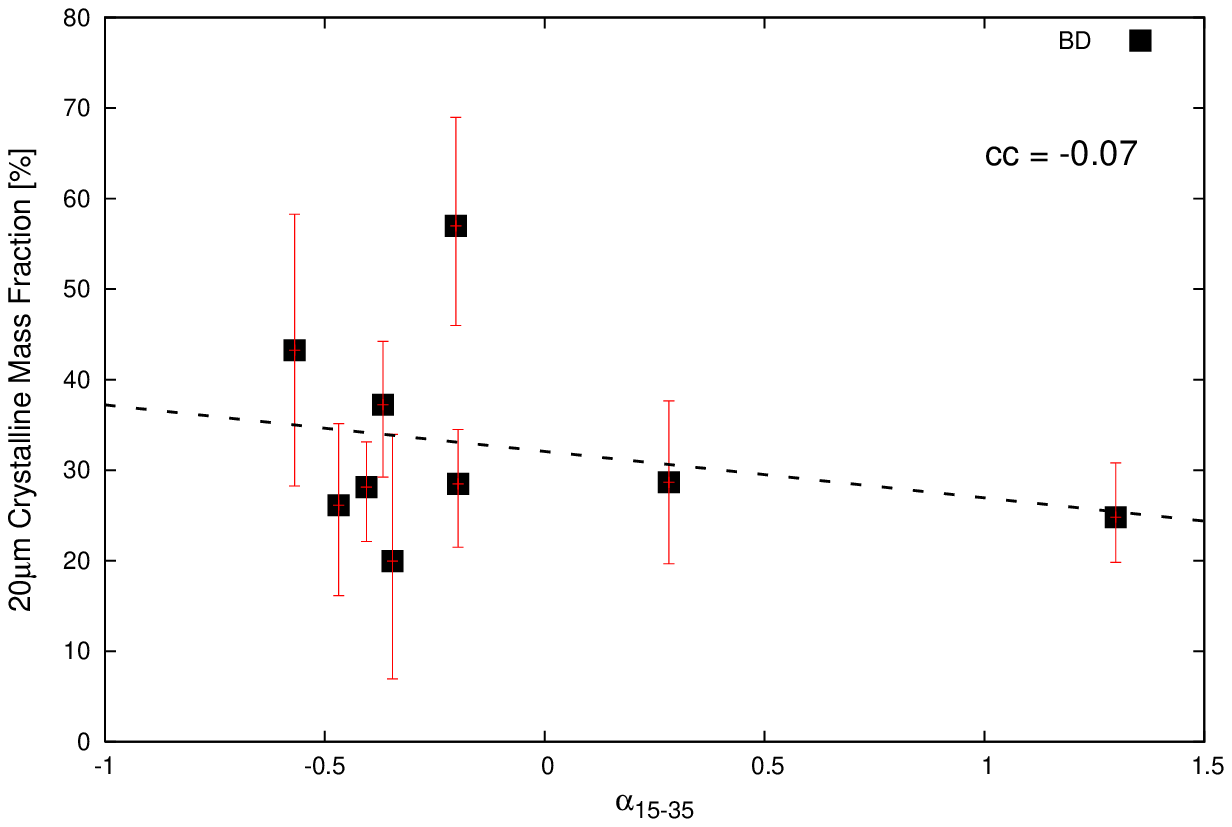} \\      
           \includegraphics[width=85mm]{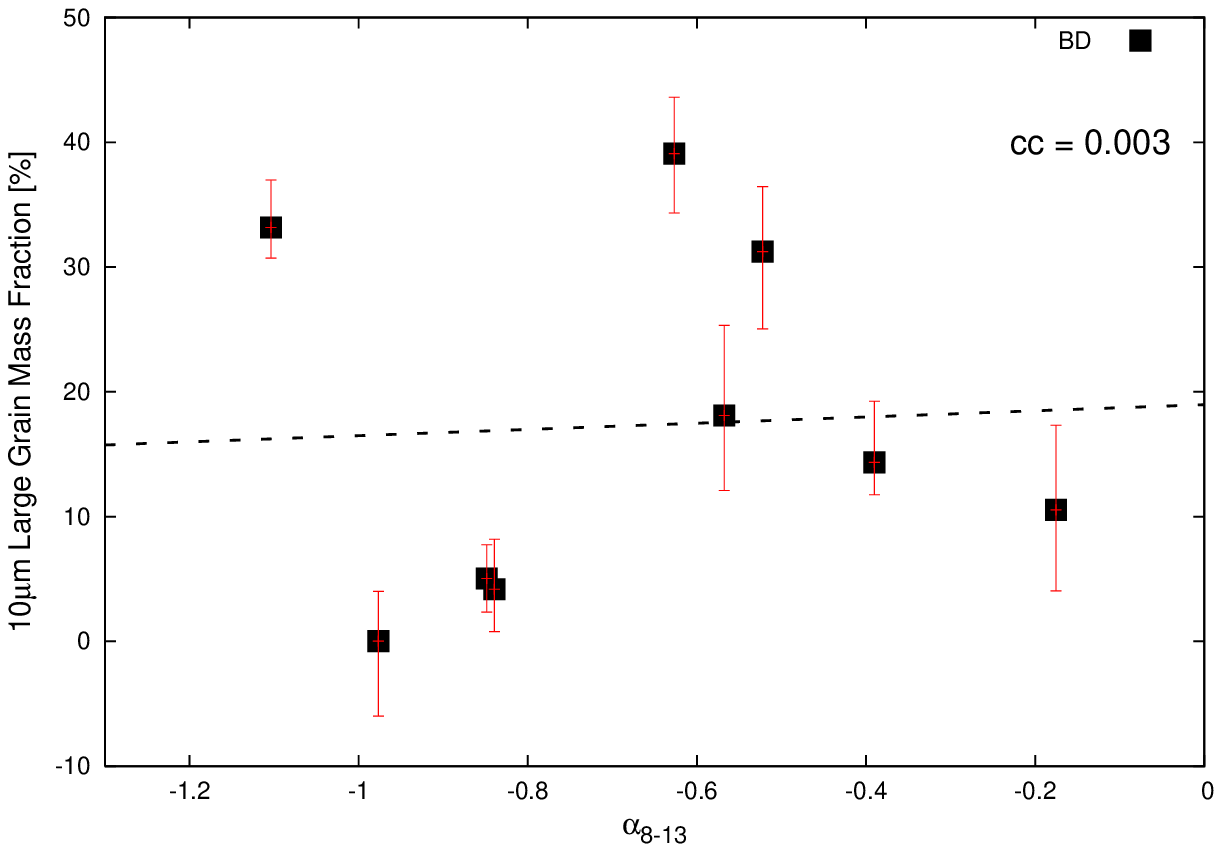} 
     \includegraphics[width=85mm]{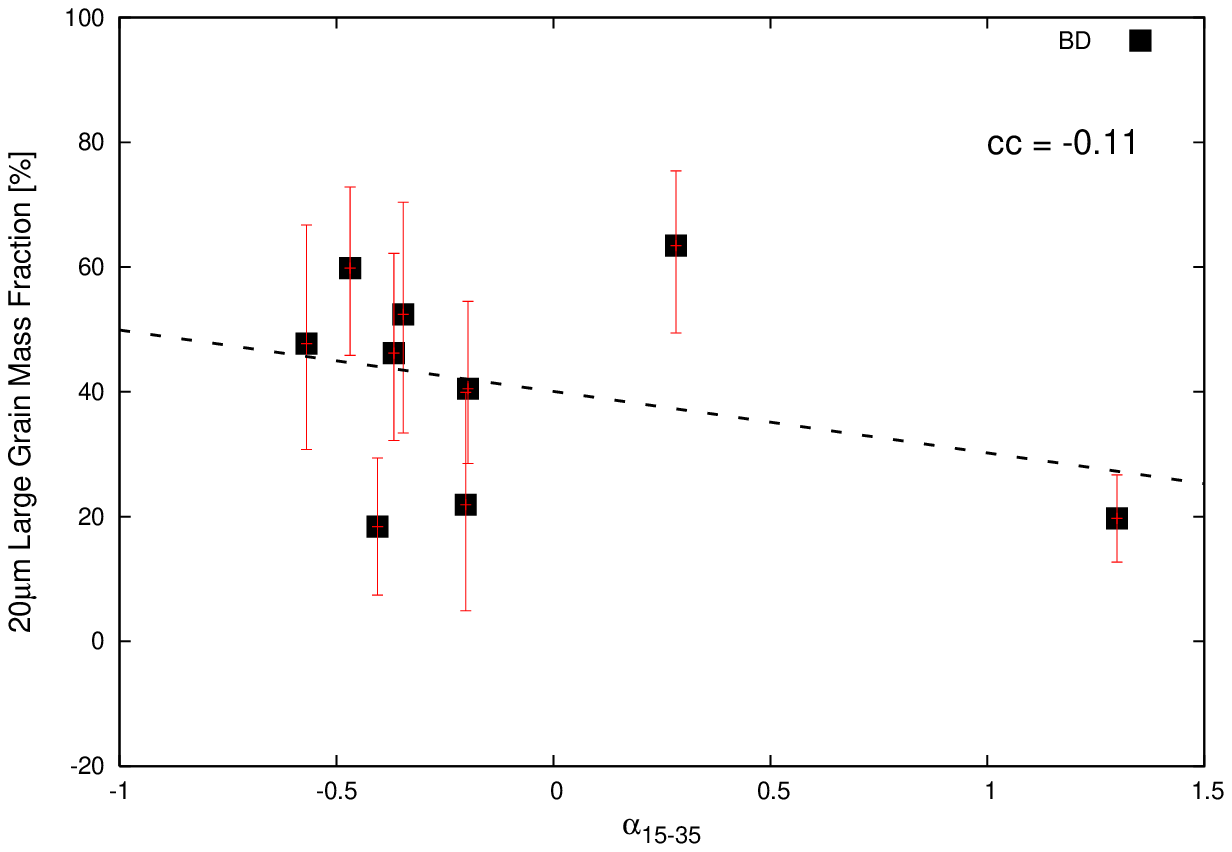} \\               
    \caption{A comparison of the crystalline and large-grain mass fractions for brown dwarfs with the slopes in the inner and outer disk regions, as indicated by the spectral index ($\alpha$) of the derived continuum from the 10 and 20$\micron$ features.  } 
    \label{flaring}  
 \end{figure*} 
 
   \begin{figure*}     
    \includegraphics[width=85mm]{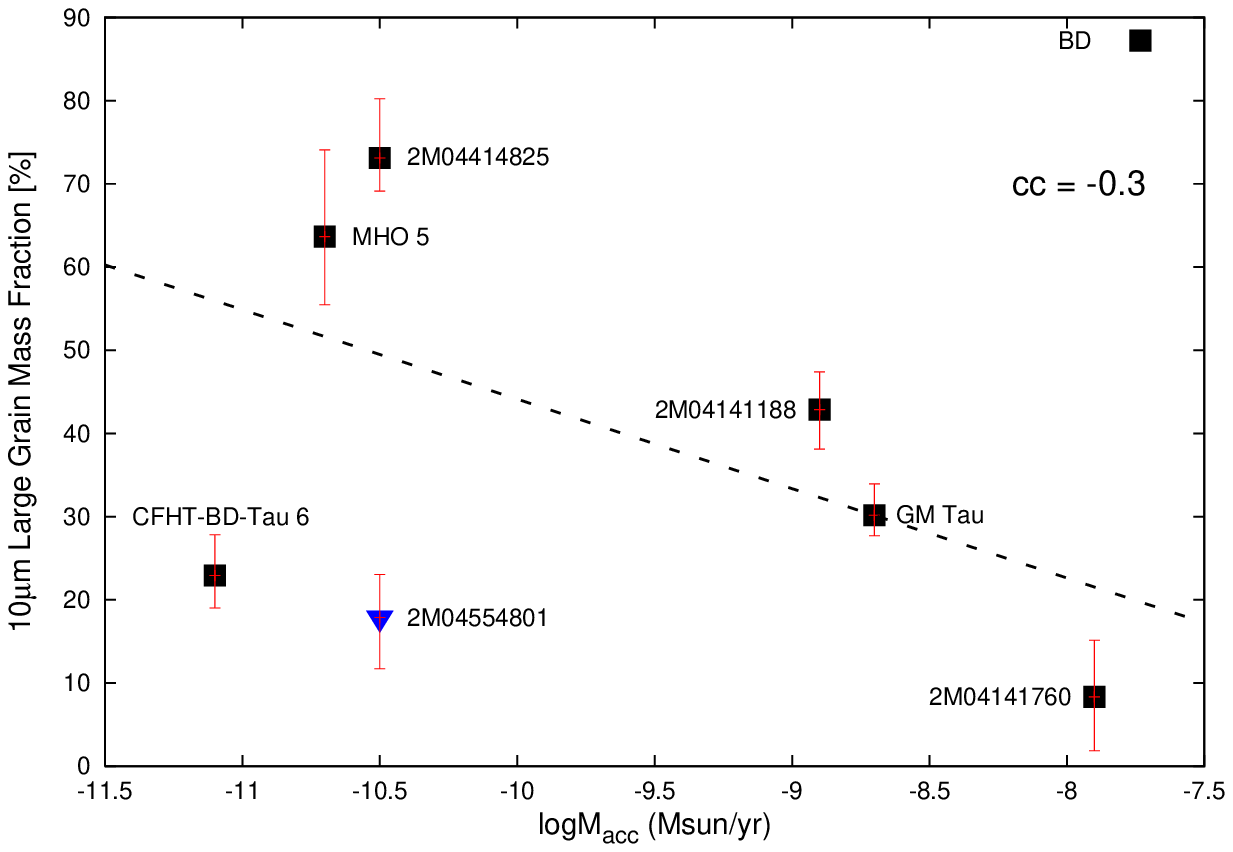} 
     \includegraphics[width=85mm]{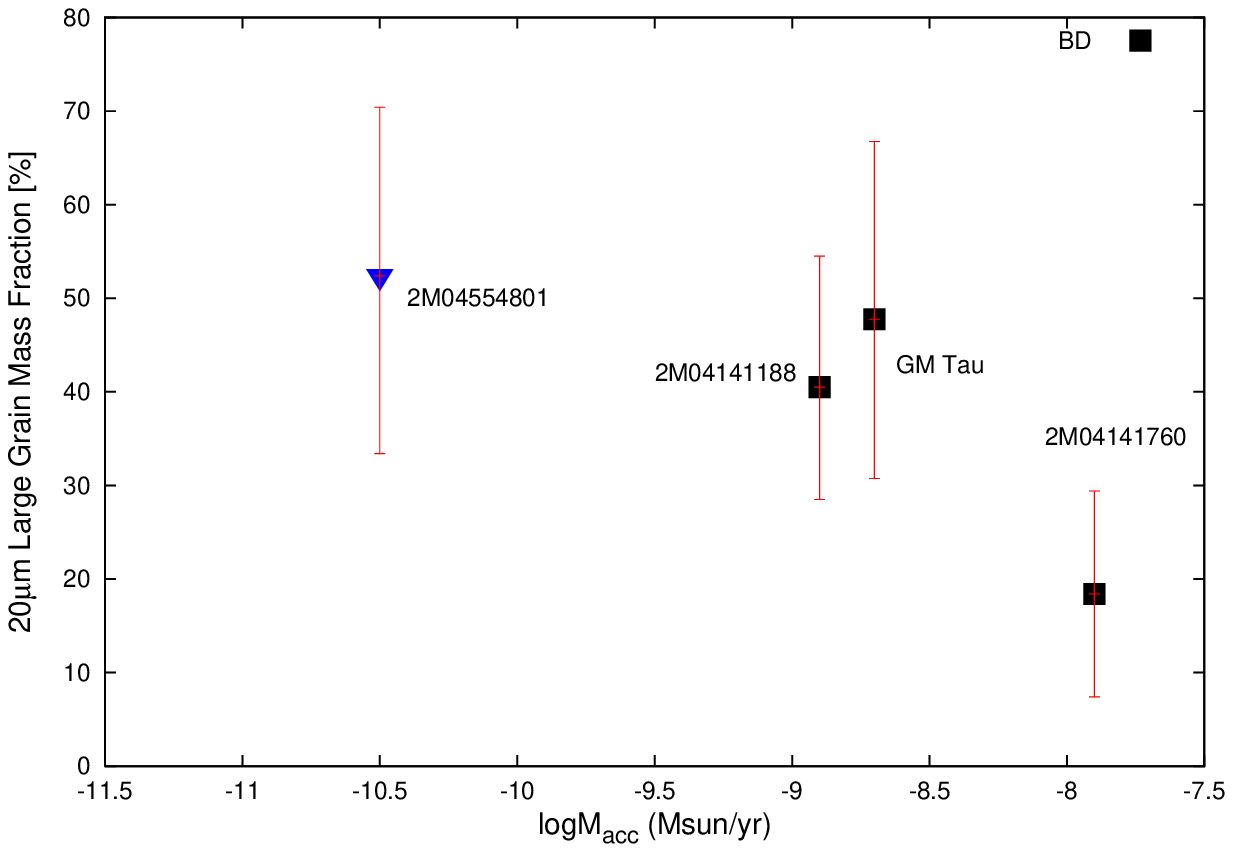} \\  
     \includegraphics[width=85mm]{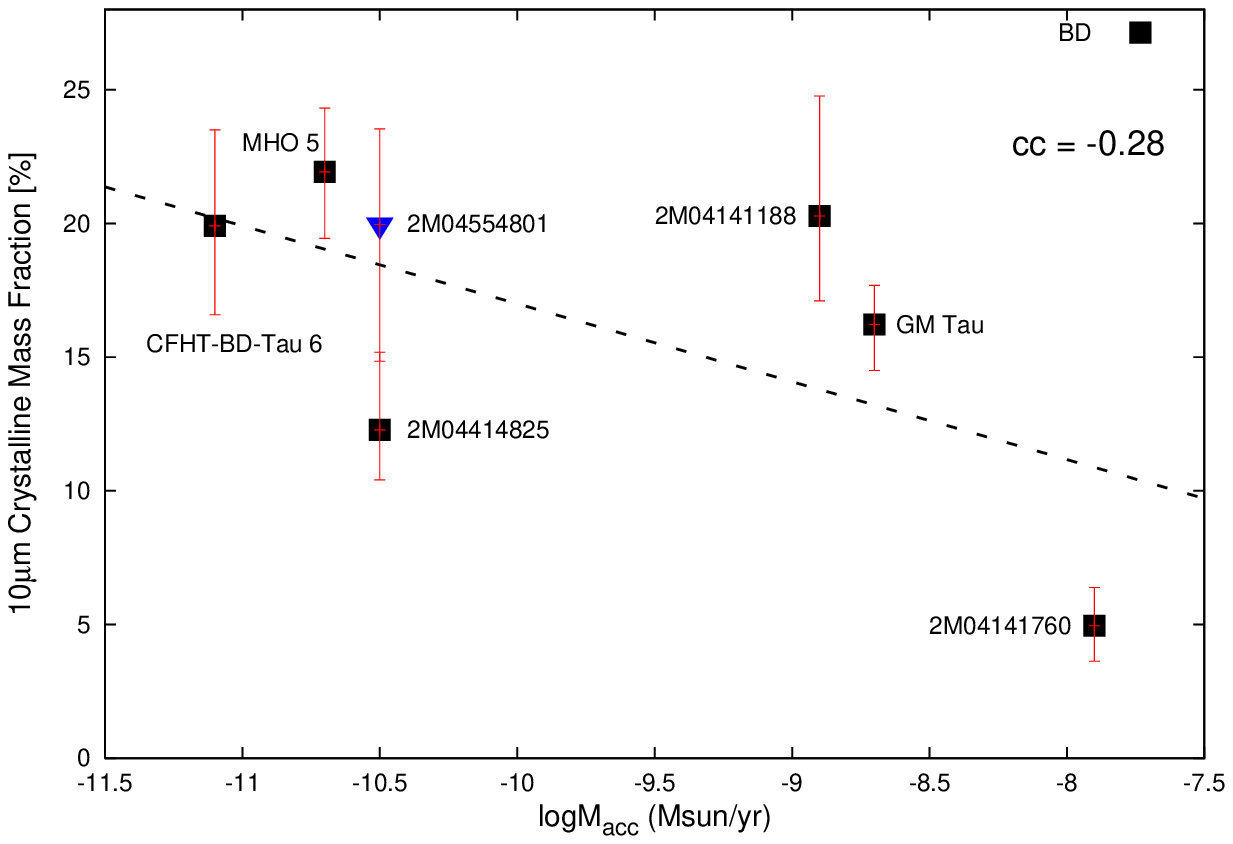}  
     \includegraphics[width=85mm]{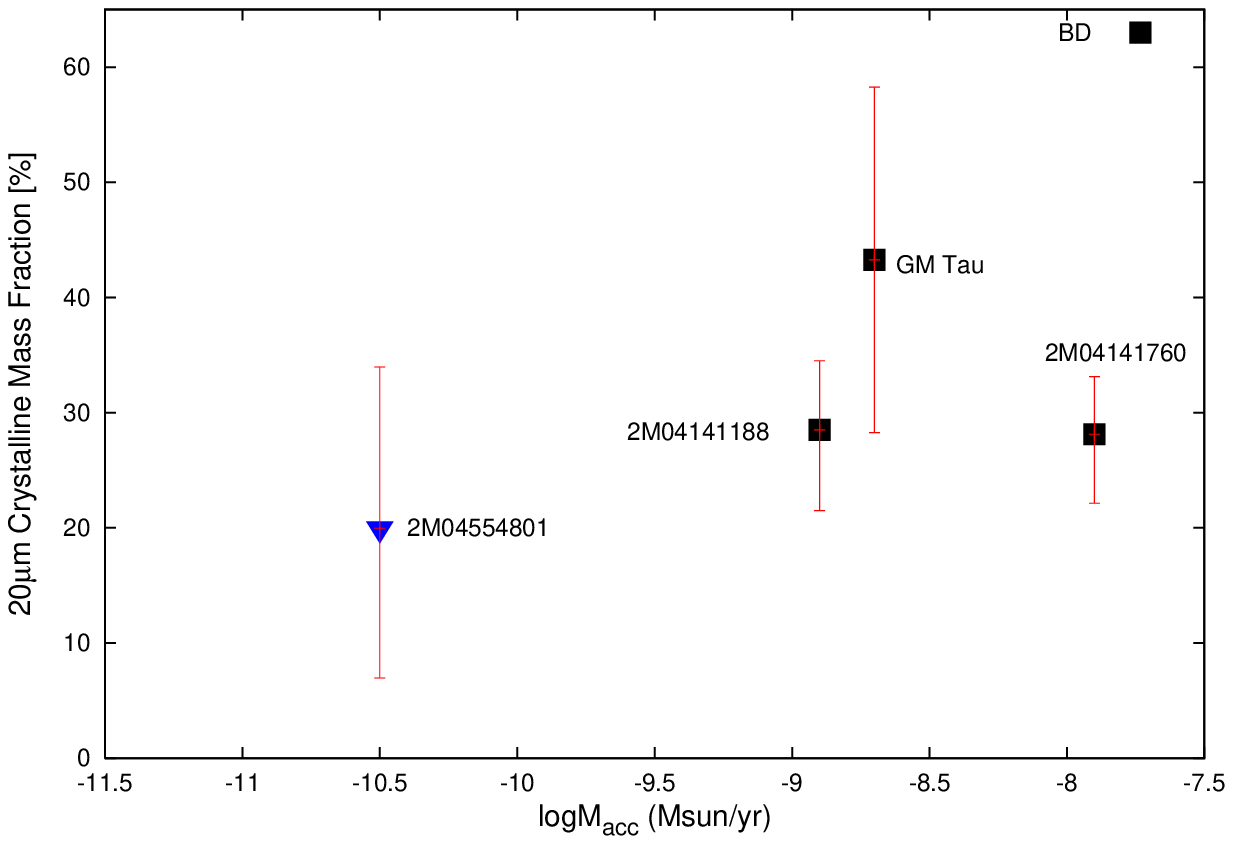} \\                   
    \caption{A comparison of the large-grain and crystalline mass fractions for brown dwarfs with the disk mass accretion rates. Upper limits are indicated by blue arrowheads.   } 
    \label{accretion}  
 \end{figure*} 

\begin{figure*}    
     \includegraphics[width=90mm]{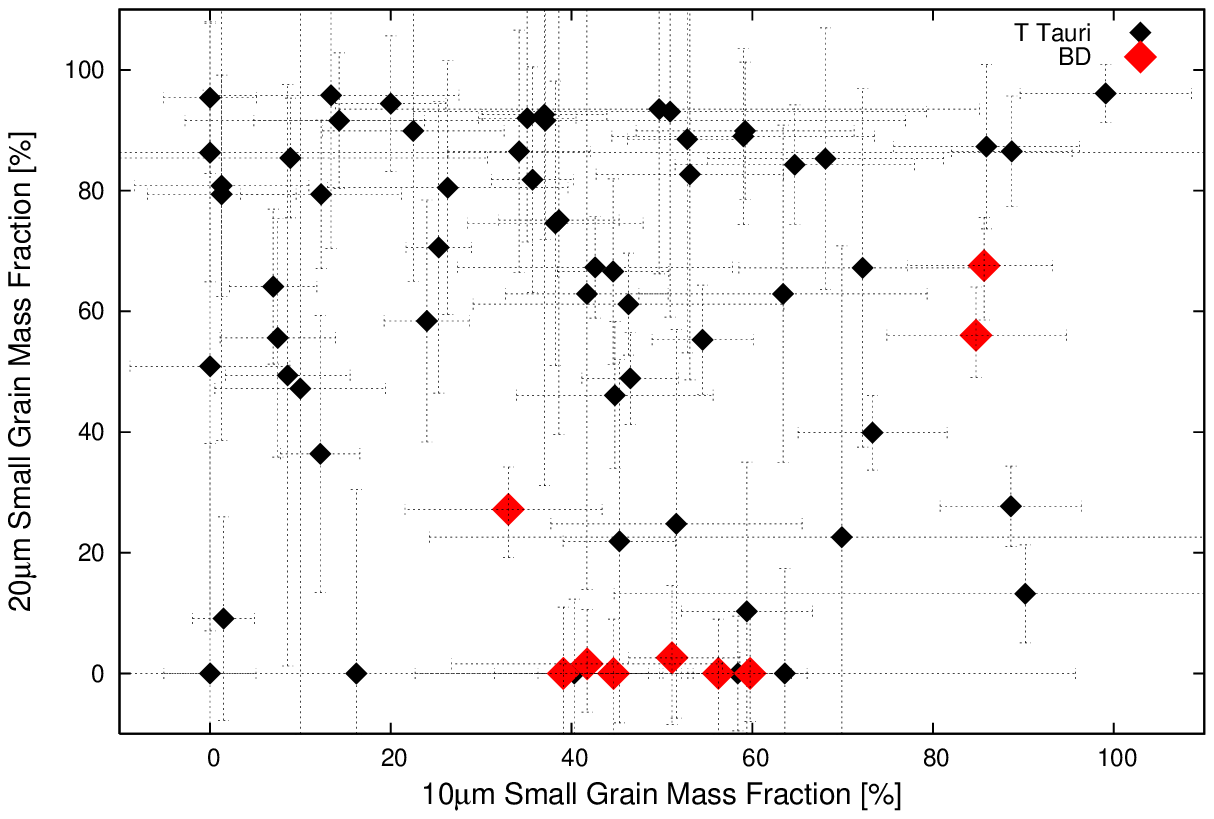} \\
     \includegraphics[width=90mm]{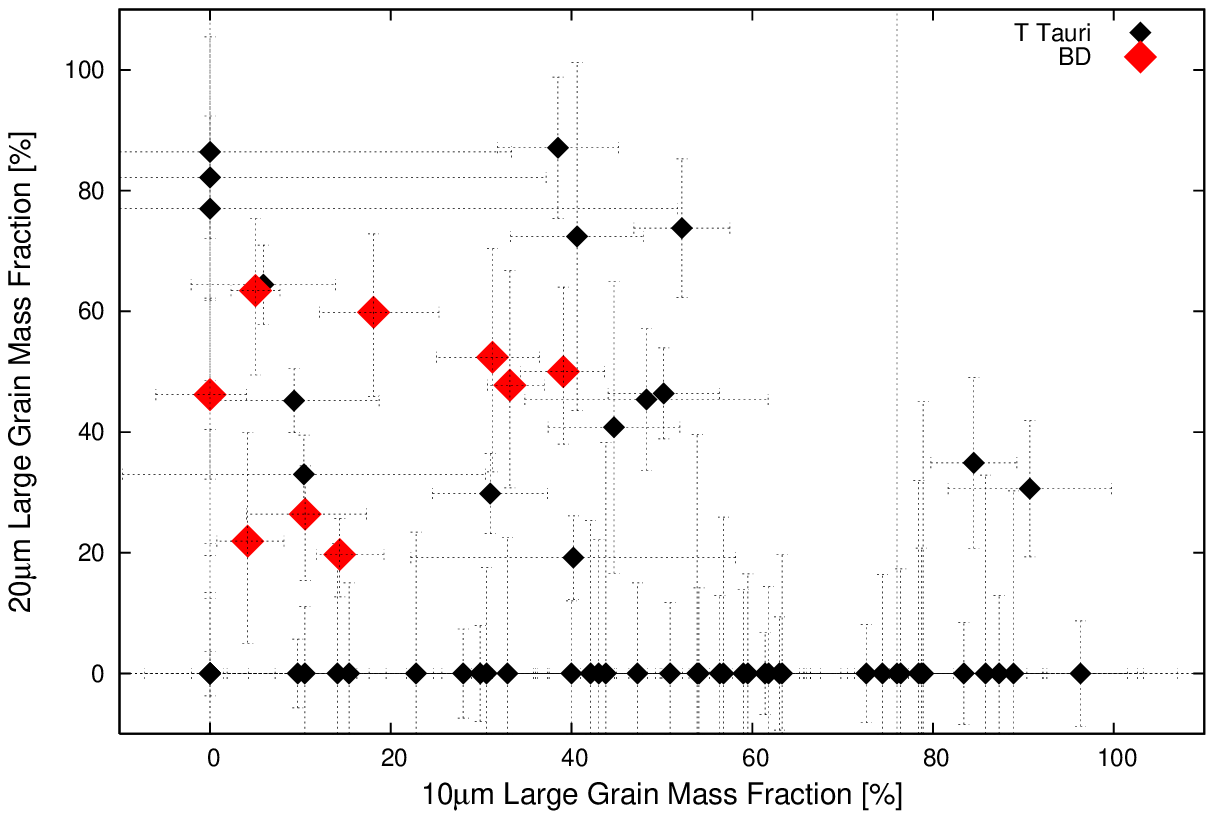} \\      
           \includegraphics[width=90mm]{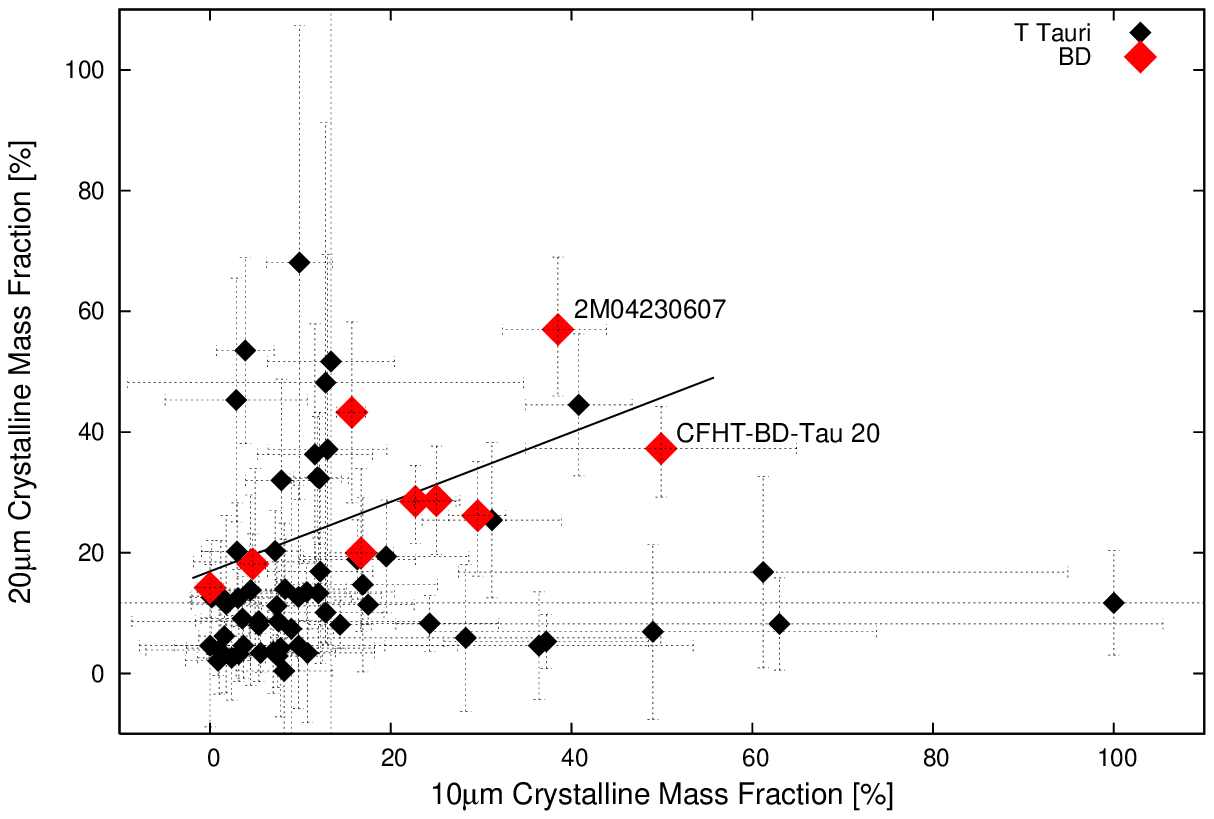} \\                                    
    \caption{A comparison of the 10 and 20$\micron$ mass fractions for the small amorphous [{\it top panel}, (a)], large amorphous [{\it middle panel}, (b)] and crystalline grains [{\it bottom panel}, (c)]. For the crystalline grains, the solid black line is a straight line fit to the brown dwarf sample. Brown dwarfs are denoted by red symbols, T Tauri stars by black. The data for T Tauri stars is from Sargent et al. (2009a).    } 
    \label{10vs20}  
 \end{figure*}

\begin{figure*}  
     \includegraphics[width=90mm]{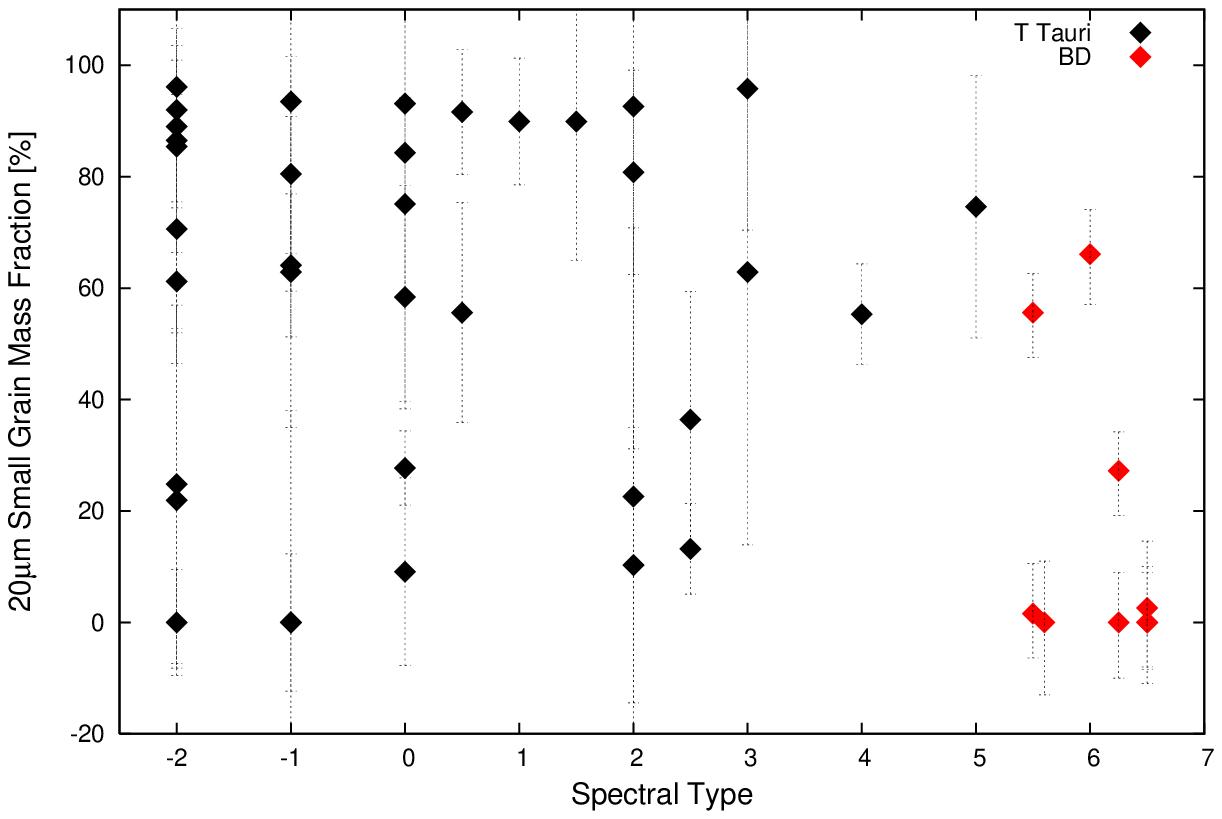} \\ 
     \includegraphics[width=90mm]{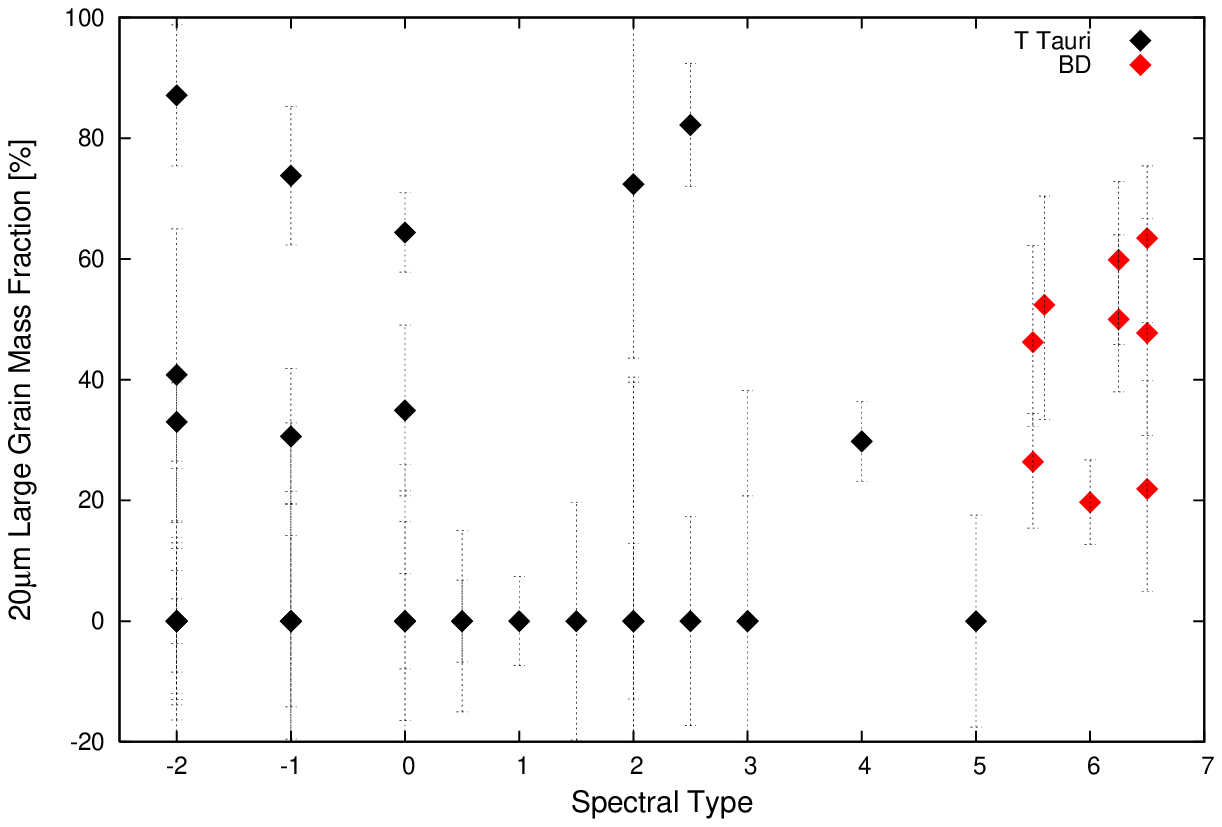} \\      
           \includegraphics[width=90mm]{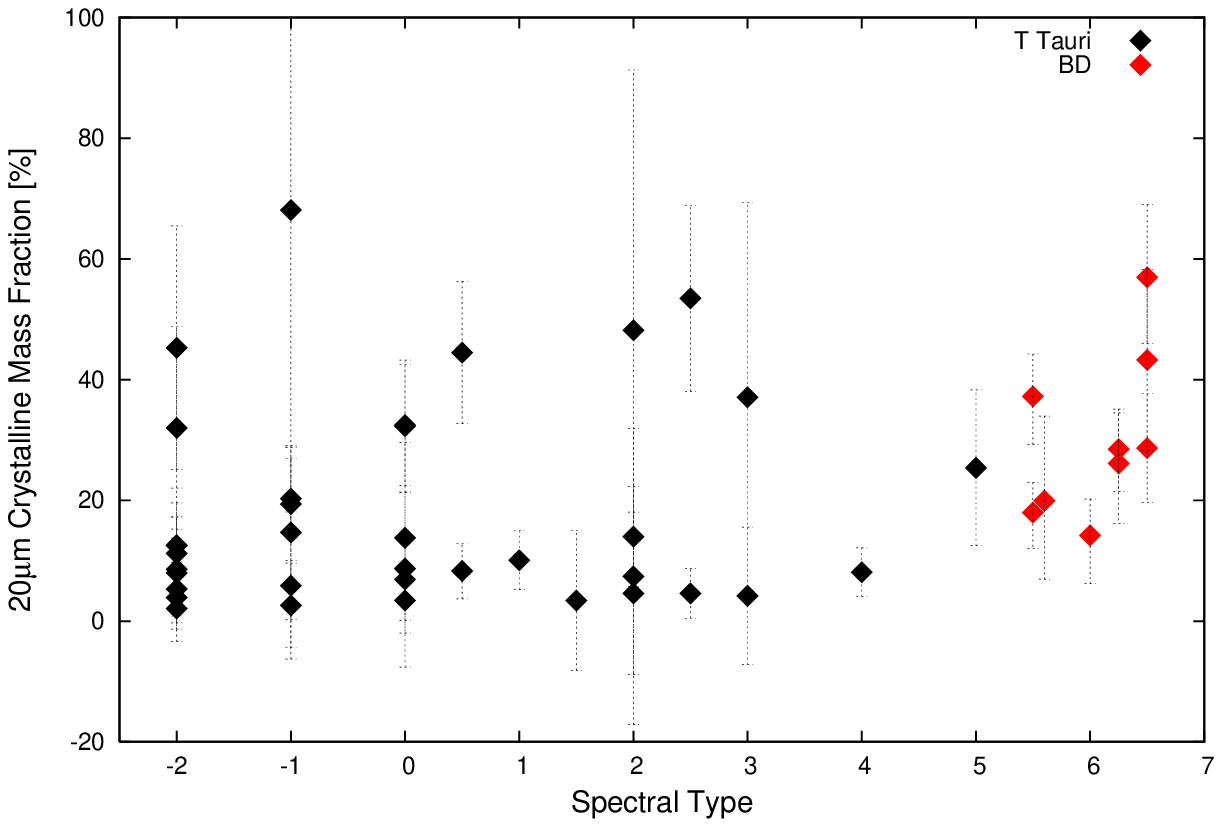} \\    
    \caption{SpT vs. the small grain [{\it top panel}, (a)], large grain [{\it middle panel}, (b)], and crystalline [{\it bottom panel}, (c)] mass fractions for T Tauri stars and brown dwarfs in Taurus. The value of -2 indicates a SpT of K7, -1 is K5, while 0-7 are M0-M7. Symbols are the same as in Fig.~\ref{10vs20}.  } 
    \label{spt-frac} 
 \end{figure*}

\begin{figure*}          
      \includegraphics[width=85mm]{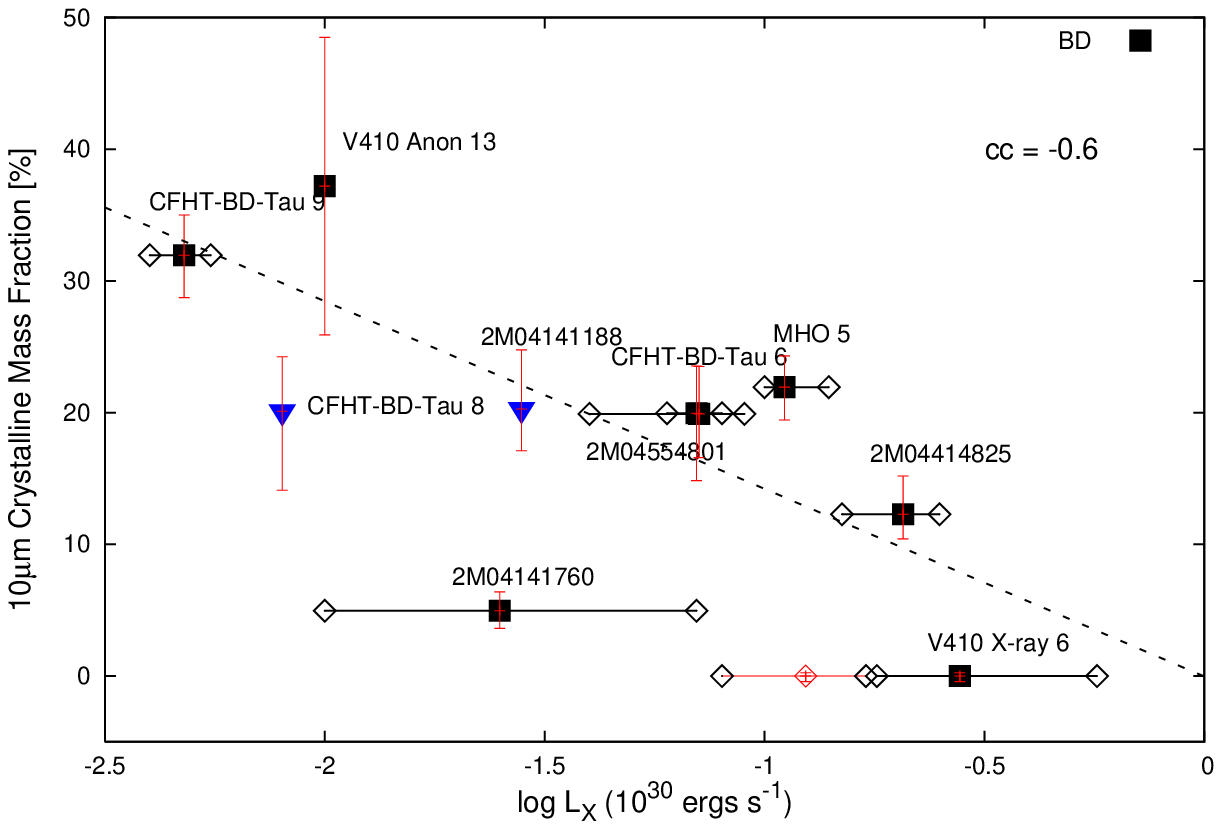}  
     \includegraphics[width=85mm]{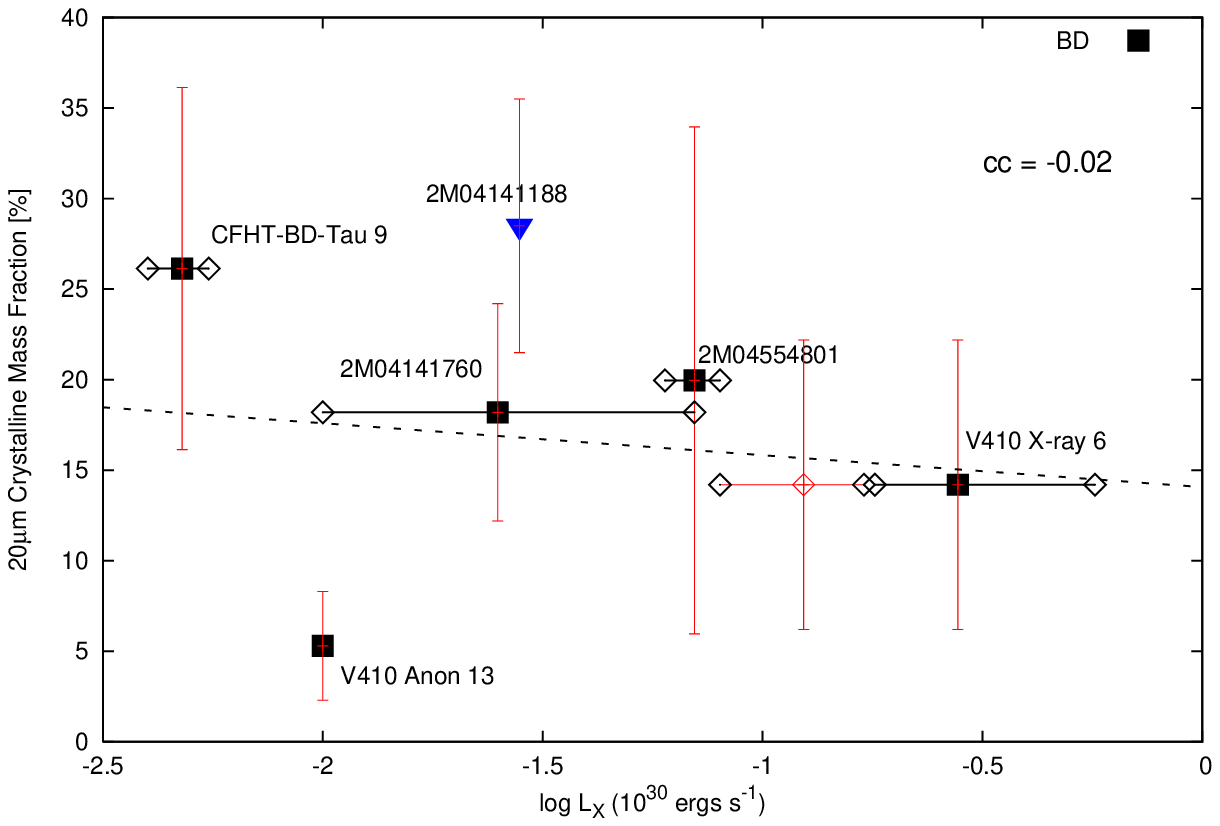} \\
      \includegraphics[width=85mm]{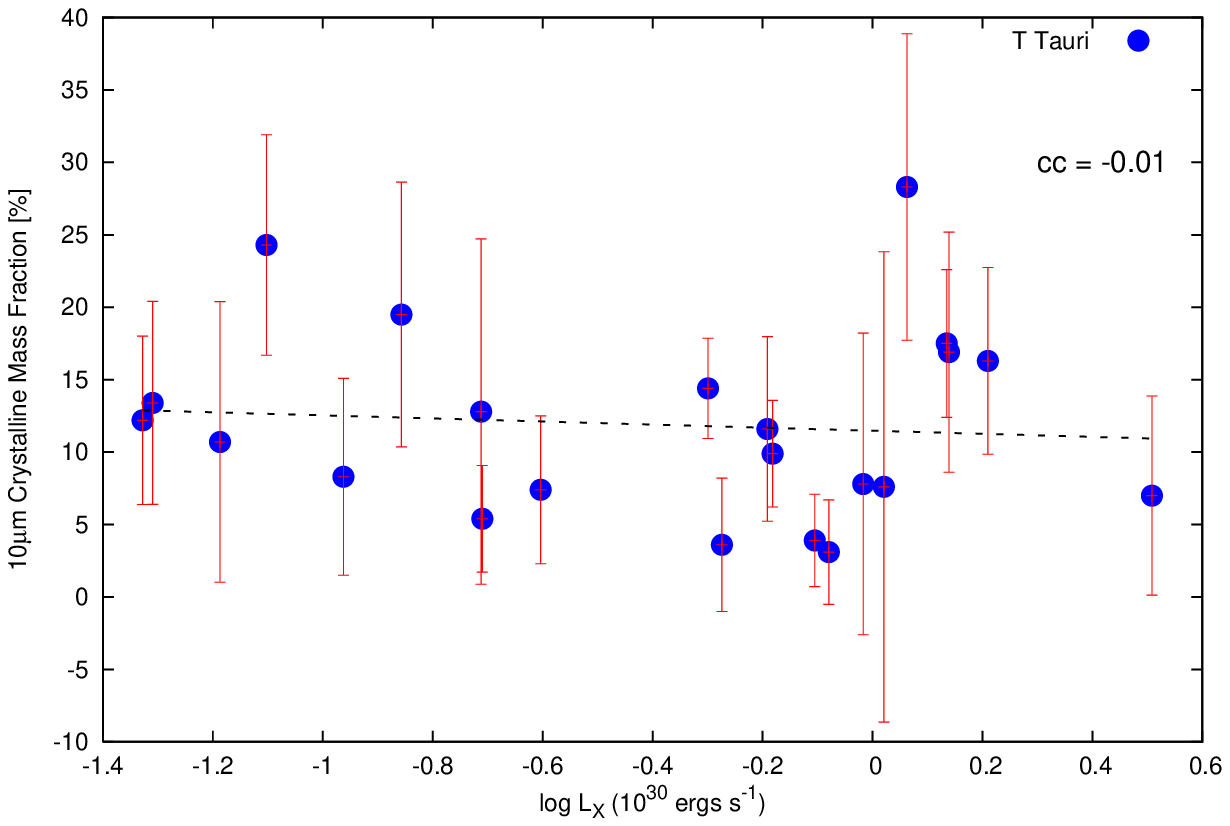}  
     \includegraphics[width=85mm]{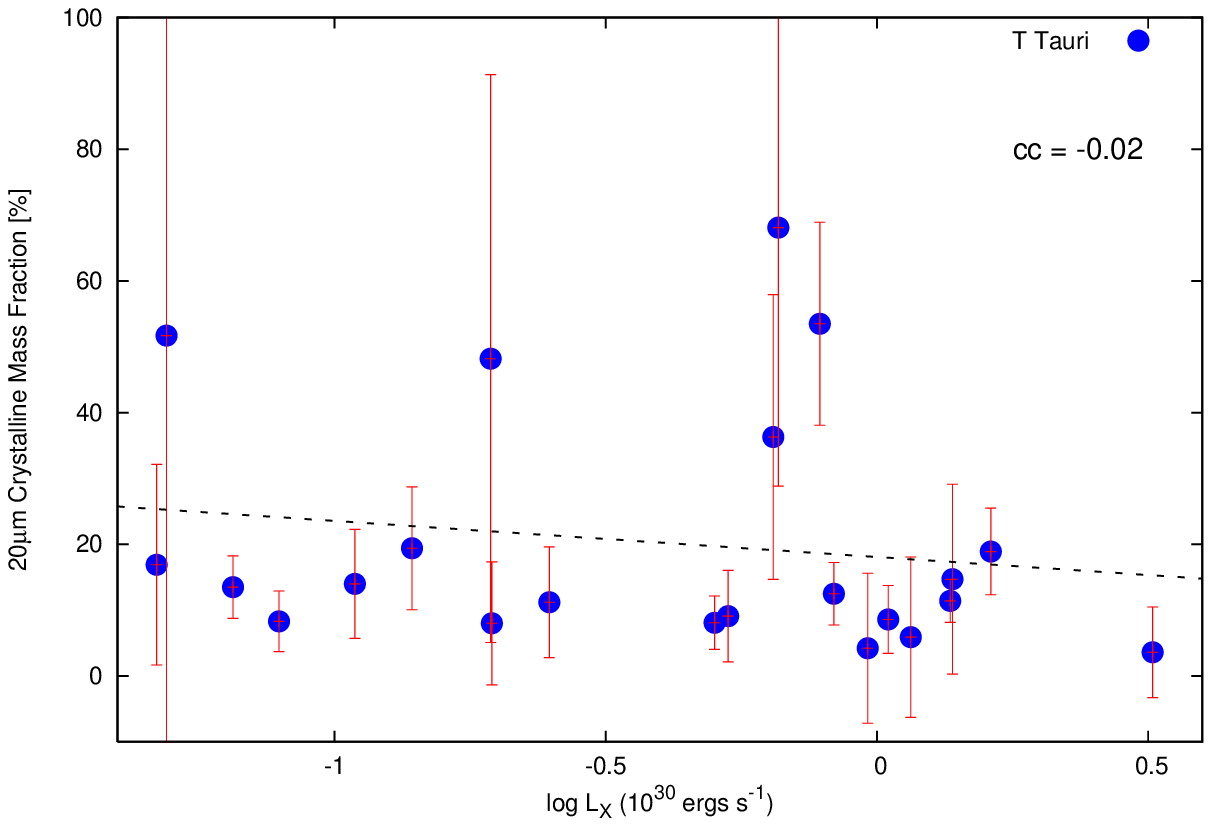} \\                  
    \caption{(a): The X-ray luminosity plotted against the crystalline mass fractions ({\it Top panel}: brown dwarfs; {\it Bottom panel}: T Tauri stars). Red open diamond denotes the quiescent state for V410 X-ray 6. Black open diamonds mark the range in X-ray emission. Upper limits are indicated by blue arrowheads.  } 
    \label{xray}  
 \end{figure*}

\begin{figure*}
\setcounter{figure}{9}          
      \includegraphics[width=85mm]{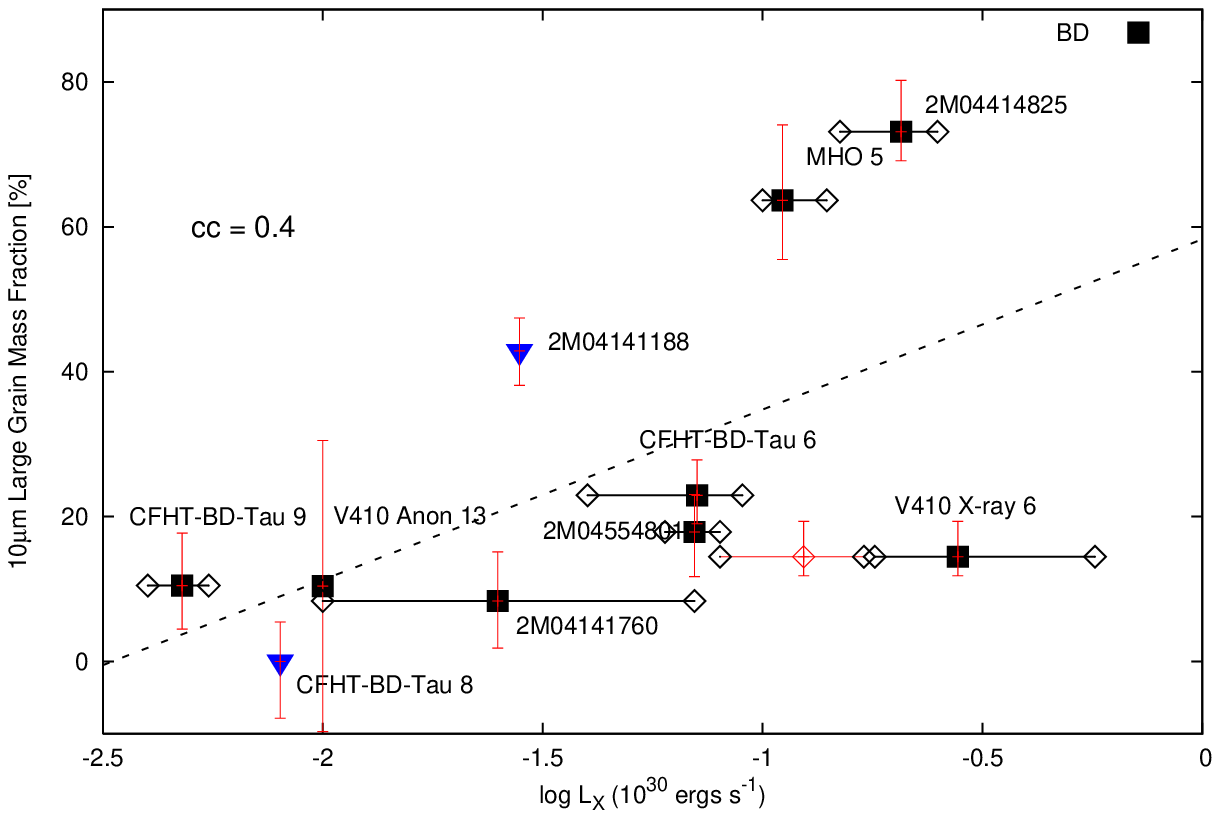}  
     \includegraphics[width=85mm]{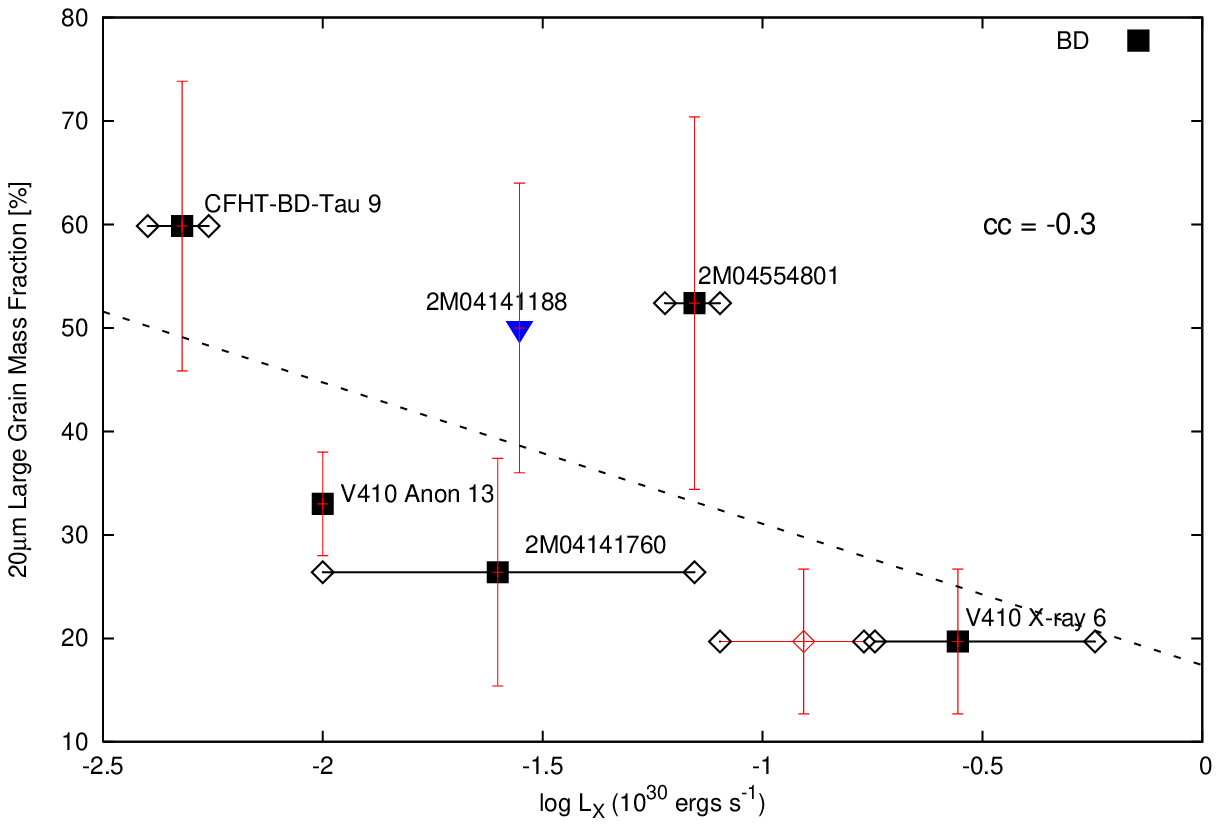} \\
      \includegraphics[width=85mm]{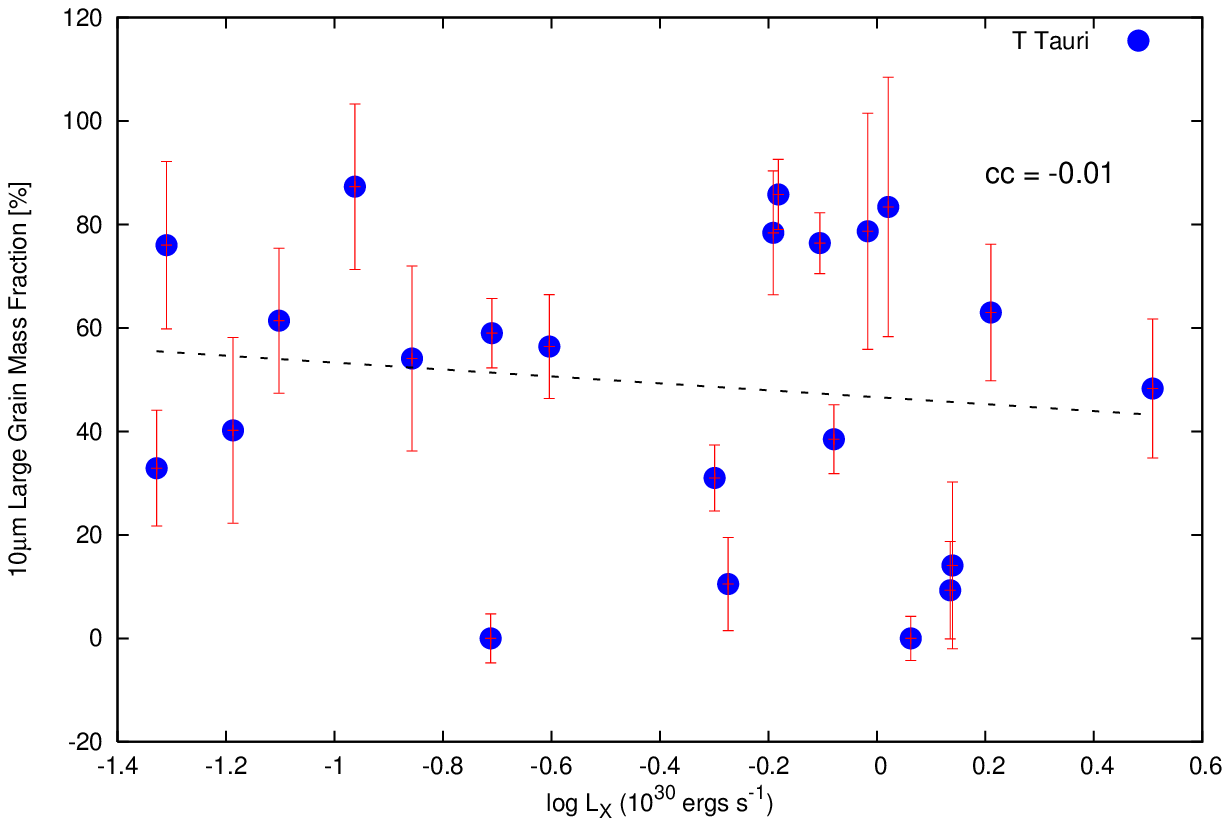}  
     \includegraphics[width=85mm]{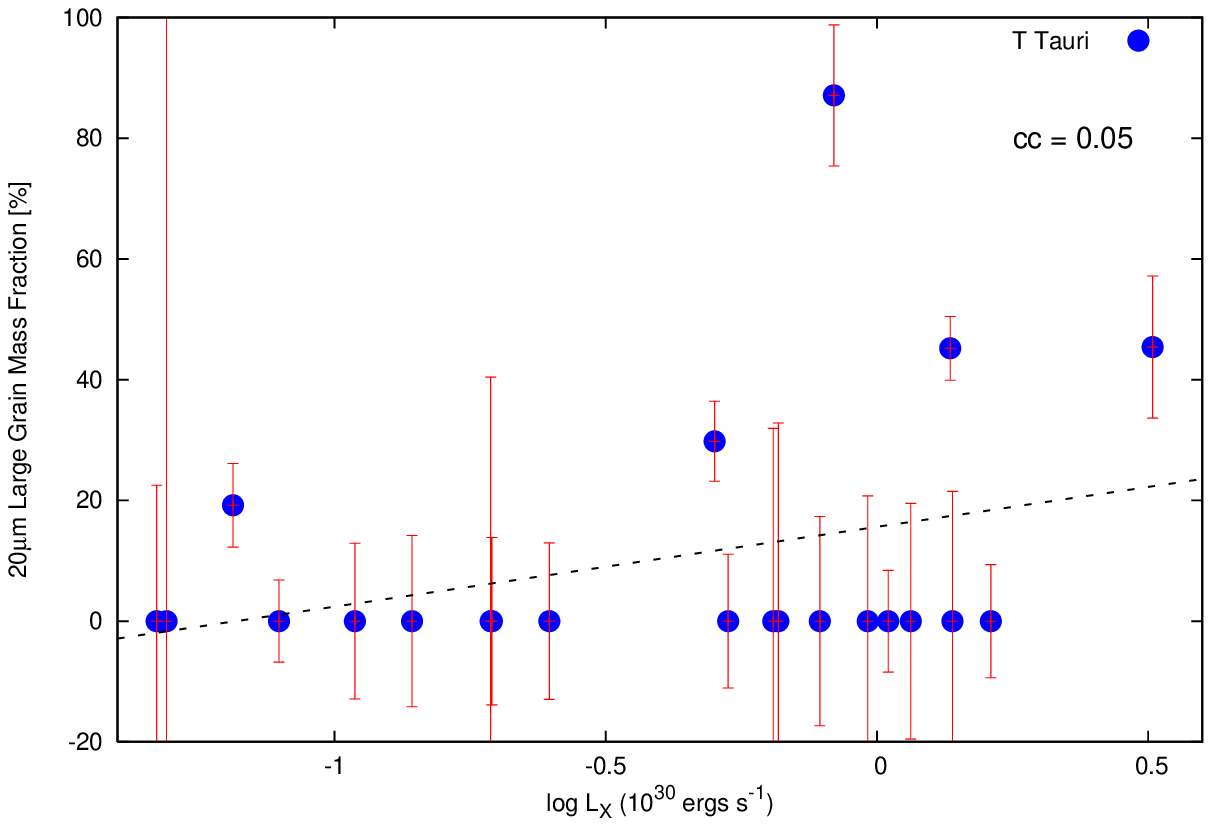} \\                
    \caption{(b): The X-ray luminosity plotted against the large-grain mass fractions ({\it Top panel}: brown dwarfs; {\it Bottom panel}: T Tauri stars). Symbols are the same as in Fig.~\ref{xray}a.  } 
    \label{xray}
 \end{figure*}

\begin{figure*}          
      \includegraphics[width=85mm]{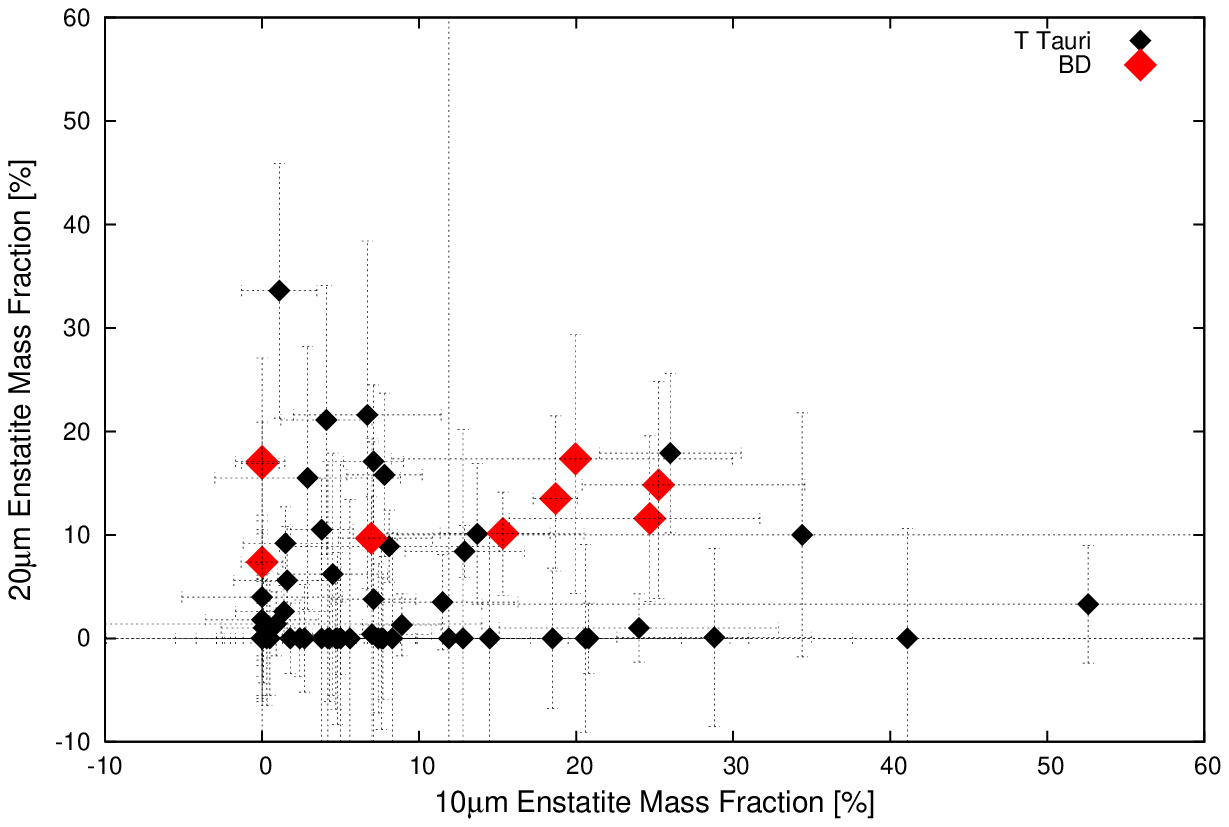} \\ 
     \includegraphics[width=85mm]{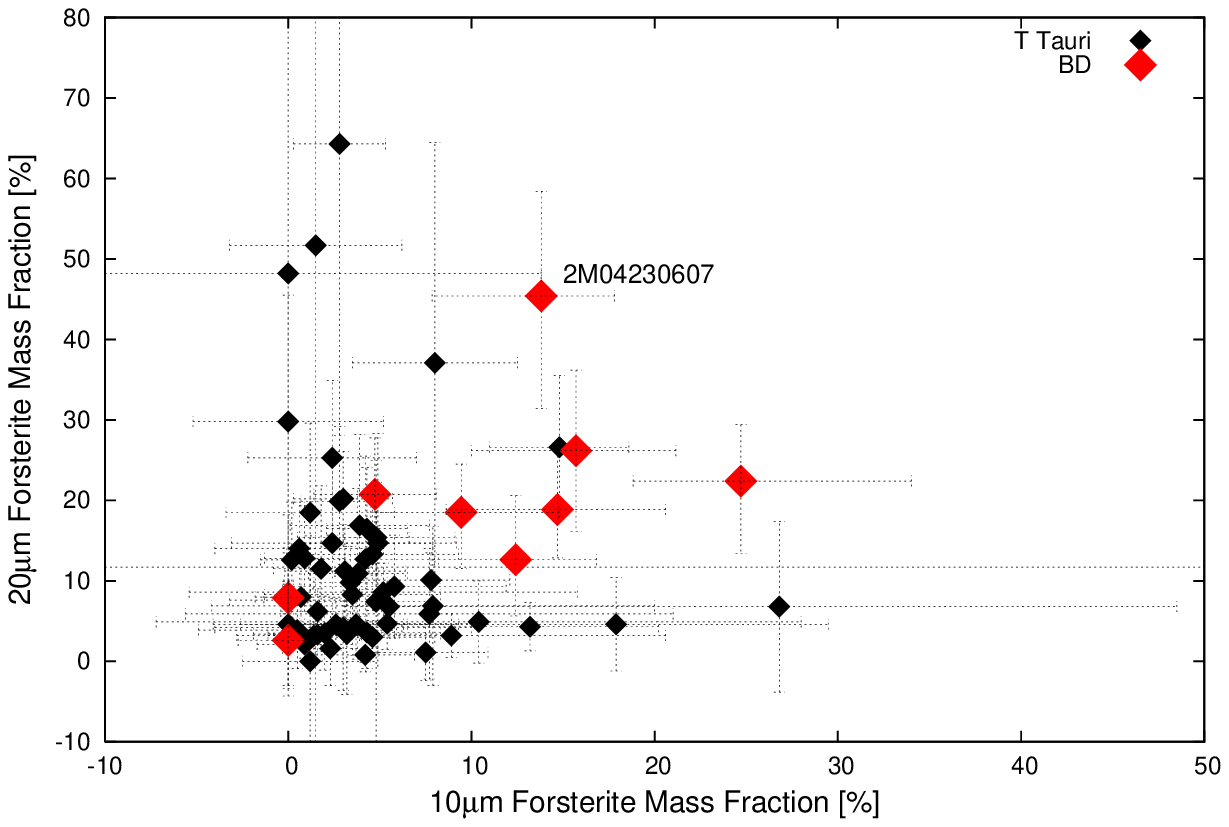} \\       
    \caption{A comparison of the 10 and 20$\micron$ mass fractions for the enstatite ({\it top panel}) and forsterite ({\it bottom panel}) crystalline silicates. Brown dwarfs are denoted by red symbols, T Tauri stars by black.   } 
    \label{ens-fors}
 \end{figure*}

\label{lastpage}

\end{document}